\documentstyle[epsf,12pt]{article}

\newlength{\figsize}
\begin{document}

\begin{titlepage}

\begin{tabbing}
\` {\sl hep-lat/9801008} \\
%    \\
\` Edinburgh preprint 98-1 \\
\` Oxford preprint OUTP-97-76P \\
\end{tabbing}
 
\vspace*{.1in}
 
\begin{center}
{\large\bf Topological Structure of the SU(3) Vacuum\\}
\vspace*{.7in}
{\large{\it UKQCD Collaboration}}\\
\vspace*{.2in}
Douglas A. Smith$^1$ and Michael J. Teper$^2$\\
\vspace*{.3in}
{$^1$Department of Physics, University of Edinburgh,\\ 
Mayfield Road, Edinburgh EH9 3JZ, U.K.\\
}
\vspace*{.2in}
{$^2$Theoretical Physics, University of Oxford,\\
1 Keble Road, Oxford, OX1 3NP, U.K.\\
}
\end{center}

\vspace*{0.8in}

\begin{center}
{\bf Abstract}
\end{center}

We investigate the topological structure of the vacuum in SU(3)
lattice gauge theory. We use under-relaxed cooling to remove
the high-frequency fluctuations and a variety of ``filters'' 
to identify the topological charges in the resulting
smoothened field configurations. We find a densely packed
vacuum with an average instanton size, in the 
continuum limit, of ${\bar\rho} \sim 0.5$ fm. 
The density at large $\rho$ decreases rapidly as $1/\rho^{\sim 11}$.
At small sizes we see some signs of a 
trend towards the asymptotic perturbative 
behaviour of $D(\rho) \propto \rho^6$. We find that an
interesting polarisation phenomenon occurs: the large topological
charges tend to have, on the average, the same sign and are over-screened 
by the smaller charges which tend to have, again on the average,
the opposite sign to the larger instantons. We also calculate
the topological susceptibility, $\chi_t$, for which we obtain
a continuum value of $\chi_t^{1/4} \sim 187 $ MeV. 
We perform the calculations for various volumes, lattice spacings
and numbers of cooling sweeps, so as to obtain some control over
the associated systematic errors.
The coupling range is $6.0 \leq \beta \leq 6.4$
and the lattice volumes range from $16^3 48$ to $32^3 64$.

\end{titlepage}

\setcounter{page}{1}
\newpage
\pagestyle{plain}

\section{Introduction}

SU(N) gauge fields in four Euclidean dimensions possess an integer 
topological charge $Q$
\cite{U1-Cole}.    
The topological fluctuations of the gauge fields are important
in QCD; for example they are the reason why the $\eta^{\prime}$ has
a mass $\sim 1$GeV rather than being a Goldstone boson
\cite{U1-H}.    
One can also argue that they have something to 
do with chiral symmetry breaking
\cite{Diakonov,Shuryak,SH-MT,ND-MT}
and that they may have a significant influence 
upon the hadron spectrum
\cite{Shuryak}.
The reason why topology might be able to do all this is that
an isolated instanton produces a zero-mode in the Dirac operator.
In the real vacuum these modes will mix with each other and shift away 
from zero. Just how they do so will determine their importance
for the physics described above. This mixing will be determined
by the topological structure of the vacuum; and in the first instance
by how large and densely packed are the component topological charges.
Although what one ultimately wants to know is what happens in the 
vacuum of QCD, the pure SU(3) gauge theory is also interesting; not
least because of its relevance to the physics of quenched QCD,
which seems to be a good approximation to the real world
\cite{Yoshie}.
Moreover in the case of the $\eta^{\prime}$ it turns out that 
it is the topological charge in the pure gauge theory that is
most relevant: one can use large-$N_c$ arguments 
\cite{U1-Wit,U1-Ven}
to relate the strength of the topological fluctuations, 
$\langle Q^2 \rangle$,
in the SU(3) gauge vacuum to $m_{\eta^{\prime}}$:
\begin{equation}
\chi_t \equiv {{\langle Q^2 \rangle}\over{volume}}
\simeq {{f_{\pi}^2}\over{2N_f}}
(m_{\eta^{\prime}}^2 + m_{\eta}^2 - 2 m_K^2)  
\sim (180 \ MeV)^4.     
\label{A1}
\end{equation}
Naturally this has long been a focus for lattice calculations
\cite{JH-MT}
and indeed it appears that eqn(\ref{A1}) is satisfied
\cite{JH-MT,MT-SU3,PISA-SU3}
as well as one could expect.

In this paper we attempt to see what one can learn about
the detailed topological structure of the SU(3) vacuum,
using simulations of the corresponding lattice theory. 
Some of our initial motivation was provided by an early
work of this kind
\cite{CM-PS}
in the SU(2) theory. Recently, several more detailed SU(2)
studies have appeared
\cite{Brower,DEG-SU2,DEF-SU2}
as well as preliminary reports
\cite{DS-MT,DEF-SU3,Boulder-SU3}
of some SU(3) work (including a brief summary of the work in this
paper). Most of these papers
appeared too recently to influence our work. For this reason
we shall not attempt to review them or to compare our
results in detail with theirs. However the reader should be aware
that there are some quite sharp disagreements within the most
recent SU(2) calculations. In particular between those studies that 
claim to find a relatively dilute gas of rather small instantons,
\cite{DEG-SU2}
and those that find a dense gas of considerably larger instantons
\cite{DEF-SU2}.
Naively, this difference would seem to be important to the physics 
that one derives from the topological structure; in particular 
the former picture fits better with instanton liquid models
\cite{Shuryak}.
This indicates that current lattice calculations of topological 
structure -- including this one -- should be regarded as exploratory. 

The work we do in this paper uses an ensemble of stored 
SU(3) field configurations that were generated by UKQCD
for other purposes. All were generated with a standard
plaquette action on periodic lattices. We shall analyse
100 $16^3 48$ and 50 $32^3 64$ lattice field configurations
at $\beta=6.0$, 100 $24^3 48$ configurations at $\beta=6.2$,
and 20 $32^3 64$ configurations at $\beta=6.4$. The
field configurations are typically separated by 800 to 2400 
Monte Carlo sweeps and therefore represent approximately
independent snapshots of the vacuum. The lattice
spacing $a$ decreases by almost a factor of two over our
range of $\beta$ and so this will allow us some control
over the continuum limit of the theory. At the same time,
the two quite different volumes at $\beta=6.0$ will
allow us some control over the thermodynamic limit.

In the next section we discuss topology on the lattice
and introduce the cooling algorithm which we use to
reveal the topological charge density, $Q(x)$. Once we 
have $Q(x)$, we then want to decompose it into a sum of
instantons and anti-instantons of various sizes. For
the densely packed vacuum that we find, this represents
a difficult pattern recognition problem. We shall
provide a sequence of procedures - one might call them
filters - which are designed to solve this problem.
These procedures are necessarily approximate and
the details can be tedious, but they are essential for
anyone who wishes to reproduce our calculations.
For this reason we shall relegate some of the
technical details to the appendix. There 
follow two sections describing the main results of our
investigation of the vacuum topological structure. This
will include the instanton size density, $D(\rho)$,
with a particular emphasis on the mean instanton size,
the functional form of the small-$\rho$ tail, where
asymptotic freedom makes asymptotic predictions, and
the large-$\rho$ tail, which is determined by analytically
incalculable infrared effects. We then investigate
the correlations between topological charges. Here
we find a quite striking long-distance polarisation
phenomenon which has not, as far as we are aware,
been remarked upon before. The next section contains
our calculation of the continuum topological susceptibility,
something which is free of the many uncertainties that
adhere to our calculations of the vacuum structure.
We finish with some conclusions. Throughout the paper we
attempt to point out how our study can and should be improved.

\section{Topology of lattice gauge fields}

Two continuous gauge fields that have different topological 
charges cannot be continuously deformed into each
other. When we discretise space-time, however, the fields are
no longer continuous and the notion of topology becomes
ambiguous. Nonetheless, because the theory is renormalisable
(and because the lattice is surely a good regulator) it must be 
the case that we recover all the usual topological properties
as the lattice spacing vanishes, $a\to 0$. (For a brief
discussion of this issue see 
\cite{MT-newton}.)
In this section we summarise some relevant properties
of continuum topology and some of the problems that arise when
gauge fields are regularised onto a space-time lattice. We 
focus on one approach to solving these problems, `cooling'
\cite{cool-MT,JH-MT},
and then motivate in some detail the particular version of cooling
that we shall use in this paper.

\subsection{topology of continuum fields}

The topological charge, $Q$, of a gauge field can be expressed
as the integral over Euclidean space-time of a topological
charge density, $Q(x)$, where
\begin{equation}
Q(x)= {1\over{32\pi^2}} \epsilon_{\mu\nu\rho\sigma}
Tr\{F_{\mu\nu}(x)F_{\rho\sigma}(x)\}.
\label{A2}
\end{equation}
The minimum action field configuration with $Q=1$ is the instanton.
The action and topological charge density are localised within
a core of size $\rho$. At the classical level the theory is scale 
invariant and so all sizes are possible and the action is independent 
of $\rho$. The gauge potential of an instanton of size $\rho$ 
centered at $x=0$ is given by 
\begin{equation}
A_\mu^I(x) = {{x^2}\over{x^2+\rho^2}}g^{-1}(x)\partial_\mu g(x)
\label{A3}
\end{equation}       
where
\begin{equation}
g(x) = {{x_0+ix_j\sigma_j}\over{{(x_{\mu}x_{\mu})}^{1/2}}}
\label{A4}
\end{equation}       
up to a gauge transformation. These expressions are for SU(2);
they can be trivially extended to SU(3) by embedding the
SU(2) fields into SU(3) fields. 

In the semiclassical limit a field of charge $Q$ 
will typically contain $n_I$ instantons and $n_{\bar I}=n_I-Q$ 
anti-instantons, all of which are well separated. In this dilute
gas approximation, the average density of instantons will depend 
on $\rho$ as 
\begin{equation}
D(\rho) d\rho = {{d\rho}\over{\rho}}{1\over{\rho^4}}
e^{-{{8\pi^2}\over{g^2(\rho)}}} ....
\label{A5}
\end{equation}          
where the `...' represents factors varying weakly with $\rho$. We 
recognise in this equation the scale-invariant integration measure; 
also a factor to account for the fact that a ball of volume $\rho^4$
can be placed in $1/\rho^4$ different ways in a unit volume;
and finally a factor arising from the classical instanton
action, $S_I=8\pi^2/g^2(\rho)$. 

Note that at this point we have departed from the classical 
calculation: perturbative fluctuations around the instanton
break scale-invariance, promoting the bare $g^2$ to a running 
$g^2(\rho)$ in the usual way. This is crucial. When we insert 
the asymptotically free form of the coupling, we
obtain
\begin{equation}
D(\rho)  \propto \Bigl({\rho\over\xi}\Bigr)^6.
\label{A6}
\end{equation}
where $\xi$ is the physical length scale of the theory.
(The corresponding power in SU(2) would be $\rho^{7/3}$.)
We observe that the number of instantons vanishes rapidly 
as $\rho\to 0$ (rather than diverging as it did in the
classical theory). This makes it plausible that the introduction 
of a lattice will not affect the physics once $a \ll \xi$. 

The behaviour of $D(\rho)$ in eqn(\ref{A6}) is only
valid for $\rho \ll \xi$ since only then is $g^2(\rho)$
small enough for perturbation theory to be applicable. For
$\rho \geq \xi$ the instantons will presumably overlap
and the density is not calculable analytically. One of the
things we want to learn from lattice calculations
is what actually happens at larger $\rho$. Note that
the characterisation of the topogcal charge density in terms 
of charges of size $\rho$ might not be possible, even to a first
approximation, in the real vacuum. Although we shall 
use that language for convenience in our discussions, we
shall make some attempt to question its validity.

\subsection{topology of lattice fields and cooling}

A lattice gauge field consists of group elements, $U_\mu(x)$, 
on the links of the lattice. A lattice `instanton' can be
constructed straightforwardly by defining
\begin{equation}
U_\mu^I(x) = {\cal P} \exp \int\limits_x^{x+a{\hat\mu}}
A_\mu^I(x)dx         
\label{A7}
\end{equation}
where the gauge potential $A_\mu^I(x)$ is as in eqn(\ref{A3}),
but with its origin translated to the centre of a hypercube.
(On a compact space, e.g. a hypertorus, we need to go to
singular gauge, using the translated version of $g(x)$ 
in eqn(\ref{A4}), before imposing the periodic boundary conditions.) 
As long as $\rho \gg a$ any reasonable definition of
topological charge will assign $Q=1$ to this lattice field.
If we are in a finite periodic volume of length $La$,
then this  $Q=1$ lattice field will be close to being a minimum 
action configuration as long as $a \ll \rho \ll La$. (Exactly how 
close will depend on the particular lattice action being used.) 
If we now smoothly decrease $\rho$ to values $\rho \ll a$ this lattice
field will become indistinguishable from a gauge singularity and
hence will have $Q=0$. Thus we explicitly see the ambiguity
in assigning a topological charge to a lattice gauge field.

Note that this ambiguity disappears as $a \to 0$. Indeed
suppose a lattice field configuration is to be smoothly
deformed from $Q=1$ to $Q=0$. This requires a topological fluctuation
to be squeezed out of the lattice, as described above.  
While we do not know much about the structure of the original 
fluctuation (it will typically be on a size scale $\sim \xi$
which is beyond the reach of our analytic techniques) we
do know that if the lattice spacing is sufficiently small
then to reach $\rho \sim a$ the `instanton' will have to pass
through sizes $\xi \gg \rho \gg a$. In this region the density
is calculable as we saw above, with a probability that is very strongly
suppressed; at least as $\sim (\rho/\xi)^6$ for SU(3). So
the changing of $Q$ is conditional upon the involvement of 
field configurations whose probability$\to 0$ as $a \to 0$. 
Thus, as we approach the continuum limit this lattice ambiguity 
vanishes very rapidly. (And much more rapidly in SU(3) than in SU(2).)
That is to say, the situation is much 
as with the calculation of any other physical quantity: one can
only trust one's results after performing
the appropriate scaling analysis.

Since we are interested in learning about the sizes of the 
topological charges we need a lattice version of the
continuum charge density $Q(x)$ defined
in eqn(\ref{A2}). Let $U_{\mu\nu}(x)$ be the ordered product of 
link matrices around the plaquette labelled by the site $x$ and
the plane $\{\mu,\nu\}$. (For brevity we will refer to this
group element as a plaquette.) As is well known, we can expand
$U_{\mu\nu}(x) = 1 + a^2 F_{\mu\nu}(x) + ....$ and so we can
define a lattice topological charge $Q_L(x)$ as follows
\cite{diV}:
\begin{equation}
Q_L(x) \equiv {1\over{32\pi^2}} \epsilon_{\mu\nu\rho\sigma}
Tr\{U_{\mu\nu}(x)U_{\rho\sigma}(x)\}
= a^4 Q(x) +O(a^6).        
\label{A8}
\end{equation}
(In fact we employ the version of this that is symmetrised
with respect to forward and backward directions, so that
the operator changes sign under reflection in any axis.)
If we apply this formula to a smooth gauge field
then we find, as expected, that the corrections are 
$O(a^2)$; for example, in the case of our instanton
$Q_L = \int Q_L(x)dx = 1 + O(a^2/\rho^2)$.

If we apply $Q_L(x)$ to the real vacuum, however, we
immediately encounter problems.
The operator is dimensionless and so $O(a^6)$ actually
means terms like $\sim a^6 F^3$, $\sim a^6 FD^2F$ etc.
For smooth fields these are indeed  $O(a^6)$. However
realistic fields (those that contribute to the path integral)
have fluctuations all the way up to frequencies of $O(1/a)$.
The contribution of these high frequency modes to the 
$O(a^6)$ terms will be
$\delta Q_L(x) \sim a^6 \times 1/a^6 \sim O(a^0)$
(up to some powers of $\beta$ that can be calculated in 
perturbation theory). Thus in the 
real world $Q_L(x)$ possesses interesting topological
contributions that are of order 
$a^4 \propto e^{-{{16\pi^2}\over{33}}\beta}$
(using the running coupling on scale $a$ for $g^2$)
and uninteresting ultraviolet contributions that are of order
$1/\beta^n$. So as we approach the continuum limit,
$\beta \to \infty$, the ultraviolet fluctuations dominate and
completely mask the interesting physics.

Actually things are somewhat worse than this. Like other
composite lattice operators, $Q_L(x)$ possesses a 
multiplicative lattice renormalisation factor:
${\bar Q}_L = Z_Q Q$ where
$Z_Q \simeq 1 - 5.451/\beta + O(1/\beta^2)$
\cite{PISA-SU3}.
This looks innocuous, and indeed in the continuum limit
it obviously is. However in
the range of values of $\beta$ where current lattice
calculations are performed, typically $\beta \sim 6$,
we see that $Z_Q \ll 1$, rendering the topological
charge virtually invisible.

To deal with these problems we shall use
the technique of `cooling'
\cite{cool-MT}
the fields. The idea rests on the observation that the
problems are all caused by the ultraviolet fluctuations 
on wavelengths $\sim a$. By contrast, 
if we are close to the continuum limit, the topology 
is on wavelengths $\rho \gg a$. One can therefore imagine
taking the lattice fields and locally smoothing them
over distances $\gg a$ but $\ll \rho$. Such a
smoothing would  erase the unwanted ultraviolet
fluctuations while not significantly disturbing the
physical topological charge fluctuations. One could then
apply the operator $Q_L(x)$ to these `cooled' fields
to reveal the topological charge distribution of the
vacuum.

How do we cool a lattice gauge field? The simplest
procedure is to take the field and generate from it
a new field by the standard Monte Carlo heat bath
algorithm subject to one important modification: we always
choose the new link matrix to locally minimise the plaquette
action. Since $Tr U_{\mu\nu}(x)$ measures the variations of
the link matrices over a distance $a$, minimising the plaquette 
action is a very efficient way to erase the ultraviolet 
fluctuations. Obviously there are many possible variations 
on this theme and we shall return to that question shortly. 

Thus the idea is that we take our ensemble of $N$ gauge 
fields, $\{U^{I=1,..,N}\}$, perform a suitable number of cooling 
sweeps on each one of these, and so obtain a corresponding
ensemble $\{U_c^{I=1,..,N}\}$ of cooled fields. We then 
extract the desired topological properties from these
cooled fields. What are the ambiguities? As we
cool, topological charges of opposite sign will gradually
annihilate. This changes the topological charge density
but not the total value of $Q$. Eventually this leads to a  
dilute gas of instantons. As we cool even further these 
isolated instantons will gradually shrink. Eventually they shrink
within a hypercube and at this point even $Q$ will change.
(This will occur if we cool with a plaquette action 
on a large enough volume:
other actions may lead to other outcomes.) Of course when
an instanton becomes narrow it has a very peaked charge
density and is impossible to miss. So we will certainly know 
when it disappears out of the lattice and can, if we think it 
appropriate, correct for that. So cooling provides a
reliable method for calculating the total topological
charge of a lattice field. However the topological
charge density changes continuously throughout the
cooling process and so what we learn from it is far more ambiguous.
For this reason we shall try to use as few cooling sweeps as 
possible and, in addition, we shall repeat the calculations for 
various numbers of cooling sweeps so as to try and
disentangle any artifacts of the cooling procedure.

\subsection{under-relaxed cooling}

Consider first the case of SU(2) and suppose we are
using a plaquette action. The part of the action that 
involves the link matrix $U_{\mu}(x)$ is proportional to
$Tr\{U_{\mu}(x){\hat\Sigma}(x;\mu)\}$ where 
${\hat\Sigma}(x;\mu)$ is an SU(2) matrix proportional to
the sum of the `staples' around the link under consideration.
So if we were choosing a new link matrix $U_{\mu}^{\prime}(x)$
to locally minimise the action, we would choose
$U_{\mu}^{\prime}(x) = {\hat\Sigma}^{\dag}(x;\mu)$.
Repeating this procedure for all the links of the lattice
would constitute the simplest type of cooling sweep.
We might however imagine generalising it to the choice
\begin{equation}
U_{\mu}^{\prime}(x) = 
c \bigl( 
\alpha U_{\mu}(x) + {\hat\Sigma}^{\dag}(x;\mu)
\bigr)
\label{A9}
\end{equation}
where $c$ is a normalisation constant ensuring that the link
matrix is unitary and $\alpha$ is a free parameter. 
(This has been called `under-relaxed' cooling
\cite{CM-PS}.)
This will smoothen the fields for $\alpha\geq 0$. We will use 
this freedom to try and choose a form of cooling that is
optimal for our purposes. 
  
To use this method in SU(3), we simply apply it to the
SU(2) subgroups that arise in the standard Cabibbo-Marinari 
algorithm.

We shall use three criteria for deciding what is the
optimal choice of $\alpha$. 

{\noindent}$\bullet$ We want to use as few cooling sweeps as possible.

{\noindent}$\bullet$ We want to disturb the topological charge
density as little as possible.

{\noindent}$\bullet$ We do not want instantons which initially have
$\rho \sim a$ to broaden as we cool.
%discuss the background to this somewhere earlier

So the first thing we wish to do is to compare the
different kinds of cooling sweeps. We calibrate them as
follows. Discretise an instanton of size $\rho =2a$
as in eqn(\ref{A7}). Now cool it until it disappears.
Our criterion for disappearance is that the action
drops below $10\%$ of the continuum instanton action.
We find that the number of cools, $n_c$, to do this
varies with $\alpha$ approximately as follows:
\begin{equation}
n_c(\alpha) = 23 + 32 \alpha.
\label{A10}
\end{equation}
(One can easily show that, at large $\alpha$, $n_c$ must
increase linearly.) This is a measure of how effectively
the different kinds of cooling erase high-frequency modes:
and as expected we need more sweeps as we increase $\alpha$. 

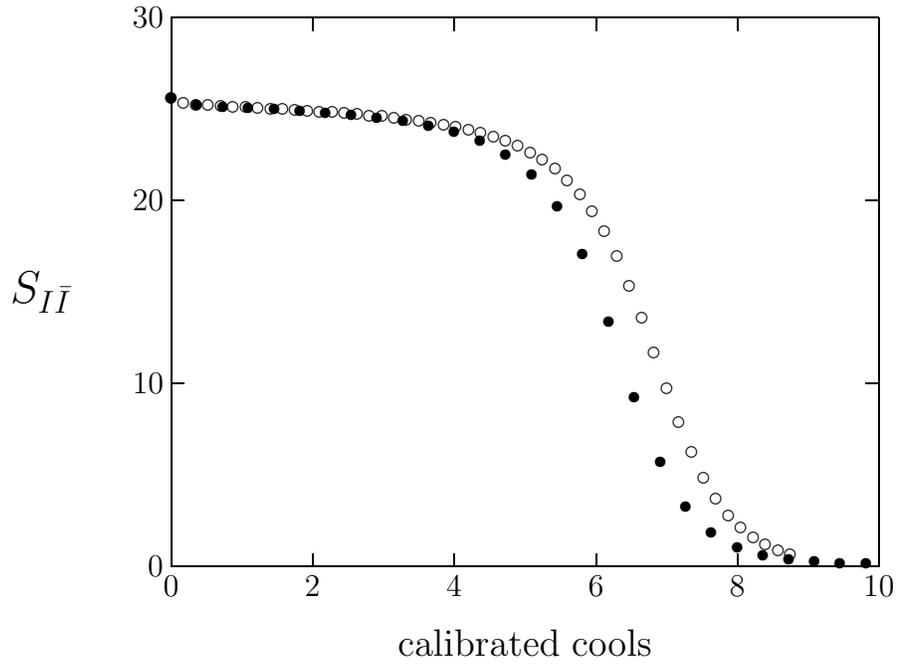
\begin	{figure}[p]
\begin	{center}
\leavevmode
% GNUPLOT: LaTeX picture
\setlength{\unitlength}{0.240900pt}
\ifx\plotpoint\undefined\newsavebox{\plotpoint}\fi
\sbox{\plotpoint}{\rule[-0.200pt]{0.400pt}{0.400pt}}%
\begin{picture}(1275,900)(0,0)
\font\gnuplot=cmr10 at 12pt
\gnuplot
\sbox{\plotpoint}{\rule[-0.200pt]{0.400pt}{0.400pt}}%
\put(120.0,31.0){\rule[-0.200pt]{4.818pt}{0.400pt}}
\put(108,31){\makebox(0,0)[r]{{$0$}}}
\put(1211.0,31.0){\rule[-0.200pt]{4.818pt}{0.400pt}}
\put(120.0,318.0){\rule[-0.200pt]{4.818pt}{0.400pt}}
\put(108,318){\makebox(0,0)[r]{{$10$}}}
\put(1211.0,318.0){\rule[-0.200pt]{4.818pt}{0.400pt}}
\put(120.0,606.0){\rule[-0.200pt]{4.818pt}{0.400pt}}
\put(108,606){\makebox(0,0)[r]{{$20$}}}
\put(1211.0,606.0){\rule[-0.200pt]{4.818pt}{0.400pt}}
\put(120.0,893.0){\rule[-0.200pt]{4.818pt}{0.400pt}}
\put(108,893){\makebox(0,0)[r]{{$30$}}}
\put(1211.0,893.0){\rule[-0.200pt]{4.818pt}{0.400pt}}
\put(120.0,31.0){\rule[-0.200pt]{0.400pt}{4.818pt}}
\put(120,19){\makebox(0,0){\shortstack{\\ \\ \\ {$0$}}}}
\put(120.0,873.0){\rule[-0.200pt]{0.400pt}{4.818pt}}
\put(342.0,31.0){\rule[-0.200pt]{0.400pt}{4.818pt}}
\put(342,19){\makebox(0,0){\shortstack{\\ \\ \\ {$2$}}}}
\put(342.0,873.0){\rule[-0.200pt]{0.400pt}{4.818pt}}
\put(564.0,31.0){\rule[-0.200pt]{0.400pt}{4.818pt}}
\put(564,19){\makebox(0,0){\shortstack{\\ \\ \\ {$4$}}}}
\put(564.0,873.0){\rule[-0.200pt]{0.400pt}{4.818pt}}
\put(787.0,31.0){\rule[-0.200pt]{0.400pt}{4.818pt}}
\put(787,19){\makebox(0,0){\shortstack{\\ \\ \\ {$6$}}}}
\put(787.0,873.0){\rule[-0.200pt]{0.400pt}{4.818pt}}
\put(1009.0,31.0){\rule[-0.200pt]{0.400pt}{4.818pt}}
\put(1009,19){\makebox(0,0){\shortstack{\\ \\ \\ {$8$}}}}
\put(1009.0,873.0){\rule[-0.200pt]{0.400pt}{4.818pt}}
\put(1231.0,31.0){\rule[-0.200pt]{0.400pt}{4.818pt}}
\put(1231,19){\makebox(0,0){\shortstack{\\ \\ \\ {$10$}}}}
\put(1231.0,873.0){\rule[-0.200pt]{0.400pt}{4.818pt}}
\put(120.0,31.0){\rule[-0.200pt]{267.640pt}{0.400pt}}
\put(1231.0,31.0){\rule[-0.200pt]{0.400pt}{207.656pt}}
\put(120.0,893.0){\rule[-0.200pt]{267.640pt}{0.400pt}}
\put(-84,462){\makebox(0,0){{\Large{$S_{I\bar{I}}$}}}}
\put(675,-89){\makebox(0,0){{\large{calibrated cools}}}}
\put(120.0,31.0){\rule[-0.200pt]{0.400pt}{207.656pt}}
\put(120,766){\circle*{18}}
\put(160,756){\circle*{18}}
\put(201,753){\circle*{18}}
\put(241,751){\circle*{18}}
\put(282,749){\circle*{18}}
\put(322,746){\circle*{18}}
\put(362,743){\circle*{18}}
\put(403,740){\circle*{18}}
\put(443,736){\circle*{18}}
\put(484,730){\circle*{18}}
\put(524,723){\circle*{18}}
\put(564,713){\circle*{18}}
\put(605,699){\circle*{18}}
\put(645,678){\circle*{18}}
\put(686,646){\circle*{18}}
\put(726,597){\circle*{18}}
\put(766,521){\circle*{18}}
\put(807,415){\circle*{18}}
\put(847,297){\circle*{18}}
\put(888,195){\circle*{18}}
\put(928,125){\circle*{18}}
\put(968,84){\circle*{18}}
\put(1009,61){\circle*{18}}
\put(1049,48){\circle*{18}}
\put(1090,42){\circle*{18}}
\put(1130,38){\circle*{18}}
\put(1170,36){\circle*{18}}
\put(1211,35){\circle*{18}}
\put(120,766){\circle{18}}
\put(139,759){\circle{18}}
\put(159,756){\circle{18}}
\put(178,755){\circle{18}}
\put(198,754){\circle{18}}
\put(217,752){\circle{18}}
\put(237,752){\circle{18}}
\put(256,751){\circle{18}}
\put(276,750){\circle{18}}
\put(295,749){\circle{18}}
\put(314,748){\circle{18}}
\put(334,746){\circle{18}}
\put(353,745){\circle{18}}
\put(373,744){\circle{18}}
\put(392,743){\circle{18}}
\put(412,741){\circle{18}}
\put(431,739){\circle{18}}
\put(451,738){\circle{18}}
\put(470,736){\circle{18}}
\put(489,733){\circle{18}}
\put(509,731){\circle{18}}
\put(528,728){\circle{18}}
\put(548,725){\circle{18}}
\put(567,721){\circle{18}}
\put(587,717){\circle{18}}
\put(606,712){\circle{18}}
\put(626,706){\circle{18}}
\put(645,699){\circle{18}}
\put(664,691){\circle{18}}
\put(684,681){\circle{18}}
\put(703,669){\circle{18}}
\put(723,655){\circle{18}}
\put(742,637){\circle{18}}
\put(762,615){\circle{18}}
\put(781,589){\circle{18}}
\put(800,557){\circle{18}}
\put(820,518){\circle{18}}
\put(839,472){\circle{18}}
\put(859,421){\circle{18}}
\put(878,366){\circle{18}}
\put(898,311){\circle{18}}
\put(917,258){\circle{18}}
\put(937,210){\circle{18}}
\put(956,170){\circle{18}}
\put(975,137){\circle{18}}
\put(995,111){\circle{18}}
\put(1014,91){\circle{18}}
\put(1034,76){\circle{18}}
\put(1053,65){\circle{18}}
\put(1073,56){\circle{18}}
\put(1092,50){\circle{18}}
\end{picture}
\end	{center}
\vskip 0.15in
\caption{
Action of $I\bar I$ pair against number of calibrated cooling sweeps,
for $\alpha=0$ ($\bullet$) and $\alpha=2$ ($\circ$).}
\label{fig-pair-cool}
\end 	{figure}

Let us now ask how rapidly the topological structure
changes under cooling. To address this question we construct
classical instanton anti-instanton pairs of various widths
and various distances apart. We then cool them and see
how many sweeps it takes to annihilate them. We express
the number of sweeps in units of the corresponding
calibrated sweeps, i.e. in units of $n_c(\alpha)$ in
eqn(\ref{A10}). An example
is shown in Fig.\ref{fig-pair-cool}. In this plot we show
how the action of an instanton anti-instanton pair, with
$\rho = 3a$ and separation $9a$, varies with the number of
cooling sweeps. We show separately what happens for $\alpha=0$
and $\alpha=2$. We observe that in units of the calibrated
sweeps, cooling with $\alpha=2$ alters the topological
structure more slowly than $\alpha = 0$ cooling. This is
not a large effect but is characteristic of what we see
with other examples.

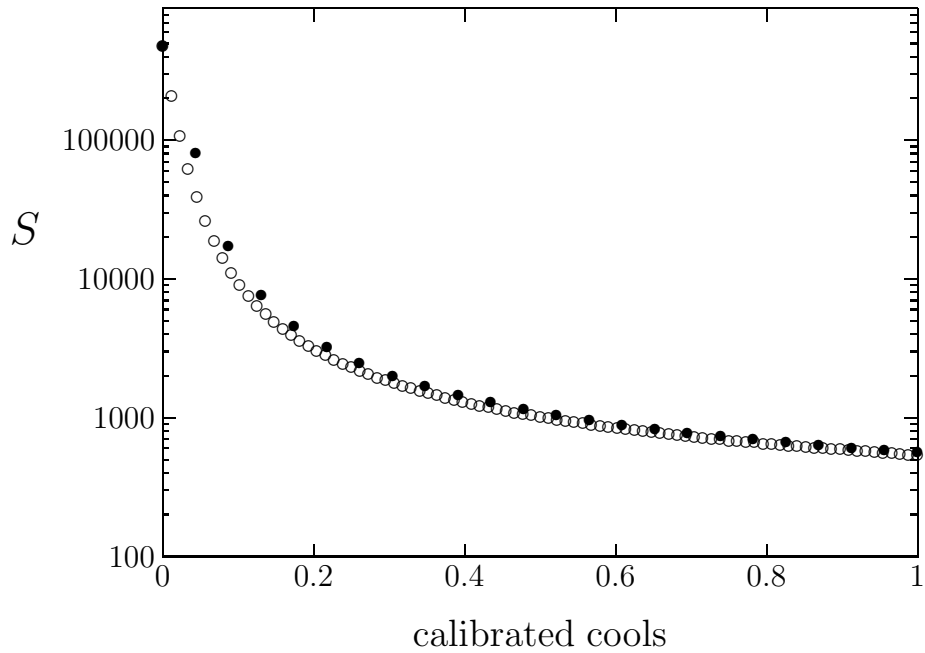
\begin	{figure}[p]
\begin	{center}
\leavevmode
% GNUPLOT: LaTeX picture
\setlength{\unitlength}{0.240900pt}
\ifx\plotpoint\undefined\newsavebox{\plotpoint}\fi
\sbox{\plotpoint}{\rule[-0.200pt]{0.400pt}{0.400pt}}%
\begin{picture}(1349,900)(0,0)
\font\gnuplot=cmr10 at 12pt
\gnuplot
\sbox{\plotpoint}{\rule[-0.200pt]{0.400pt}{0.400pt}}%
\put(120.0,31.0){\rule[-0.200pt]{4.818pt}{0.400pt}}
\put(108,31){\makebox(0,0)[r]{{$100$}}}
\put(1285.0,31.0){\rule[-0.200pt]{4.818pt}{0.400pt}}
\put(120.0,97.0){\rule[-0.200pt]{2.409pt}{0.400pt}}
\put(1295.0,97.0){\rule[-0.200pt]{2.409pt}{0.400pt}}
\put(120.0,135.0){\rule[-0.200pt]{2.409pt}{0.400pt}}
\put(1295.0,135.0){\rule[-0.200pt]{2.409pt}{0.400pt}}
\put(120.0,162.0){\rule[-0.200pt]{2.409pt}{0.400pt}}
\put(1295.0,162.0){\rule[-0.200pt]{2.409pt}{0.400pt}}
\put(120.0,183.0){\rule[-0.200pt]{2.409pt}{0.400pt}}
\put(1295.0,183.0){\rule[-0.200pt]{2.409pt}{0.400pt}}
\put(120.0,201.0){\rule[-0.200pt]{2.409pt}{0.400pt}}
\put(1295.0,201.0){\rule[-0.200pt]{2.409pt}{0.400pt}}
\put(120.0,215.0){\rule[-0.200pt]{2.409pt}{0.400pt}}
\put(1295.0,215.0){\rule[-0.200pt]{2.409pt}{0.400pt}}
\put(120.0,228.0){\rule[-0.200pt]{2.409pt}{0.400pt}}
\put(1295.0,228.0){\rule[-0.200pt]{2.409pt}{0.400pt}}
\put(120.0,239.0){\rule[-0.200pt]{2.409pt}{0.400pt}}
\put(1295.0,239.0){\rule[-0.200pt]{2.409pt}{0.400pt}}
\put(120.0,249.0){\rule[-0.200pt]{4.818pt}{0.400pt}}
\put(108,249){\makebox(0,0)[r]{{$1000$}}}
\put(1285.0,249.0){\rule[-0.200pt]{4.818pt}{0.400pt}}
\put(120.0,315.0){\rule[-0.200pt]{2.409pt}{0.400pt}}
\put(1295.0,315.0){\rule[-0.200pt]{2.409pt}{0.400pt}}
\put(120.0,353.0){\rule[-0.200pt]{2.409pt}{0.400pt}}
\put(1295.0,353.0){\rule[-0.200pt]{2.409pt}{0.400pt}}
\put(120.0,380.0){\rule[-0.200pt]{2.409pt}{0.400pt}}
\put(1295.0,380.0){\rule[-0.200pt]{2.409pt}{0.400pt}}
\put(120.0,401.0){\rule[-0.200pt]{2.409pt}{0.400pt}}
\put(1295.0,401.0){\rule[-0.200pt]{2.409pt}{0.400pt}}
\put(120.0,419.0){\rule[-0.200pt]{2.409pt}{0.400pt}}
\put(1295.0,419.0){\rule[-0.200pt]{2.409pt}{0.400pt}}
\put(120.0,433.0){\rule[-0.200pt]{2.409pt}{0.400pt}}
\put(1295.0,433.0){\rule[-0.200pt]{2.409pt}{0.400pt}}
\put(120.0,446.0){\rule[-0.200pt]{2.409pt}{0.400pt}}
\put(1295.0,446.0){\rule[-0.200pt]{2.409pt}{0.400pt}}
\put(120.0,457.0){\rule[-0.200pt]{2.409pt}{0.400pt}}
\put(1295.0,457.0){\rule[-0.200pt]{2.409pt}{0.400pt}}
\put(120.0,467.0){\rule[-0.200pt]{4.818pt}{0.400pt}}
\put(108,467){\makebox(0,0)[r]{{$10000$}}}
\put(1285.0,467.0){\rule[-0.200pt]{4.818pt}{0.400pt}}
\put(120.0,533.0){\rule[-0.200pt]{2.409pt}{0.400pt}}
\put(1295.0,533.0){\rule[-0.200pt]{2.409pt}{0.400pt}}
\put(120.0,571.0){\rule[-0.200pt]{2.409pt}{0.400pt}}
\put(1295.0,571.0){\rule[-0.200pt]{2.409pt}{0.400pt}}
\put(120.0,598.0){\rule[-0.200pt]{2.409pt}{0.400pt}}
\put(1295.0,598.0){\rule[-0.200pt]{2.409pt}{0.400pt}}
\put(120.0,619.0){\rule[-0.200pt]{2.409pt}{0.400pt}}
\put(1295.0,619.0){\rule[-0.200pt]{2.409pt}{0.400pt}}
\put(120.0,637.0){\rule[-0.200pt]{2.409pt}{0.400pt}}
\put(1295.0,637.0){\rule[-0.200pt]{2.409pt}{0.400pt}}
\put(120.0,651.0){\rule[-0.200pt]{2.409pt}{0.400pt}}
\put(1295.0,651.0){\rule[-0.200pt]{2.409pt}{0.400pt}}
\put(120.0,664.0){\rule[-0.200pt]{2.409pt}{0.400pt}}
\put(1295.0,664.0){\rule[-0.200pt]{2.409pt}{0.400pt}}
\put(120.0,675.0){\rule[-0.200pt]{2.409pt}{0.400pt}}
\put(1295.0,675.0){\rule[-0.200pt]{2.409pt}{0.400pt}}
\put(120.0,685.0){\rule[-0.200pt]{4.818pt}{0.400pt}}
\put(108,685){\makebox(0,0)[r]{{$100000$}}}
\put(1285.0,685.0){\rule[-0.200pt]{4.818pt}{0.400pt}}
\put(120.0,751.0){\rule[-0.200pt]{2.409pt}{0.400pt}}
\put(1295.0,751.0){\rule[-0.200pt]{2.409pt}{0.400pt}}
\put(120.0,789.0){\rule[-0.200pt]{2.409pt}{0.400pt}}
\put(1295.0,789.0){\rule[-0.200pt]{2.409pt}{0.400pt}}
\put(120.0,816.0){\rule[-0.200pt]{2.409pt}{0.400pt}}
\put(1295.0,816.0){\rule[-0.200pt]{2.409pt}{0.400pt}}
\put(120.0,837.0){\rule[-0.200pt]{2.409pt}{0.400pt}}
\put(1295.0,837.0){\rule[-0.200pt]{2.409pt}{0.400pt}}
\put(120.0,855.0){\rule[-0.200pt]{2.409pt}{0.400pt}}
\put(1295.0,855.0){\rule[-0.200pt]{2.409pt}{0.400pt}}
\put(120.0,869.0){\rule[-0.200pt]{2.409pt}{0.400pt}}
\put(1295.0,869.0){\rule[-0.200pt]{2.409pt}{0.400pt}}
\put(120.0,882.0){\rule[-0.200pt]{2.409pt}{0.400pt}}
\put(1295.0,882.0){\rule[-0.200pt]{2.409pt}{0.400pt}}
\put(120.0,893.0){\rule[-0.200pt]{2.409pt}{0.400pt}}
\put(1295.0,893.0){\rule[-0.200pt]{2.409pt}{0.400pt}}
\put(120.0,31.0){\rule[-0.200pt]{0.400pt}{4.818pt}}
\put(120,19){\makebox(0,0){\shortstack{\\ \\ \\ {$0$}}}}
\put(120.0,873.0){\rule[-0.200pt]{0.400pt}{4.818pt}}
\put(357.0,31.0){\rule[-0.200pt]{0.400pt}{4.818pt}}
\put(357,19){\makebox(0,0){\shortstack{\\ \\ \\ {$0.2$}}}}
\put(357.0,873.0){\rule[-0.200pt]{0.400pt}{4.818pt}}
\put(594.0,31.0){\rule[-0.200pt]{0.400pt}{4.818pt}}
\put(594,19){\makebox(0,0){\shortstack{\\ \\ \\ {$0.4$}}}}
\put(594.0,873.0){\rule[-0.200pt]{0.400pt}{4.818pt}}
\put(831.0,31.0){\rule[-0.200pt]{0.400pt}{4.818pt}}
\put(831,19){\makebox(0,0){\shortstack{\\ \\ \\ {$0.6$}}}}
\put(831.0,873.0){\rule[-0.200pt]{0.400pt}{4.818pt}}
\put(1068.0,31.0){\rule[-0.200pt]{0.400pt}{4.818pt}}
\put(1068,19){\makebox(0,0){\shortstack{\\ \\ \\ {$0.8$}}}}
\put(1068.0,873.0){\rule[-0.200pt]{0.400pt}{4.818pt}}
\put(1305.0,31.0){\rule[-0.200pt]{0.400pt}{4.818pt}}
\put(1305,19){\makebox(0,0){\shortstack{\\ \\ \\ {$1$}}}}
\put(1305.0,873.0){\rule[-0.200pt]{0.400pt}{4.818pt}}
\put(120.0,31.0){\rule[-0.200pt]{285.466pt}{0.400pt}}
\put(1305.0,31.0){\rule[-0.200pt]{0.400pt}{207.656pt}}
\put(120.0,893.0){\rule[-0.200pt]{285.466pt}{0.400pt}}
\put(-96,546){\makebox(0,0){{\Large{$S$}}}}
\put(712,-89){\makebox(0,0){{\large{calibrated cools}}}}
\put(120.0,31.0){\rule[-0.200pt]{0.400pt}{207.656pt}}
\put(120,833){\circle*{18}}
\put(172,665){\circle*{18}}
\put(223,519){\circle*{18}}
\put(275,441){\circle*{18}}
\put(326,393){\circle*{18}}
\put(378,360){\circle*{18}}
\put(429,335){\circle*{18}}
\put(481,315){\circle*{18}}
\put(532,299){\circle*{18}}
\put(584,285){\circle*{18}}
\put(635,273){\circle*{18}}
\put(687,263){\circle*{18}}
\put(738,253){\circle*{18}}
\put(790,245){\circle*{18}}
\put(841,238){\circle*{18}}
\put(893,231){\circle*{18}}
\put(944,225){\circle*{18}}
\put(996,220){\circle*{18}}
\put(1047,215){\circle*{18}}
\put(1099,210){\circle*{18}}
\put(1150,206){\circle*{18}}
\put(1202,202){\circle*{18}}
\put(1253,198){\circle*{18}}
\put(1305,195){\circle*{18}}
\put(120,833){\circle{18}}
\put(134,755){\circle{18}}
\put(147,692){\circle{18}}
\put(160,639){\circle{18}}
\put(174,595){\circle{18}}
\put(187,558){\circle{18}}
\put(201,526){\circle{18}}
\put(214,500){\circle{18}}
\put(228,477){\circle{18}}
\put(241,457){\circle{18}}
\put(255,440){\circle{18}}
\put(268,425){\circle{18}}
\put(282,412){\circle{18}}
\put(295,400){\circle{18}}
\put(309,389){\circle{18}}
\put(322,379){\circle{18}}
\put(335,370){\circle{18}}
\put(349,362){\circle{18}}
\put(362,354){\circle{18}}
\put(376,347){\circle{18}}
\put(389,340){\circle{18}}
\put(403,334){\circle{18}}
\put(416,328){\circle{18}}
\put(430,323){\circle{18}}
\put(443,317){\circle{18}}
\put(457,312){\circle{18}}
\put(470,308){\circle{18}}
\put(484,303){\circle{18}}
\put(497,299){\circle{18}}
\put(510,295){\circle{18}}
\put(524,291){\circle{18}}
\put(537,287){\circle{18}}
\put(551,284){\circle{18}}
\put(564,280){\circle{18}}
\put(578,277){\circle{18}}
\put(591,274){\circle{18}}
\put(605,271){\circle{18}}
\put(618,268){\circle{18}}
\put(632,265){\circle{18}}
\put(645,263){\circle{18}}
\put(659,260){\circle{18}}
\put(672,257){\circle{18}}
\put(686,255){\circle{18}}
\put(699,253){\circle{18}}
\put(713,250){\circle{18}}
\put(726,248){\circle{18}}
\put(739,246){\circle{18}}
\put(753,244){\circle{18}}
\put(766,242){\circle{18}}
\put(780,240){\circle{18}}
\put(793,238){\circle{18}}
\put(807,236){\circle{18}}
\put(820,234){\circle{18}}
\put(834,233){\circle{18}}
\put(847,231){\circle{18}}
\put(861,229){\circle{18}}
\put(874,228){\circle{18}}
\put(888,226){\circle{18}}
\put(901,225){\circle{18}}
\put(915,223){\circle{18}}
\put(928,222){\circle{18}}
\put(941,220){\circle{18}}
\put(955,219){\circle{18}}
\put(968,217){\circle{18}}
\put(982,216){\circle{18}}
\put(995,215){\circle{18}}
\put(1009,213){\circle{18}}
\put(1022,212){\circle{18}}
\put(1036,211){\circle{18}}
\put(1049,210){\circle{18}}
\put(1063,208){\circle{18}}
\put(1076,207){\circle{18}}
\put(1090,206){\circle{18}}
\put(1103,205){\circle{18}}
\put(1116,204){\circle{18}}
\put(1130,203){\circle{18}}
\put(1143,202){\circle{18}}
\put(1157,201){\circle{18}}
\put(1170,200){\circle{18}}
\put(1184,199){\circle{18}}
\put(1197,198){\circle{18}}
\put(1211,197){\circle{18}}
\put(1224,196){\circle{18}}
\put(1238,195){\circle{18}}
\put(1251,194){\circle{18}}
\put(1265,193){\circle{18}}
\put(1278,192){\circle{18}}
\put(1291,191){\circle{18}}
\put(1305,190){\circle{18}}
\end{picture}
\end	{center}
\vskip 0.15in
\caption{Action against number of calibrated cooling sweeps, for
$\alpha=0$ ($\bullet$) and $\alpha=2$ ($\circ$).}
\label{fig-action-cool}
\end 	{figure}

In addition to the above studies we have also compared
the effects of the different kind of cooling on 
thermalised field configurations. To be specific,
on five $16^3 48$ lattice fields generated at $\beta=6.0$.
We cooled these with $\alpha = 0.0, 0.5, 1.0, 1.5, 2.0,
2.5,$ and 3.0. In Fig.\ref{fig-action-cool} we show
how the action decreases with the number of (calibrated)
sweeps. We see that the curves corresponding to $\alpha=0$
and $\alpha=2$ almost fall on top of each other.
This confirms the fact that to a good approximation
a calibrated sweep has the same effect on the ultraviolet
lattice modes for any $\alpha$. 

As an aside we note that the action drops much further
in the first $\alpha=0$ cooling sweep than in the first 
$\alpha=2$ (uncalibrated) sweep. To the extent that
there is a worry about what might be happening at this
early stage, there might be some advantage in the
smoother under-relaxed cooling procedure.

Finally we note that while the value of $Q_L$ after one
calibrated cooling sweep is independent of $\alpha$ on 
three of the five configurations, it differed by 
1 on two of them. To be precise, the fields cooled
with $\alpha \geq 0.5$ all agreed with each other
and the disagreement was with the $\alpha=0$
case. It could be readily traced to a single `instanton'
that was narrow during the first few cooling sweeps,
and which rapidly shrank out of the lattice with
further $\alpha \geq 0.5$ cooling, but which broadened
under $\alpha =0$ cooling. This anomalous broadening
must be a result of the non-trivial environment
in which the narrow instanton is sitting. It is
however something we would wish to suppress as
much as possible and this is an argument for not using
$\alpha =0$.

We have seen that in appropriate units we can
disturb the topological charge density less by
using $\alpha \neq 0$ and that we simultaneously 
reduce the probability of $\rho \sim a$ artifacts
surviving the cooling. Our studies are far from
definitive and because of their low statistics
might even be misleading; but they do serve to illustrate
the criteria that it would make sense to use.
Motivated by what we have found we shall use
under-relaxed cooling with $\alpha =1$ for the remainder
of this paper.

\section{Pattern recognition}

Since the cooling algorithm is
local it will erase the highest frequency modes first.
Ideally we would like to stop the cooling once it has erased
all the modes on scales $\lambda \ll \xi$ but before it
has significantly affected the physically interesting modes
on scales $\lambda \sim O(\xi)$. Such a clean separation is
not possible in practice and by the time we have cooled
enough to reveal the long-distance structure of $Q(x)$ 
we have certainly deformed that structure. Thus one 
has to perform the calculations
for various numbers of cooling sweeps and attempt to
identify those features that are relatively robust.

Because our cooling algorithm gradually deforms a field 
configuration towards the minimum of the action, the
topological charge density will increasingly resemble
a set of overlapping instantons and anti-instantons.
As we cool further, those that are strongly overlapping will
annihilate and the vacuum will become less densely
packed. So in order to identify the structure of $Q(x)$
we shall assume that it is given by an overlapping
set of (anti)instantons of various sizes. This is
of course a crude approximation. It also raises
a fundamental question: how much of this structure
is driven by the cooling and how much of it is
intrinsic to the original uncooled field configuration?
One way to try and answer this question is to
increase $\beta$ so that the separation between
the physical and ultraviolet modes becomes better
defined. We have therefore included calculations
up to $\beta=6.4$.

In this section we describe how we extract and categorise
the topological structure of the cooled field configurations.

The first step is to use the peaks of $Q(x)$ to locate the
centres of the topological charges and to provide a first
estimate of their sizes. We then need to correct for
the influence of the charges on each other.
The next step is designed to reduce the number of
false identifications. These may arise, for example,
from secondary ripples on very large instantons.
We implement two filters for this purpose. That there
are in fact many mis-identifications is easy to show.
We have $n_I$ candidate instantons and $n_{\bar I}$
candidate anti-instantons. The total topological
charge is therefore predicted to be $n_I - n_{\bar I}$.
At the same time we can calculate $Q$ directly from
$\int Q_L(x) d^4 x$. The quantity 
\begin{equation}
\delta_Q \equiv |Q - (n_I - n_{\bar I})|
\label{A11}
\end{equation}
provides a direct measure of the mis-identification,
and typically turns out to be substantial. At the
same time this provides us with a criterion for
choosing the parameters in our filters: they are
chosen so as to minimise the value of $\langle \delta_Q \rangle$.

Our discussion so far has been based upon the 
topological charge density. Clearly there is information
carried by the action density as well and one might ask
whether it would be useful to incorporate that.
We investigate this question in the last subsection
and find that the action density has little new to tell us 
about the smaller charges that are easy to identify
anyway, and is not able to resolve the larger charges
where all the uncertainties lie. Thus for the 
remainder of the paper our analysis will be entirely
based upon the topological charge density. 

\subsection{peaks and neighbours}

Once we have cooled a field configuration we calculate
the topological charge density using eqn(\ref{A8}). The peaks
in this density are candidate locations of instantons.
However it is our experience in dealing with smooth
discretised instantons that it is dangerous to define
a peak only with respect to the sites that are $\pm a$
away in any one direction. One instanton can readily produce 
peaks on sites across diagonals of a hypercube. 

We therefore define $Q_L(x)$ to have a peak at $x_0$ if its 
value at $x=x_0$ is greater than at all the $3^4$ sites belonging 
to the corresponding hyperbox centered on $x_0$. (With an
obvious modification to account for negative maxima.) Of course if 
two instantons happen to be close enough together, then we will 
miss one of them by using this criterion. However the probability
of this occuring will decrease rapidly as $a$ decreases. So, once 
again, as long as we perform a scaling analysis there is no
ambiguity.

At this stage we have candidate positive charges at 
$\{x_i^+ ; i=1,...,n_+\}$ and candidate negative charges at 
$\{x_i^- ; i=1,...,n_-\}$. We shall make the customary assumption
that only charges are with $Q=\pm 1$ are present. (It is non-trivial
matter to test this assumption and we do not attempt 
to do so in this paper.)
To obtain a first estimate of the sizes of these charges we 
can use the classical instanton relation between
the topological charge density at the peak and the width:
\begin{equation}
Q_p = {{6}\over{\pi^2\rho^4}}.
\label{A12}
\end{equation}
This relation is for a continuum instanton. It applies
equally well for a large lattice instanton, but will
become inaccurate for smaller instantons where 
$O({{a^2}\over{\rho^2}})$ 
lattice corrections become significant. In practice
we use a lattice corrected version of eqn(\ref{A12})
as described in the Appendix. 

We now have a first estimate for the positions, $x_i^{\pm}$, 
and sizes, $\rho_i^{\pm}$, of the (anti)instantons.
However we know that the value of $Q_L(x)$
at  $x=x_i^{\pm}$ will receive contributions from the
tails of all the other (anti)instantons and so may
not be an accurate reflection of the peak value
of the topological charge that is centered there.
To correct for this we have implemented the following
iterative procedure.

We shall make two main approximations. First we shall
assume that the topological charge is additive.
Secondly we shall only attempt to calculate the
corrections to the sizes, $\rho_i^\pm$, and not to 
the locations, $x_i^\pm$. These are approximations
that should be improved upon. Under these assumptions we
can write
\begin{equation}
Q(x_i^\pm) = Q_p(\rho_i^\pm) + \sum_{x_j^{\pm}\not=x_i^{\pm}}
Q_I(|x_j^{\pm}-x_i^{\pm}|;\rho_j^{\pm})
\label{A13}
\end{equation}
where $Q_p(\rho)$ is the peak value of a topological charge of
radius $\rho$ (as given in eqn(\ref{A12}) and with the lattice
corrections as in the Appendix) and $Q_I(|x-x_0|;\rho)$ is 
the contribution to
the charge density at $x$ from an instanton of size $\rho$
located at $x_0$. We use the continuum expression for this
\begin{equation}
Q_I(|x-x_0|;\rho) = 
{6\over{\pi^2}} {{\rho^4}\over{(x^2+\rho^2)^4}}
\label{A14}
\end{equation}
with the opposite sign for anti-instantons. While one should
improve upon this expression by including lattice
corrections at small-$\rho$, this is not necessary to
a first approximation, because the corrections to $\rho$
that are embodied in eqn(\ref{A14}) turn out to be modest.
Note that to avoid cluttering the equations we
have dropped the subscript on $Q_L$.

What we know in eqn(\ref{A13}) are the values of the
$Q(x_i^\pm)$ and what we want to solve for are the $\rho_i^\pm$.
One can attempt to do this by iteration, using eqn(\ref{A12})
and $Q_p(\rho_i^\pm)=Q(x_i^\pm)$ to provide us with our 
starting values of  $\rho_i^\pm$. We pick say the
charge at $x_1^+$ and calculate the contribution of all the
other peaks using eqns(\ref{A13},\ref{A14}). From the
renormalised peak value we extract a corrected value of
$\rho_1^+$ to replace our first guess. We go to the next
charge and repeat the same procedure there except that
we use the updated value of $\rho_1^+$ in calculating $Q_I$.
Repeating this procedure at each relevant site constitutes
one iteration. We perform as many iterations as are required to
reach convergence. Our criterion for convergence is that
the change in $\rho$ during the final iteration
should satisfy $\delta\rho \leq 0.001\rho$ for all the charges.

In practice we have applied the above procedure with
a slight modification: if at any stage the apparent
sign of a charge changes when we take into account the
influence of the other charges, then this charge
is removed and plays no further part in the analysis.
The reason for doing this is that given the approximate
nature of the correction, if it makes such a large
difference then we cannot be confident that there is
in fact a charge at that location. To throw the charge
away is of course an arbitrary choice. Fortunately
this arises very infrequently. For example on
20 $16^3 48$ lattices after 23 cooling sweeps 
one peak was removed from 6 configurations and 
four peaks from one configuration; despite the
fact that the average configuration contained
169 peaks. In this sense the modification is 
indeed slight.

We remark that the above procedure has always converged; 
presumably because our starting point is always close enough 
to the final solution. We have explicitly checked that 
the final solution does not depend on the order in which the
peaks are considered, and, more to the point, that neither
do the peaks that are thrown out because they change sign.

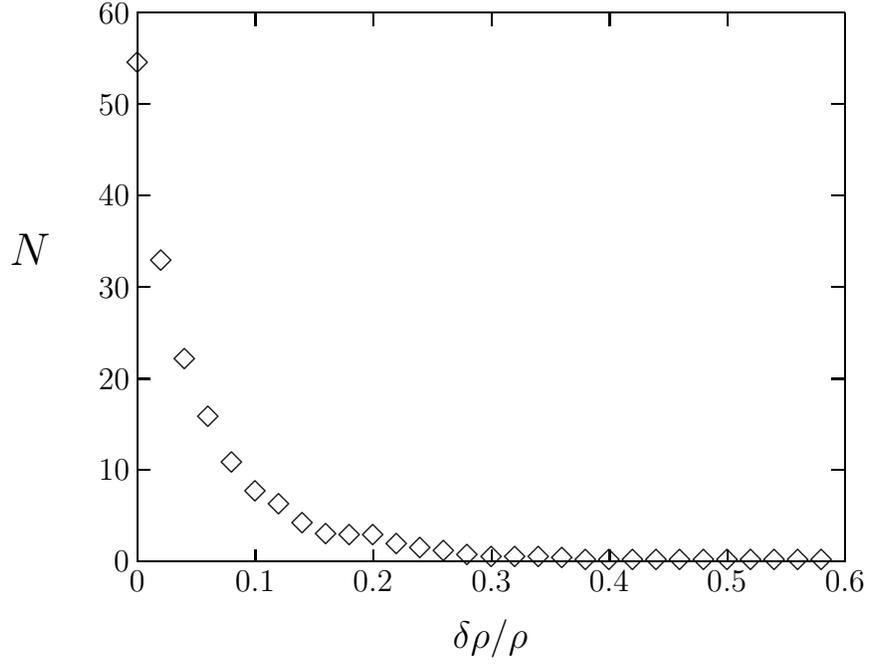
\begin	{figure}[p]
\begin	{center}
\leavevmode
% GNUPLOT: LaTeX picture
\setlength{\unitlength}{0.240900pt}
\ifx\plotpoint\undefined\newsavebox{\plotpoint}\fi
\sbox{\plotpoint}{\rule[-0.200pt]{0.400pt}{0.400pt}}%
\begin{picture}(1275,900)(0,0)
\font\gnuplot=cmr10 at 12pt
\gnuplot
\sbox{\plotpoint}{\rule[-0.200pt]{0.400pt}{0.400pt}}%
\put(120.0,31.0){\rule[-0.200pt]{4.818pt}{0.400pt}}
\put(108,31){\makebox(0,0)[r]{{$0$}}}
\put(1211.0,31.0){\rule[-0.200pt]{4.818pt}{0.400pt}}
\put(120.0,175.0){\rule[-0.200pt]{4.818pt}{0.400pt}}
\put(108,175){\makebox(0,0)[r]{{$10$}}}
\put(1211.0,175.0){\rule[-0.200pt]{4.818pt}{0.400pt}}
\put(120.0,318.0){\rule[-0.200pt]{4.818pt}{0.400pt}}
\put(108,318){\makebox(0,0)[r]{{$20$}}}
\put(1211.0,318.0){\rule[-0.200pt]{4.818pt}{0.400pt}}
\put(120.0,462.0){\rule[-0.200pt]{4.818pt}{0.400pt}}
\put(108,462){\makebox(0,0)[r]{{$30$}}}
\put(1211.0,462.0){\rule[-0.200pt]{4.818pt}{0.400pt}}
\put(120.0,606.0){\rule[-0.200pt]{4.818pt}{0.400pt}}
\put(108,606){\makebox(0,0)[r]{{$40$}}}
\put(1211.0,606.0){\rule[-0.200pt]{4.818pt}{0.400pt}}
\put(120.0,749.0){\rule[-0.200pt]{4.818pt}{0.400pt}}
\put(108,749){\makebox(0,0)[r]{{$50$}}}
\put(1211.0,749.0){\rule[-0.200pt]{4.818pt}{0.400pt}}
\put(120.0,893.0){\rule[-0.200pt]{4.818pt}{0.400pt}}
\put(108,893){\makebox(0,0)[r]{{$60$}}}
\put(1211.0,893.0){\rule[-0.200pt]{4.818pt}{0.400pt}}
\put(120.0,31.0){\rule[-0.200pt]{0.400pt}{4.818pt}}
\put(120,19){\makebox(0,0){\shortstack{\\ \\ \\ {$0$}}}}
\put(120.0,873.0){\rule[-0.200pt]{0.400pt}{4.818pt}}
\put(305.0,31.0){\rule[-0.200pt]{0.400pt}{4.818pt}}
\put(305,19){\makebox(0,0){\shortstack{\\ \\ \\ {$0.1$}}}}
\put(305.0,873.0){\rule[-0.200pt]{0.400pt}{4.818pt}}
\put(490.0,31.0){\rule[-0.200pt]{0.400pt}{4.818pt}}
\put(490,19){\makebox(0,0){\shortstack{\\ \\ \\ {$0.2$}}}}
\put(490.0,873.0){\rule[-0.200pt]{0.400pt}{4.818pt}}
\put(676.0,31.0){\rule[-0.200pt]{0.400pt}{4.818pt}}
\put(676,19){\makebox(0,0){\shortstack{\\ \\ \\ {$0.3$}}}}
\put(676.0,873.0){\rule[-0.200pt]{0.400pt}{4.818pt}}
\put(861.0,31.0){\rule[-0.200pt]{0.400pt}{4.818pt}}
\put(861,19){\makebox(0,0){\shortstack{\\ \\ \\ {$0.4$}}}}
\put(861.0,873.0){\rule[-0.200pt]{0.400pt}{4.818pt}}
\put(1046.0,31.0){\rule[-0.200pt]{0.400pt}{4.818pt}}
\put(1046,19){\makebox(0,0){\shortstack{\\ \\ \\ {$0.5$}}}}
\put(1046.0,873.0){\rule[-0.200pt]{0.400pt}{4.818pt}}
\put(1231.0,31.0){\rule[-0.200pt]{0.400pt}{4.818pt}}
\put(1231,19){\makebox(0,0){\shortstack{\\ \\ \\ {$0.6$}}}}
\put(1231.0,873.0){\rule[-0.200pt]{0.400pt}{4.818pt}}
\put(120.0,31.0){\rule[-0.200pt]{267.640pt}{0.400pt}}
\put(1231.0,31.0){\rule[-0.200pt]{0.400pt}{207.656pt}}
\put(120.0,893.0){\rule[-0.200pt]{267.640pt}{0.400pt}}
\put(-48,522){\makebox(0,0){{\Large{$N$}}}}
\put(675,-89){\makebox(0,0){{\large{$\delta\rho/\rho$}}}}
\put(120.0,31.0){\rule[-0.200pt]{0.400pt}{207.656pt}}
\put(120,813){\raisebox{-.8pt}{\makebox(0,0){$\Diamond$}}}
\put(157,502){\raisebox{-.8pt}{\makebox(0,0){$\Diamond$}}}
\put(194,347){\raisebox{-.8pt}{\makebox(0,0){$\Diamond$}}}
\put(231,257){\raisebox{-.8pt}{\makebox(0,0){$\Diamond$}}}
\put(268,185){\raisebox{-.8pt}{\makebox(0,0){$\Diamond$}}}
\put(305,139){\raisebox{-.8pt}{\makebox(0,0){$\Diamond$}}}
\put(342,119){\raisebox{-.8pt}{\makebox(0,0){$\Diamond$}}}
\put(379,89){\raisebox{-.8pt}{\makebox(0,0){$\Diamond$}}}
\put(416,73){\raisebox{-.8pt}{\makebox(0,0){$\Diamond$}}}
\put(453,70){\raisebox{-.8pt}{\makebox(0,0){$\Diamond$}}}
\put(490,71){\raisebox{-.8pt}{\makebox(0,0){$\Diamond$}}}
\put(527,56){\raisebox{-.8pt}{\makebox(0,0){$\Diamond$}}}
\put(564,51){\raisebox{-.8pt}{\makebox(0,0){$\Diamond$}}}
\put(601,45){\raisebox{-.8pt}{\makebox(0,0){$\Diamond$}}}
\put(638,39){\raisebox{-.8pt}{\makebox(0,0){$\Diamond$}}}
\put(676,36){\raisebox{-.8pt}{\makebox(0,0){$\Diamond$}}}
\put(713,37){\raisebox{-.8pt}{\makebox(0,0){$\Diamond$}}}
\put(750,37){\raisebox{-.8pt}{\makebox(0,0){$\Diamond$}}}
\put(787,35){\raisebox{-.8pt}{\makebox(0,0){$\Diamond$}}}
\put(824,32){\raisebox{-.8pt}{\makebox(0,0){$\Diamond$}}}
\put(861,32){\raisebox{-.8pt}{\makebox(0,0){$\Diamond$}}}
\put(898,32){\raisebox{-.8pt}{\makebox(0,0){$\Diamond$}}}
\put(935,32){\raisebox{-.8pt}{\makebox(0,0){$\Diamond$}}}
\put(972,32){\raisebox{-.8pt}{\makebox(0,0){$\Diamond$}}}
\put(1009,32){\raisebox{-.8pt}{\makebox(0,0){$\Diamond$}}}
\put(1046,32){\raisebox{-.8pt}{\makebox(0,0){$\Diamond$}}}
\put(1083,32){\raisebox{-.8pt}{\makebox(0,0){$\Diamond$}}}
\put(1120,31){\raisebox{-.8pt}{\makebox(0,0){$\Diamond$}}}
\put(1157,32){\raisebox{-.8pt}{\makebox(0,0){$\Diamond$}}}
\put(1194,31){\raisebox{-.8pt}{\makebox(0,0){$\Diamond$}}}
\end{picture}
\end	{center}
\vskip 0.15in
\caption{Number of times a given fractional change in width 
occurs when using eqn(\ref{A13}): at $\beta=6.0$ after 23 cools.}
\label{fig-iter1}
\end 	{figure}

How much of a difference does it make to estimate the 
instanton sizes using eqn(\ref{A13}) rather than
just applying eqn(\ref{A12}) to the observed peak heights?
In Fig.\ref{fig-iter1} we show what happens on a test sample
of 20 $\beta=6.0$ $16^3 48$ field configurations after 23
cooling sweeps. On the x-axis we plot the quantity
\begin{equation}
{{\delta\rho}\over{\rho}} =
\biggl\vert {{\rho_{final}-\rho_{orig}}\over{\rho_{orig}}} 
\biggr\vert
\label{A15}
\end{equation}
where $\rho_{orig}$ is the initial estimate of the size using
eqn(\ref{A12}), and  $\rho_{final}$ is the value obtained
after solving  eqn(\ref{A13}). On the y-axis we plot the
average number of times a value of ${{\delta\rho}\over{\rho}}$
occurs per field configuration. We observe that the fractional
change in $\rho$ is typically at the $\sim 5\%$ level; that is
to say, small but significant.

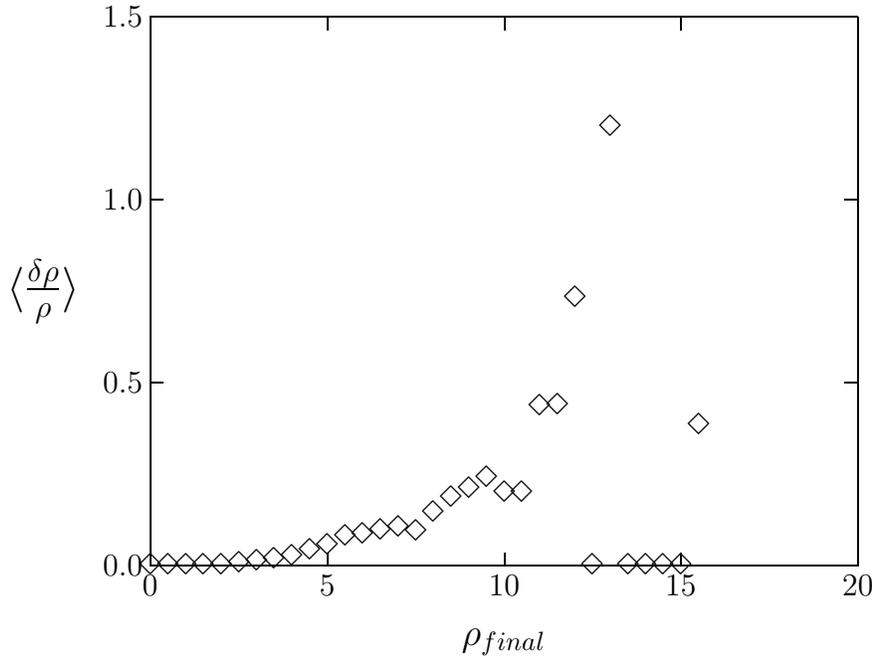
\begin	{figure}[p]
\begin	{center}
\leavevmode
% GNUPLOT: LaTeX picture
\setlength{\unitlength}{0.240900pt}
\ifx\plotpoint\undefined\newsavebox{\plotpoint}\fi
\sbox{\plotpoint}{\rule[-0.200pt]{0.400pt}{0.400pt}}%
\begin{picture}(1275,900)(0,0)
\font\gnuplot=cmr10 at 12pt
\gnuplot
\sbox{\plotpoint}{\rule[-0.200pt]{0.400pt}{0.400pt}}%
\put(120.0,31.0){\rule[-0.200pt]{4.818pt}{0.400pt}}
\put(108,31){\makebox(0,0)[r]{{$0.0$}}}
\put(1211.0,31.0){\rule[-0.200pt]{4.818pt}{0.400pt}}
\put(120.0,318.0){\rule[-0.200pt]{4.818pt}{0.400pt}}
\put(108,318){\makebox(0,0)[r]{{$0.5$}}}
\put(1211.0,318.0){\rule[-0.200pt]{4.818pt}{0.400pt}}
\put(120.0,606.0){\rule[-0.200pt]{4.818pt}{0.400pt}}
\put(108,606){\makebox(0,0)[r]{{$1.0$}}}
\put(1211.0,606.0){\rule[-0.200pt]{4.818pt}{0.400pt}}
\put(120.0,893.0){\rule[-0.200pt]{4.818pt}{0.400pt}}
\put(108,893){\makebox(0,0)[r]{{$1.5$}}}
\put(1211.0,893.0){\rule[-0.200pt]{4.818pt}{0.400pt}}
\put(120.0,31.0){\rule[-0.200pt]{0.400pt}{4.818pt}}
\put(120,19){\makebox(0,0){\shortstack{\\ \\ \\ {$0$}}}}
\put(120.0,873.0){\rule[-0.200pt]{0.400pt}{4.818pt}}
\put(398.0,31.0){\rule[-0.200pt]{0.400pt}{4.818pt}}
\put(398,19){\makebox(0,0){\shortstack{\\ \\ \\ {$5$}}}}
\put(398.0,873.0){\rule[-0.200pt]{0.400pt}{4.818pt}}
\put(676.0,31.0){\rule[-0.200pt]{0.400pt}{4.818pt}}
\put(676,19){\makebox(0,0){\shortstack{\\ \\ \\ {$10$}}}}
\put(676.0,873.0){\rule[-0.200pt]{0.400pt}{4.818pt}}
\put(953.0,31.0){\rule[-0.200pt]{0.400pt}{4.818pt}}
\put(953,19){\makebox(0,0){\shortstack{\\ \\ \\ {$15$}}}}
\put(953.0,873.0){\rule[-0.200pt]{0.400pt}{4.818pt}}
\put(1231.0,31.0){\rule[-0.200pt]{0.400pt}{4.818pt}}
\put(1231,19){\makebox(0,0){\shortstack{\\ \\ \\ {$20$}}}}
\put(1231.0,873.0){\rule[-0.200pt]{0.400pt}{4.818pt}}
\put(120.0,31.0){\rule[-0.200pt]{267.640pt}{0.400pt}}
\put(1231.0,31.0){\rule[-0.200pt]{0.400pt}{207.656pt}}
\put(120.0,893.0){\rule[-0.200pt]{267.640pt}{0.400pt}}
\put(-48,462){\makebox(0,0){{\Large{$\bigl\langle {{\delta\rho}\over{\rho}} \bigr\rangle$}}}}
\put(675,-89){\makebox(0,0){{\large{$\rho_{final}$}}}}
\put(120.0,31.0){\rule[-0.200pt]{0.400pt}{207.656pt}}
\put(120,31){\raisebox{-.8pt}{\makebox(0,0){$\Diamond$}}}
\put(148,31){\raisebox{-.8pt}{\makebox(0,0){$\Diamond$}}}
\put(176,31){\raisebox{-.8pt}{\makebox(0,0){$\Diamond$}}}
\put(203,32){\raisebox{-.8pt}{\makebox(0,0){$\Diamond$}}}
\put(231,32){\raisebox{-.8pt}{\makebox(0,0){$\Diamond$}}}
\put(259,34){\raisebox{-.8pt}{\makebox(0,0){$\Diamond$}}}
\put(287,37){\raisebox{-.8pt}{\makebox(0,0){$\Diamond$}}}
\put(314,41){\raisebox{-.8pt}{\makebox(0,0){$\Diamond$}}}
\put(342,45){\raisebox{-.8pt}{\makebox(0,0){$\Diamond$}}}
\put(370,54){\raisebox{-.8pt}{\makebox(0,0){$\Diamond$}}}
\put(398,63){\raisebox{-.8pt}{\makebox(0,0){$\Diamond$}}}
\put(426,76){\raisebox{-.8pt}{\makebox(0,0){$\Diamond$}}}
\put(453,79){\raisebox{-.8pt}{\makebox(0,0){$\Diamond$}}}
\put(481,86){\raisebox{-.8pt}{\makebox(0,0){$\Diamond$}}}
\put(509,91){\raisebox{-.8pt}{\makebox(0,0){$\Diamond$}}}
\put(537,85){\raisebox{-.8pt}{\makebox(0,0){$\Diamond$}}}
\put(564,114){\raisebox{-.8pt}{\makebox(0,0){$\Diamond$}}}
\put(592,137){\raisebox{-.8pt}{\makebox(0,0){$\Diamond$}}}
\put(620,151){\raisebox{-.8pt}{\makebox(0,0){$\Diamond$}}}
\put(648,169){\raisebox{-.8pt}{\makebox(0,0){$\Diamond$}}}
\put(676,145){\raisebox{-.8pt}{\makebox(0,0){$\Diamond$}}}
\put(703,146){\raisebox{-.8pt}{\makebox(0,0){$\Diamond$}}}
\put(731,281){\raisebox{-.8pt}{\makebox(0,0){$\Diamond$}}}
\put(759,283){\raisebox{-.8pt}{\makebox(0,0){$\Diamond$}}}
\put(787,451){\raisebox{-.8pt}{\makebox(0,0){$\Diamond$}}}
\put(814,31){\raisebox{-.8pt}{\makebox(0,0){$\Diamond$}}}
\put(842,720){\raisebox{-.8pt}{\makebox(0,0){$\Diamond$}}}
\put(870,31){\raisebox{-.8pt}{\makebox(0,0){$\Diamond$}}}
\put(898,31){\raisebox{-.8pt}{\makebox(0,0){$\Diamond$}}}
\put(925,31){\raisebox{-.8pt}{\makebox(0,0){$\Diamond$}}}
\put(953,31){\raisebox{-.8pt}{\makebox(0,0){$\Diamond$}}}
\put(981,252){\raisebox{-.8pt}{\makebox(0,0){$\Diamond$}}}
\end{picture}
\end	{center}
\vskip 0.15in
\caption{Fractional change in width, against the
final width, after using eqn(\ref{A13}) to account 
for presence of other charges:
at $\beta=6.0$ after 23 cools.}
\label{fig-iter2}
\end 	{figure}

We note from eqn(\ref{A13}) that subtracting a small constant 
$\delta Q$ from the peak value of $Q(x)$, leads to a fractional
change in the width $\delta\rho/\rho \propto \rho^4 \delta Q$.
We would therefore expect that charges with small $\rho$ would be 
practically unaffected by the corrections in eqn(\ref{A13}),
but that the fractional change would rapidly increase
with $\rho$ and, at some point, would cease to be reliable.
In Fig.\ref{fig-iter2} we show how the average value of
the fractional change in $\rho$ depends on the final value of 
$\rho$, in our test sample of configurations. We see that
$\langle \delta\rho/\rho\rangle$ is very small up to $\rho \sim 5$,
which, as we shall shortly see, is roughly where the
charge density, $D(\rho)$, has its maximum at $\beta=6.0$. 
It then grows rapidly with $\rho$ but remains small enough to
be credible up to $\rho \sim 10$. Thereafter it becomes
large and our approximations are presumably inadequate. 
However, as we shall see, there are almost no instantons
for $\rho \geq 10$ and so we believe that our procedure
provides a reasonable first approximation for the range of $\rho$
relevant to our calculations.

\subsection{filtering the peaks}

At this stage we have a set of candidate charges. We 
claim to know their positions and their widths. 
If this was all that was needed then we would expect that
the value of $\delta_Q$ in eqn(\ref{A11}) would be zero.
We show in Table~\ref{table-deltaQ} what the average
values of this quantity actually are for the $24^3 48$
configurations at $\beta=6.2$. We also show, for comparison, 
the value of $\surd \langle Q^2\rangle$ and the average number 
of charges, $\langle N_{tot} \rangle \equiv 
\langle n_I + n_{\bar I} \rangle$. 
We do this for various numbers of cooling sweeps. 

We observe that there is a substantial mismatch between
the value of $Q$ as calculated directly and that obtained
from the peaks of $Q(x)$. The former is certainly reliable  
(up to errors in the lattice corrections, which are 
negligible relative to $\delta_Q$). So either some
charges do not show up as peaks, and so we have missed them,
or some of the peaks in $Q(x)$ are not topological charges.
We cannot deal with the first possibility without
a much more sophisticated correction procedure than that
embodied in eqn(\ref{A13}). This is beyond the scope of
the present paper. To address the second possibility we shall 
calculate some further properties of the topological charge 
density around the peaks and use these to ``filter out'' the
peaks that are most likely not to be instantons.

\subsubsection{a width filter}

For an instanton of size $\rho$ the charge within a radius $R$
is given by
\begin{equation}
Q(|x|\leq R) = 1
-3\Biggl({1\over{1+{{R^2}\over{\rho^2}}}}\Biggr)^2 
+2\Biggl({1\over{1+{{R^2}\over{\rho^2}}}}\Biggr)^3
\label{A16}
\end{equation}
(This will require significant lattice corrections for small
$\rho$, as discussed in the Appendix.) 

We can use eqn(\ref{A16}) to calculate $\rho$ from
$Q(|x|\leq R)$. For an isolated classical instanton
we will get the same value of $\rho$ whatever value of $R$
we choose. In an environment where instantons overlap
this will not be the case. If we correct for this overlap  
by using an obvious generalisation of eqn(\ref{A13}) 
%(see the Appendix) 
then the extracted $\rho$ should become independent of $R$.

Our filter is therefore as follows. As described earlier, for 
each peak in $Q(x)$ we calculate a value of the width, $\rho$,
using the (corrected) value of $Q(x)$ at the peak. We then choose 
some value of $R$ and calculate the corresponding widths 
$\rho_R$ from the (corrected) values of $Q(|x|\leq R)$, as
described above. If the peak represents a real instanton
then we expect that the values $\rho$ and $\rho_R$ should be  
similar. We therefore impose the condition
\begin{equation}
Max\biggl({ {{\rho_R}\over{\rho}}, 
{{\rho}\over{\rho_R}}
} \biggr) - 1 < \epsilon_R
\label{A17}
\end{equation}
where $\epsilon_R$ is a small number that will be fixed
by minimising the quantity $\delta_Q$ in eqn(\ref{A11}).
Only if a peak satisfies this condition will it be
counted as a genuine topological charge.
In practice we shall use $R=2$ in our
later calculations. (Note that we shall switch between
physical and lattice units as convenient, when there
is no ambiguity.)

\subsubsection{a distance filter}

Very broad instantons are likely to be significantly 
distorted and so one needs a reasonably generous
value of $\epsilon_R$ in eqn(\ref{A17}) if one is not 
to run the risk of filtering out too many genuine topological
charges. It is therefore useful to supplement the previous 
filter with an additional one.

We choose to focus on the possibility that a very broad
instanton might possess a long-wavelength ripple across its
surface which then leads to a misidentification of the structure
as containing two (or more) broad instantons. (This is a 
possibility because the small number of cooling sweeps that
we shall be using will not affect long-wavelength modes.)

Our filter consists of the following steps. 

{\noindent}1. Consider a randomly chosen peak of $Q(x)$
at position $x_0$ with width $\rho_0$.

{\noindent}2. Identify the peak nearest to it. If this has
the opposite sign accept the original peak. If it has the same
sign, follow the steps below.

{\noindent}3. Let $\rho_n$, $x_n$ be the width and position
of the nearest neighbour.
Let $\rho_c$ be a cut-off value to be chosen. Then we
accept the original peak if either it or the nearest neighbour
is narrower than $\rho_c$ 
i.e. if $\rho_0\leq\rho_c$ or $\rho_n\leq\rho_c$.

{\noindent}4. If both peaks are broader than $\rho_c$
and if the distance between them is
small compared to their sizes then we
reject the peak under consideration. The detailed
criterion is
$|x_0 - x_n| \leq \epsilon_c (\rho_0 + \rho_n)$
where $\epsilon_c$ is a small number to be chosen.

We consider each peak on the lattice in this way. The peaks are
considered in a random order and therefore the choice of 
which of two broad nearby peaks gets thrown out is in reality
random.

\subsubsection{using the filters}

In practice we apply the distance filter first and apply the
width filter to those peaks that survive. We have
the parameters $\epsilon_R$, $\rho_c$ and  $\epsilon_c$
to fix. This is done by minimising $\delta_Q$, 
in eqn(\ref{A11}), with respect to
variations in all three parameters simultaneously. This is
a time-consuming calculation and we typically perform
it on a subset of $\sim 20$ of the configurations, and
then use the parameters so determined to analyse the
whole ensemble.

The quantity $\delta_Q$ will often have several
minima that are not significantly higher than the
absolute minimum. In such situations we choose the minimum
that leads to fewer peaks being rejected. This is to avoid
loose cuts that lead to the loss of too many real instantons 
along with the false peaks. In Table~\ref{table-deltaQ-filter1}
we list the filter values we use in the calculations of this
paper.

We now give an example of the application of the above filters,
using our $24^3 48$ lattice fields at 
$\beta=6.2$. We consider the three ensembles obtained
after 23,32 and 46 cooling sweeps. 
In Table~\ref{table-deltaQ-filter1} we see the filter
parameters and the corresponding values of $\delta_Q$. We
observe that $\delta_Q$ is dramatically reduced when 
compared to the unfiltered values in Table~\ref{table-deltaQ}.
The width filter used here involved $R=2$. The corresponding
values with $R=3$ are shown in Table~\ref{table-deltaQ-filter2}.
The results are not dissimilar.

\begin	{figure}[p]
\begin	{center}
\leavevmode
\input	{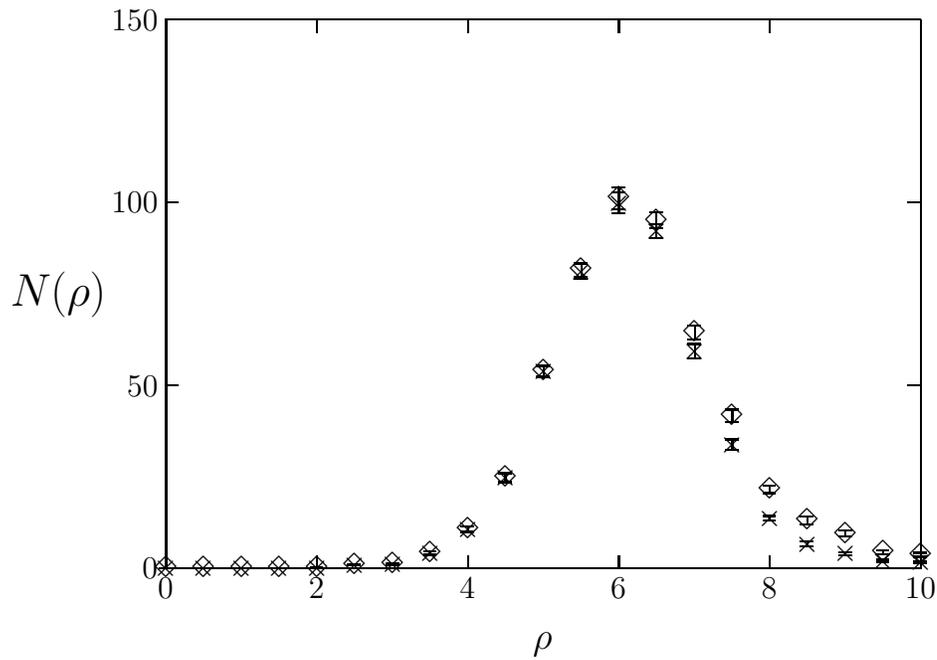}
\end	{center}
\vskip 0.15in
\caption{Filtered ($\times$) and unfiltered ($\diamond$)
size distributions after 23 cooling sweeps at $\beta=6.2$.}
\label{fig-filter}
\end 	{figure}

In order to achieve these acceptably small mismatches
between $Q$ and $n_I-n_{\bar I}$, how severely do we need
to change the distribution of charges? Not very much is
the answer. In Table~\ref{table-num-filter} we show
the number of peaks before and after the filters are applied.
Even for the smallest number of cooling sweeps,
we only lose $\sim10\%$ of the peaks. 
In Fig.\ref{fig-filter} we show the  
number of instantons per configuration as a function of
the size $\rho$, before and after applying the filters. 
Here we see that the
change is concentrated amongst the very largest
instantons. This is as it should be: it is these charges, with 
their very small charge densities and their large overlaps
with many other charges, that are the hardest to extract 
reliably.

Although the purpose of our filters is to reject false
peaks, it is inevitable that occasionally they will
reject real charges. This is especially so with
the distance filter: two broad instantons may be 
close together just by chance. It would be useful to have
some crude estimate of this. One could do this by throwing 
the charges into our space-time box, with the 
observed size density, and seeing how often they
would be rejected by the distance filter. In throwing
the charges into the box, one should incorporate
some broad features of the correlations. As we shall see
one such feature is that the nearest neighbour tends to
be of the opposite sign; and a second is that there is
a strong suppression of (anti)instantons very close to 
each other. We have not implemented such a realistic
model, but have simply thrown the instantons into
the box entirely at random. In that case we find that
the number of real rejected instantons is close to 
the actual number we reject. It should be clear that
the qualitative effect of modifying the random
distribution to include the features we just
described, will be to markedly reduce the number
of mistaken rejections. Thus we anticipate that
only a small fraction of the charges rejected are
real ones. However this is only a qualitative argument
and it is certainly no substitute for an explicit
and careful ``background'' calculation: this still 
needs to be done.

We have seen that by the addition of two physically
motivated filters we are able to reduce the discrepancy,
$\delta_Q$, quite dramatically and that this only involves the
rejection of a small percentage of the peaks. Moreover
the rejected peaks are concentrated amonst the very broadest
charges, as they should be. In the remainder of our work these 
are the filters that we shall employ.

\subsection{the action density}

Before moving onto our results, we briefly ask whether
there is much to be gained by using the action density,
$S(x)$, in addition to or in place of the topological 
charge density. 

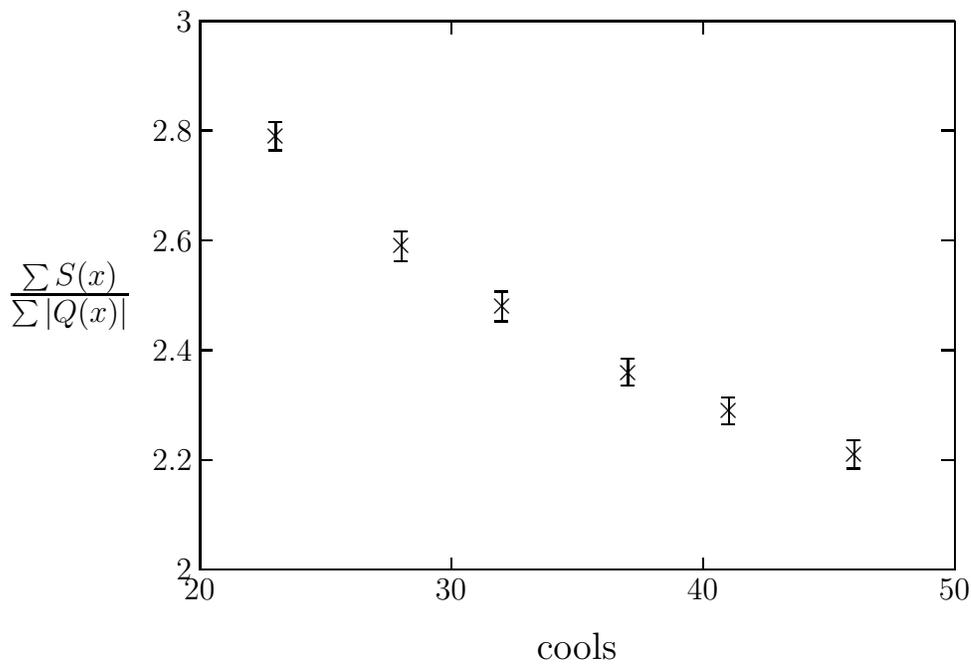
\begin	{figure}[p]
\begin	{center}
\leavevmode
% GNUPLOT: LaTeX picture
\setlength{\unitlength}{0.240900pt}
\ifx\plotpoint\undefined\newsavebox{\plotpoint}\fi
\sbox{\plotpoint}{\rule[-0.200pt]{0.400pt}{0.400pt}}%
\begin{picture}(1349,900)(0,0)
\font\gnuplot=cmr10 at 12pt
\gnuplot
\sbox{\plotpoint}{\rule[-0.200pt]{0.400pt}{0.400pt}}%
\put(120.0,31.0){\rule[-0.200pt]{4.818pt}{0.400pt}}
\put(108,31){\makebox(0,0)[r]{{$2$}}}
\put(1285.0,31.0){\rule[-0.200pt]{4.818pt}{0.400pt}}
\put(120.0,203.0){\rule[-0.200pt]{4.818pt}{0.400pt}}
\put(108,203){\makebox(0,0)[r]{{$2.2$}}}
\put(1285.0,203.0){\rule[-0.200pt]{4.818pt}{0.400pt}}
\put(120.0,376.0){\rule[-0.200pt]{4.818pt}{0.400pt}}
\put(108,376){\makebox(0,0)[r]{{$2.4$}}}
\put(1285.0,376.0){\rule[-0.200pt]{4.818pt}{0.400pt}}
\put(120.0,548.0){\rule[-0.200pt]{4.818pt}{0.400pt}}
\put(108,548){\makebox(0,0)[r]{{$2.6$}}}
\put(1285.0,548.0){\rule[-0.200pt]{4.818pt}{0.400pt}}
\put(120.0,721.0){\rule[-0.200pt]{4.818pt}{0.400pt}}
\put(108,721){\makebox(0,0)[r]{{$2.8$}}}
\put(1285.0,721.0){\rule[-0.200pt]{4.818pt}{0.400pt}}
\put(120.0,893.0){\rule[-0.200pt]{4.818pt}{0.400pt}}
\put(108,893){\makebox(0,0)[r]{{$3$}}}
\put(1285.0,893.0){\rule[-0.200pt]{4.818pt}{0.400pt}}
\put(120.0,31.0){\rule[-0.200pt]{0.400pt}{4.818pt}}
\put(120,19){\makebox(0,0){\shortstack{\\ \\ \\ {$20$}}}}
\put(120.0,873.0){\rule[-0.200pt]{0.400pt}{4.818pt}}
\put(515.0,31.0){\rule[-0.200pt]{0.400pt}{4.818pt}}
\put(515,19){\makebox(0,0){\shortstack{\\ \\ \\ {$30$}}}}
\put(515.0,873.0){\rule[-0.200pt]{0.400pt}{4.818pt}}
\put(910.0,31.0){\rule[-0.200pt]{0.400pt}{4.818pt}}
\put(910,19){\makebox(0,0){\shortstack{\\ \\ \\ {$40$}}}}
\put(910.0,873.0){\rule[-0.200pt]{0.400pt}{4.818pt}}
\put(1305.0,31.0){\rule[-0.200pt]{0.400pt}{4.818pt}}
\put(1305,19){\makebox(0,0){\shortstack{\\ \\ \\ {$50$}}}}
\put(1305.0,873.0){\rule[-0.200pt]{0.400pt}{4.818pt}}
\put(120.0,31.0){\rule[-0.200pt]{285.466pt}{0.400pt}}
\put(1305.0,31.0){\rule[-0.200pt]{0.400pt}{207.656pt}}
\put(120.0,893.0){\rule[-0.200pt]{285.466pt}{0.400pt}}
\put(-84,462){\makebox(0,0){{\Large{${{\sum S(x)}\over{\sum |Q(x)|}}$}}}}
\put(712,-89){\makebox(0,0){{\large{cools}}}}
\put(120.0,31.0){\rule[-0.200pt]{0.400pt}{207.656pt}}
\put(239,712){\makebox(0,0){$\times$}}
\put(436,540){\makebox(0,0){$\times$}}
\put(594,445){\makebox(0,0){$\times$}}
\put(792,341){\makebox(0,0){$\times$}}
\put(950,281){\makebox(0,0){$\times$}}
\put(1147,212){\makebox(0,0){$\times$}}
\put(239.0,690.0){\rule[-0.200pt]{0.400pt}{10.600pt}}
\put(229.0,690.0){\rule[-0.200pt]{4.818pt}{0.400pt}}
\put(229.0,734.0){\rule[-0.200pt]{4.818pt}{0.400pt}}
\put(436.0,516.0){\rule[-0.200pt]{0.400pt}{11.322pt}}
\put(426.0,516.0){\rule[-0.200pt]{4.818pt}{0.400pt}}
\put(426.0,563.0){\rule[-0.200pt]{4.818pt}{0.400pt}}
\put(594.0,421.0){\rule[-0.200pt]{0.400pt}{11.322pt}}
\put(584.0,421.0){\rule[-0.200pt]{4.818pt}{0.400pt}}
\put(584.0,468.0){\rule[-0.200pt]{4.818pt}{0.400pt}}
\put(792.0,320.0){\rule[-0.200pt]{0.400pt}{10.359pt}}
\put(782.0,320.0){\rule[-0.200pt]{4.818pt}{0.400pt}}
\put(782.0,363.0){\rule[-0.200pt]{4.818pt}{0.400pt}}
\put(950.0,260.0){\rule[-0.200pt]{0.400pt}{10.118pt}}
\put(940.0,260.0){\rule[-0.200pt]{4.818pt}{0.400pt}}
\put(940.0,302.0){\rule[-0.200pt]{4.818pt}{0.400pt}}
\put(1147.0,190.0){\rule[-0.200pt]{0.400pt}{10.600pt}}
\put(1137.0,190.0){\rule[-0.200pt]{4.818pt}{0.400pt}}
\put(1137.0,234.0){\rule[-0.200pt]{4.818pt}{0.400pt}}
\end{picture}
\end	{center}
\vskip 0.15in
\caption{
$\frac{\sum_x S(x)}{\sum_x|Q(x)|}$ against number of cooling 
sweeps, at $\beta=6.0$.}
\label{fig-S-Q}
\end 	{figure}

What do we expect? Generally $S(x) \geq |Q(x)|$,
if we use a normalisation where  $S(x) = |Q(x)|$
for a self-dual field. As we cool we shall eventually
be driven to such a self-dual solution (up to lattice
corrections). It is only when $S(x)\simeq |Q(x)|$
that one can use analogues of  eqn(\ref{A13}) for the
action so as to estimate widths from the action densities.
In Fig.\ref{fig-S-Q} we show how the ratio
$\sum S(x)/\sum |Q(x)|$ varies with the number of
cooling sweeps. (This comes from 5 $16^3 48$ 
configurations at $\beta = 6.0$.)
We see that the fields are far from being
self-dual.

If we ignore the non-self dual nature of the fields
and extract widths from the peaks in the action density, 
then we obtain the size distributions shown in 
Fig.\ref{fig-size-SQ23} and Fig.\ref{fig-size-SQ46}. 
We also show, for comparison, the
corresponding distributions that we obtain from  
the topological charge density. (Note that both analyses
simply use the peak height, with no added filters
of the kind described above.) We observe that for
small $\rho$ the distributions are essentially
identical while at large $\rho$ the distribution from
the action is suppressed, and that this effect is
stronger for fewer cools.

This can be qualitatively understood in the approximation
where we think of the extra non-selfdual action, 
$\delta S = \sum S(x) -\sum |Q(x)|$,
as being smoothly distributed over the whole volume.
If we calculate $\rho$ from the peak action density,
then this increment will shift instanton sizes to smaller 
values. Narrow instantons have large peaks that will
be little changed by this addition. On the other
hand the action density will never be smaller
than $S(x) = \delta S/volume$ and this provides
an upper limit on the $\rho$ that one extracts.
This effect should be weaker for a larger number 
of cooling sweeps because $\delta S/volume$
decreases -- see  Fig.\ref{fig-S-Q}. This certainly
provides a first approximation to what we observe
in Fig.\ref{fig-size-SQ23} and Fig.\ref{fig-size-SQ46}. 
At small $\rho$ no change; at large $\rho$ a quite sharp
cut-off; at medium $\rho$ an enhancement in the
size density from the action as one would expect if larger
peaks had been shifted to smaller ones. The numbers roughly 
fit too, except it is clear that only a small fraction 
of the broad charges have been shifted to smaller values:
most of them have apparently disappeared. This is to be
expected. If we have a very broad instanton overlapping
with a very broad anti-instanton we can see two peaks
in $Q(x)$ because of the sign difference. $S(x)$
however is always positive and is quite likely not to 
show two peaks - just a single broad peak covering
the pair.

One can go a step further and ask
whether the peaks in $S(x)$ are in fact
associated with peaks in $Q(x)$. In Table~\ref{table-peak-SQ}
we show the number of peaks obtained from $S(x)$ and
$Q(x)$ for 23 and 46 cooling sweeps. We also show
how many of the peaks in $S(x)$ are associated
with peaks in $Q(x)$: either because they are at the same site
or because they are within 2 lattice spacings. We observe that
the latter accounts for nearly all the action peaks.

We conclude that as long as we work with a small number 
of cooling sweeps the action density loses most of the 
information about the larger topological charges,
although it does reproduce the narrower topological
charges that we find using $Q(x)$. Thus we shall
ignore the action in the remainder of this paper in the 
expectation that including it would yield marginal 
benefits.

\section{Size distribution of instantons}

The size distribution of the topological charges, $D(\rho)$, 
is the simplest quantity characterising the vacuum topological 
structure. In this section we shall explore it in some detail.

\subsection{general features}

\begin	{figure}[p]
\begin	{center}
\leavevmode
\input	{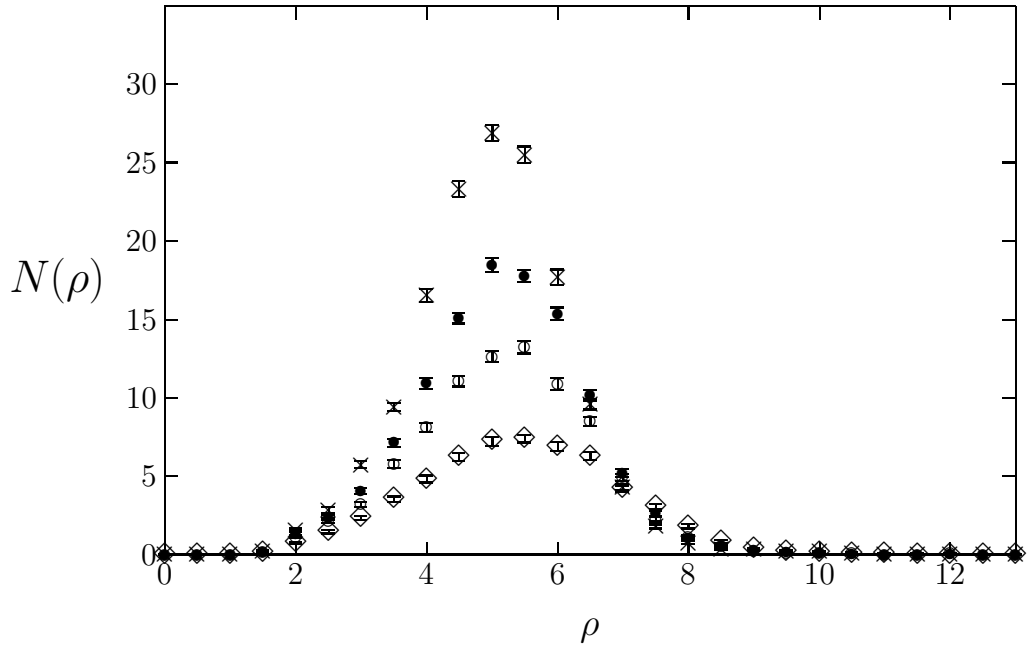}
\end	{center}
\vskip 0.15in
\caption{The number of charges of different sizes; at $\beta = 6.0$
for 23($\times$), 28($\bullet$), 32($\circ$) and 46($\diamond$) cools.}
\label{fig-size-cool60}
\end 	{figure}

In  Fig.\ref{fig-size-cool60} we show the size distribution
as obtained on the $16^3 48$ lattice fields at $\beta = 6.0$
for various number of cooling sweeps, $n_c$.
The quantity plotted is the average number of charges, $N(\rho)$,
in each bin, $\Delta\rho$ of $\rho$. Thus 
$N(\rho) \simeq V D(\rho)\Delta\rho$, where $V$ is the space-time
volume. 

We see that there is a rapid decrease in the total number of 
charges as we cool the fields. This is presumably the result 
of nearby charges of opposite sign annihilating. Other
features, such as the location of the 
maximum of the distribution, appear to
vary much more weakly which suggests that they are robust
features of the fields prior to cooling.

It will be useful to choose a few quantities by which we can
characterise the size distributions. An obvious measure is
the average value of the size, ${\bar\rho}$. Since the distributions
are not grossly asymmetric, this will nearly coincide with 
the maximum. Another quantity we can use is the half-width, 
$\sigma_{\rho}$, of the distribution. Finally there is
the total number of charges, $\bar N_{tot}$. In
Table~\ref{table-sizes-all} we list the values of these
quantites for all our values of $\beta$ and $n_c$.

When discussing the scaling properties of these various
quantites we will need to know how the lattice spacing varies
over our range of $\beta$. For this we need some physical
quantity expressed in lattice units. We choose the confining
string tension, $\sigma$, because that has been calculated very
accurately. The relevant values are
\cite{string-tension},    
\begin{equation}
a\surd\sigma = \cases {
0.2187(12) &$ \ \ \ \ \beta=6.0$ \cr
0.1608(10) &$ \ \ \ \ \beta=6.2$ \cr
0.1216(11) &$ \ \ \ \ \beta=6.4$. \cr} 
\label{A18}
\end{equation}
Wherever we discuss lengths or volumes in physical units,
it will be by using eqn(\ref{A18}) to set the scale.

Occasionally it will be useful (or illuminating) to express
things in $MeV$ units. There are, of course, all kinds of 
ambiguities in introducing $MeV$ units into a theory
which, unlike QCD, does not describe the real world.
This is discussed in
\cite{MT-newton}
where an analysis of the hadron spectrum in the quenched 
approximation is found to lead to an estimate
\begin{equation}
 \surd\sigma = 440 \pm 15 \pm 35 MeV .
\label{A19}
\end{equation}       
Here the first error is statistical and the second
is a systematic error that reflects in part the fact that
Quenched QCD does not in fact represent the real world.
Wherever we present quantites in $MeV$ or $fm$ units
it will be through using eqn(\ref{A19}).

Before moving to a detailed consideration of the size distribution 
there is at least one qualitative conclusion we can immediately draw.
We see from Fig.\ref{fig-size-cool60} and
eqn(\ref{A18}) that 
${\bar\rho} \sim 5a \sim 1/\surd\sigma \sim 0.5fm$.
Thus the typical instanton size is quite large.
Given that the average charge has a diameter of $2{\bar\rho} \simeq 10a$.
and that  there are about 180 charges at 23 cools, it is 
clear that the $16^3 48$ lattice must be densely packed.
This is so even after 46 cools. Thus our first qualitative
conclusion is that instantons are large and strongly 
overlapping. This is a different picture to the one
that apparently underlies typical instanton liquid model
calculations
\cite{Shuryak}.    

\subsection{packing fraction}

As we have just seen, our instanton gas is dense. Since
the largest instantons are more difficult to identify
unambiguously, it is interesting to ask if the gas
is dense even if we exclude such instantons.

To address this question we define a packing fraction
$f(\rho)$ by
\begin{equation}
f(\rho) = {1 \over V}
\int\limits_0^{\rho} n(\rho) v_I(\rho) d\rho
\label{A19b}
\end{equation}       
Here $ n(\rho)$ is the number of instantons of size
$\rho$, $v_I(\rho)$ is the space-time volume
occupied by an instanton of this size and $V$ is the
total space-time volume. That is to say, $f(\rho)$ is
the fraction of space-time occupied by instantons of
size $\leq \rho$.

Since the instanton core is smooth, there is some
ambiguity about defining $v_I(\rho)$. We shall choose
to define it as a 4-sphere of radius $\rho$ : a
conservative choice. So  $v_I(\rho) = \pi^2\rho^4/2$.

\begin	{figure}[p]
\begin	{center}
\leavevmode
\input	{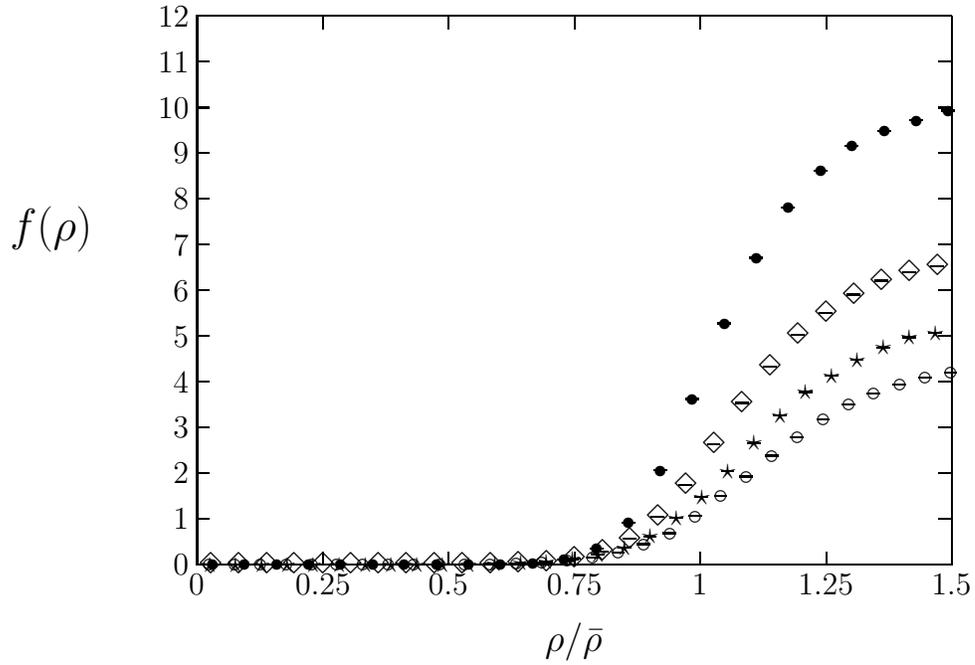}
\end	{center}
\vskip 0.15in
\caption{Packing fraction of instantons of size $\leq\rho$
against $\rho$ in units of $\bar\rho$. For 30($\bullet$),
50($\diamond$),70($\star$)
and 80($\circ$) cooling sweeps at $\beta=6.4$.}
\label{fig-pack64}
\end 	{figure}

Using our calculated size distributions, and this
definition of the instanton volume, we can calculate
$f(\rho)$. In Fig.\ref{fig-pack64} we plot $f$ 
against $\rho/\bar{\rho}$ as calculated at $\beta=6.4$ 
after $n_c=30,50,70$ and 80 cooling sweeps. We observe that
$f(\rho=\bar\rho) \geq 1$: so even if we include only 
those instantons that are of below average size,
the gas is still dense. Moreover, even though
the average instanton size decreases as $n_c$ decreases
(see Table~\ref{table-sizes-all}), the total number of charges
increases sufficiently rapidly that the packing
fraction itself gets larger. Thus it is difficult to avoid
the conclusion that the `instanton gas' in the real vacuum 
is a dense one, irrespective of any uncertainties concerning
the identification of the larger instantons.

\subsection{variation with volume}
 
\begin	{figure}[p]
\begin	{center}
\leavevmode
\input	{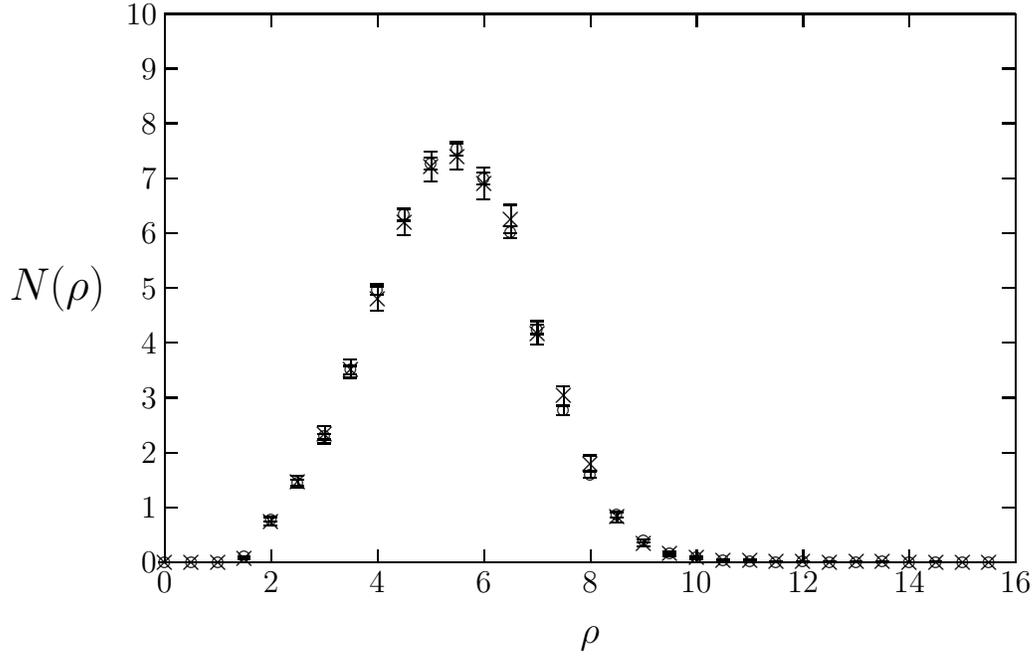}
\end	{center}
\vskip 0.15in
\caption{Comparison of size distributions on $16^3 48$ ($\times$)
and $32^3 64$ ($\circ$) lattices at $\beta=6.0$ after 46 cools.}
\label{fig-size-volume}
\end 	{figure}

Given that our instantons are large, it is important to
check if our size distribution is not distorted by finite
volume effects. In Fig.\ref{fig-size-volume} we
compare the size distributions as obtained on
the $16^3 48$ lattices and on the very much larger 
$32^3 64$ lattices, both generated at $\beta=6.0$ and
both after 46 cooling sweeps.
(The distribution on the larger lattice has been normalised 
to the volume of the smaller.) 
We observe that there are no statistically compelling
differences between the two distributions. In particular, at 
very large $\rho$, where any differences should be most
pronounced, the distributions are virtually identical.
We conclude that our $16^3 48$ lattice at $\beta = 6.0$
suffers from no significant finite volume effects.
Since the $24^3 48$ lattice at $\beta=6.2$ and 
the $32^3 64$ lattice at $\beta=6.4$ have approximately
the same volume in physical units as this lattice,
we shall assume that none of our distributions
suffer significant finite volume corrections.

\subsection{scaling with $\beta$}

The next question, whether the size distribution scales as 
$a \to 0$, is less straightforward. The reason, seen
in Fig.\ref{fig-size-cool60}, is that the number of charges 
varies rapidly with the number of cooling sweeps. 
However a cooling sweep is not
a procedure that scales; 23 cooling sweeps at $\beta =6.0$
are certainly not equivalent to 23 cools at $\beta=6.2$
or 6.4. So at what level of cooling should we compare
the size distributions at different values of $\beta$?  

\begin	{figure}[p]
\begin	{center}
\leavevmode
\input	{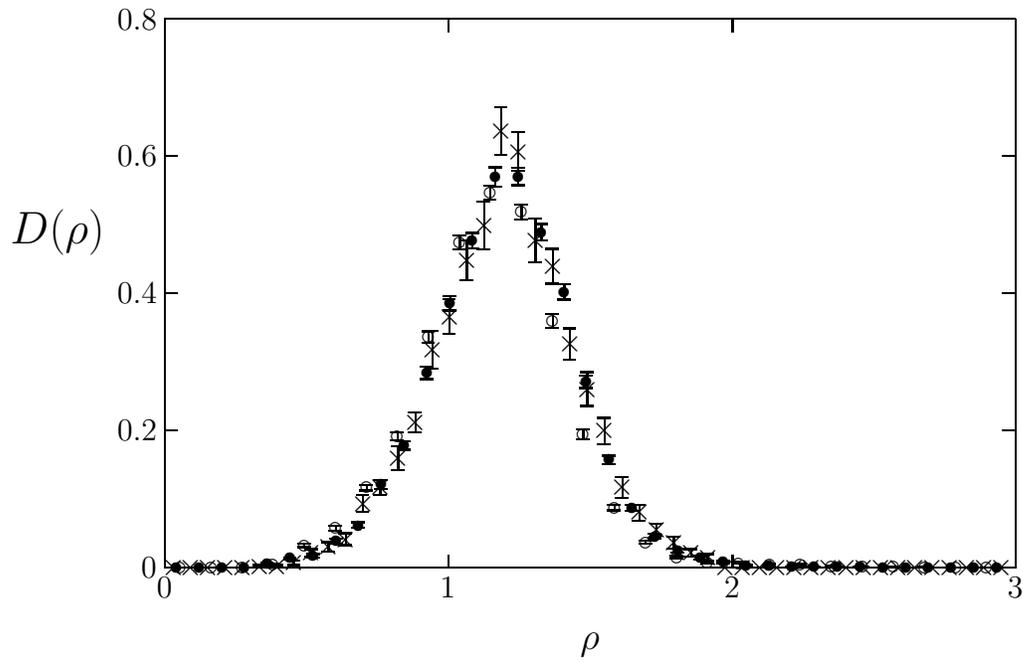}
\end	{center}
\vskip 0.15in
\caption{Density of charges against $\rho$:
at $\beta=6.0$ after 23 cools($\circ$), at $\beta=6.2$ after 46 cools 
($\bullet$) and at $\beta=6.4$ after 80 cools ($\times$). All in
physical units of $1/\surd\sigma$.}
\label{fig-size-beta}
\end 	{figure}

Indeed we can start with a more basic question: is there any
evidence that one can choose the number of cools so that
the distributions scale? The answer to this question
appears to be in the affirmative. In Fig.\ref{fig-size-beta}
we show the size distributions after 23 cools at $\beta=6.0$,
46 cools at $\beta=6.2$ and 80 cools at $\beta=6.4$.
The densities have been scaled by the physical volume,
and $\rho$ is expressed in units of the string tension.
So exact scaling would imply that for some choice of the number
of cools the distributions
should coincide. What we infer from Fig.\ref{fig-size-beta}
is that an approximate coincidence does indeed appear
to be possible.

To be more quantitative we need to set up an equivalence
between the number of cooling sweeps at $\beta=6.0, \ 6.2$
and 6.4. We do so as follows. If the distribution scales
then so does the number density. Let the average number of 
charges per unit physical volume be $N(\beta;n_c)$, 
where $n_c$ is the number of cooling sweeps and eqn(\ref{A18})
is used to define the unit physical volume. Then
$n_c$ cooling sweeps at $\beta$ are defined to be
equivalent to $n_c^{\prime}$ cools at $\beta^{\prime}$
if the number densities are equal:
\begin{equation}
N(\beta;n_c) = N(\beta^{\prime};n_c^{\prime})
\label{A20}
\end{equation}
In Table~\ref{table-sizes-all} we show how the total number 
of charges, $N_{tot}$ varies with $\beta$ and $n_c$. 
The volume of an $L_s^3 L_t$ lattice is 
$V = \{L_s a\surd\sigma\}^3 L_t a\surd\sigma$ 
in physical units, and using the string tensions 
in eqn(\ref{A18}) we can calculate $N=N_{tot}/V$ in each
case and that is also given in  Table~\ref{table-sizes-all}.

At each $\beta$ we can interpolate between the values
in  Table~\ref{table-sizes-all}, so as to obtain
the number density as a function of $n_c$. These
interpolations can then be used in eqn(\ref{A20}) 
to find equivalent sets of $n_c$ at different values of
$\beta$. 

\begin	{figure}[p]
\begin	{center}
\leavevmode
\input	{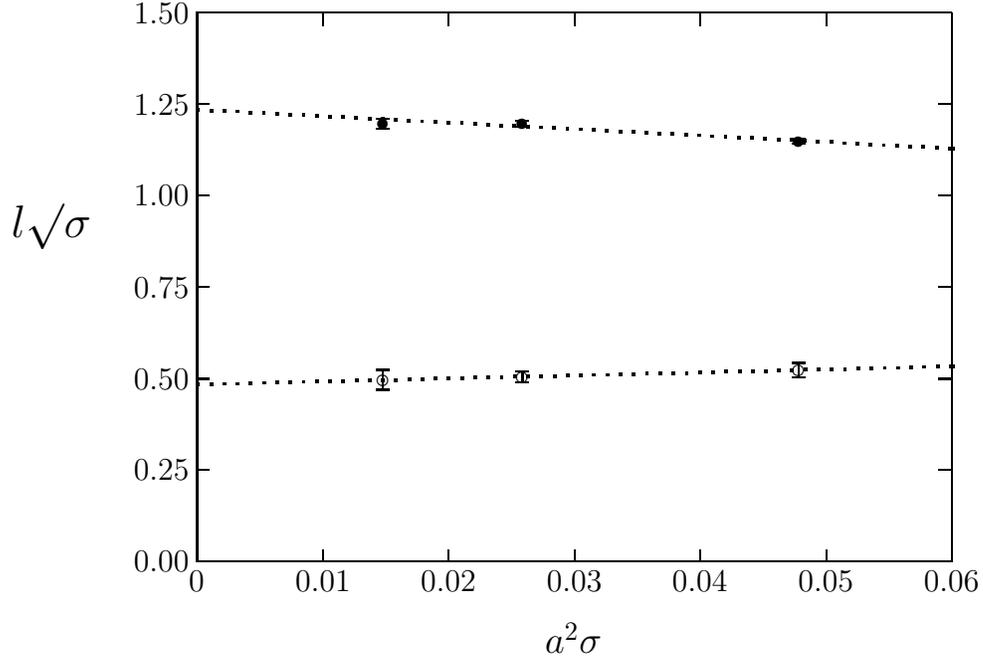}
\end	{center}
\vskip 0.15in
\caption{Average, $l=\bar\rho$($\bullet$), and full-width,
$l=2\sigma_{\rho}$($\circ$), of the instanton size
distributions. Lines are continuum extrapolations.}
\label{fig-rho-scaling}
\end 	{figure}

In fact we immediately see from Table~\ref{table-sizes-all} that
\begin{equation}
n_c = \cases {
23 &$ \ \ \ \ \beta=6.0$ \cr
46 &$ \ \ \ \ \beta=6.2$ \cr
80 &$ \ \ \ \ \beta=6.4$. \cr} 
\label{A20a}
\end{equation}
are, within errors, equivalent at the indicated values of $\beta$.
(This is no accident of course: the number of cooling sweeps
was chosen after a preliminary study designed to produce
such an equivalence.) We note the corresponding values 
of ${\bar\rho}$ and $\sigma_{\rho}$, form dimensionless
ratios with $a\surd\sigma$, and plot these against
$a^2\sigma$ in Fig.\ref{fig-rho-scaling}. The reason for 
plotting things this way is that we expect the leading 
lattice corrections to such dimensionless ratios of physical
quantities to be $O(a^2)$. (We assume that ${\bar\rho}$ and 
$\sigma_{\rho}$ are physical quantities in this sense.)
That is to say, for small enough
$a$ we can extrapolate to the continuum limit using
\begin{equation}
{{{\bar\rho}(a)}{\surd\sigma(a)}}
= {{{\bar\rho}(0)}{\surd\sigma(0)}} + ca^2\sigma      
\label{A21}
\end{equation}
with a similar expression for $\sigma_{\rho}$. These will
be straight lines in  Fig.\ref{fig-rho-scaling} and the
best fits are shown there. As we can see eqn(\ref{A21}) is
compatible with our data. From these fits we obtain the 
continuum predictions:
\begin{equation}
{\bar\rho} = 1.235(20) {1\over{\surd\sigma}} \simeq 0.56(5) fm
\label{A22}
\end{equation}
and 
\begin{equation}
{\sigma_{\rho}} = 0.242(16) {1\over{\surd\sigma}} \simeq 0.11(1) fm
\label{A23}
\end{equation}
where we have used eqn(\ref{A19}) to introduce fermi units.

We note that we are not able to derive a continuum limit for 
other (equivalent) sets of cooling sweeps, because the largest
number of cools at $\beta=6.4$ corresponds, roughly, to the
smallest number at $\beta=6.0$.  

As far as the density of charges is concerned, the
continuum limit is trivially obtained, because
eqn(\ref{A20}) ensures that the number density 
at an equivalent number of cooling sweeps
will be independent of $\beta$. Since this density
varies so rapidly with the number of cools, it
is probably not useful to attempt any conclusion
other than the qualitative one that the charges
are densely packed.

\subsection{variation with cooling}

As we have seen, most quantities that we calculate vary
to some extent with the number of cooling sweeps. Since we are
interested in the physics of the uncooled vacuum, the
logical procedure would be to try and take the $n_c \to 0$
limit of our calculated values. However, to do so would be
to ignore the fact that our procedures become increasingly
unreliable in that limit. For example, the way we correct
the instanton peak height in eqn(\ref{A13}) involves
assumptions that will break down as the instanton
gas becomes increasingly dense, as it does when $n_c$
decreases. Thus it might be that the observed decrease
of, say, $\bar\rho$ as $n_c$ decreases merely reflects this
increasing unreliability.

In the face of this uncertainty, our approach is as follows.
Where we wish to draw a qualitative conclusion, we check
whether the effect becomes more pronounced as $n_c$ decreases.
If that is the case, we take it to be evidence that the
effect under consideration is indeed a property of the uncooled 
vacuum. An example of this is our conclusion that the instanton gas
is dense. If, on the other hand,
we wish to make a statement that is quantitative, then
we pick some small number of cooling sweeps at some $\beta$
and then extrapolate to the continuum limit at an `equivalent'
$n_c(\beta)$ as described above. If the variation with $n_c$
of the quantity under consideration is small enough to be compatible
with the errors of our pattern recognition algorithm, then there
is some reason to believe that our calculation is relevant to
the uncooled vacuum. An example is our calculation above of the
average instanton width. 

To illustrate the uncertainties, we show in
Table~\ref{table-cool-b64} how the properties of a {\it single}
configuration, taken from our $\beta=6.4$ ensemble, vary with the
number of cooling sweeps, $n_c$. (Note that this configuration has
not been subjected to any filtering procedure.) The total number of 
charges, $N_{tot}$, varies so rapidly with $n_c$ that we
cannot hazard any guess at all about the number in the uncooled
vacuum. This is as it should be: perturbative fluctuations in
$F\tilde{F}$ can always be interpreted as a suitable ensemble of 
strongly overlapping topological charges, rendering the question 
of the total number fundamentally ambiguous. The average width,
$\bar\rho$, and typical fluctuations about this average, 
$\sigma_\rho$, vary much less and one might feel 
entitled to infer, for example, that the average width in
the uncooled vacuum is $\bar\rho \sim 9 \pm 1$ in lattice units. 
The decrease in $\bar\rho$ as $n_c \downarrow$ is what one would
naively expect: perturbative fluctuations will, on the average, 
increase the peak heights in $|Q(x)|$ and this will translate
into smaller values of $\rho$ via eqn(\ref{A12}). The ratio
$\sigma_\rho/\bar\rho$ shows little variation with $n_c$
and it seems safe to infer a value of $\sim 0.20 \pm 0.02$
for it. Finally, the total packing fraction $f$ is always
large and for small $n_c$ increases with decreasing $n_c$,
suggesting that it is safe to infer that the instanton gas 
is dense in the uncooled vacuum. (Note that 
the increase of $f$ for large $n_c$ is presumably an artifact
of the lack of filtering -- compare with Fig.\ref{fig-pack64}.) 

\subsection{small $\rho$ and large $\rho$}

In addition to the global features of $D(\rho)$, such
as ${\bar\rho}$, the tails of the distributions
are also of interest. We recall that at small $\rho$ we have 
the prediction from eqns(\ref{A5},\ref{A6}) that 
$N(\rho) \propto \rho^6$. This simple form 
neglects powers of $\log\rho$ (there are factors of $1/g^2(\rho)$
in $D(\rho)$ that arise from the symmetries and which are
subsumed into the `...' in eqn(\ref{A5})) so that it is only
at $very$ small $\rho$ that we would expect it to hold.
And, of course, at very small $\rho$ the cooling
will erase and alter the distribution. So although we shall
fit 
\begin{equation}
N(\rho) \propto {\rho^{\gamma_s}} \ \ \ \ \ :\rho < {\bar\rho} 
\label{A24}
\end{equation}
we are only looking for a trend: that as $a$ and the fitted
range are reduced, and the
number of cooling sweeps becomes small, $\gamma_s$ should
approach the predicted value of $\gamma_s=6$.

Because there are no analytic predictions at large values of 
$\rho$ the behaviour there is of particular interest.
We have tried both exponential and power like fits to the
large-$\rho$ tails of our distributions. In practice the latter
have significantly better $\chi^2$ and are therefore the ones
we present here. That is to say we fit 
\begin{equation}
N(\rho) \propto {1\over{\rho^{\gamma_l}}} \ \ \ \ \ :\rho > {\bar\rho} 
\label{A25}
\end{equation}
for the power $\gamma_l$.

In Table~\ref{table-tail-small} we present some power fits to
the small-$\rho$ tails of our various size distributions.
We show the range fitted (in units of $1/\surd\sigma$). 
The $\chi^2$ of the fit is generally reasonable;
indeed this served as one criterion for which range of $\rho$
to fit. We observe that while the value of $\gamma_s$
does vary a great deal, there does appear to be a trend that as 
we go to smaller $a$ and to a smaller number of cooling
sweeps the value is closer to the asymptotic 
prediction of $\gamma_s = 6$.

In Table~\ref{table-tail-large} we present similar fits to
the large-$\rho$ tails. The values seem quite consistent,
suggesting a power that gradually decreases from
$\gamma_l \sim 12$ to $\gamma_l \sim 10$ as we increase
the number of cools over our range. We also find that there appears 
to be a trend for this power to increase if we shift our fitting
range to larger $\rho$ -- but we cannot be certain of this
with our statistical accuracy. In any case, it is clear that
the suppression at large $\rho$ is
much more severe than the $D(\rho) \propto 1/\rho^5$ that
one would obtain with a coupling that freezes to some
constant value at large distances. This shows that
the full non-perturbative vacuum imposes a sharp infrared 
cut-off on the sizes of instantons.

\section{Correlations of the instantons}

In this section we investigate the correlations between the
topological charges in the vacuum. We shall begin with
the simplest question: how close are nearest neighbour
charges and how does this depend on their relative signs.
This will confirm our picture of a densely packed vacuum,
and so naturally leads to the question whether these charges
show any aspects of a dilute gas. We shall see that the
smallest charges do and the very large ones don't. However
the medium-sized charges show an unexpected behaviour 
which leads us to investigate the charge correlations in
much greater detail. We find long range charge correlations
amongst the smaller charges and, separately, amongst the
larger charges, which is related to an anti-correlation
between the smaller and the larger charges. This effect
weakens as we increase the number of cooling sweeps, so
suggesting that it reflects a property of the uncooled fields.

\subsection{(nearest) neighbours}

We begin by calculating the number of charges that are
a distance $R$ from a given charge. We do so 
separately for the case where the charges have the same
sign (`like') and where they have the opposite sign (`unlike').   
These distributions are calculated by counting the
number of (un)like charges in the spherical shell of
width $\delta R$ a distance $R$ away from each charge.
The distributions are then normalised by the volume of each shell
(for the lattice under consideration and taking the periodicity
into account). So at larger $R$, as the correlations die away, 
we would expect each of these two distributions to go to 
a constant value and that this value should be the same.

\begin	{figure}[p]
\begin	{center}
\leavevmode
\input	{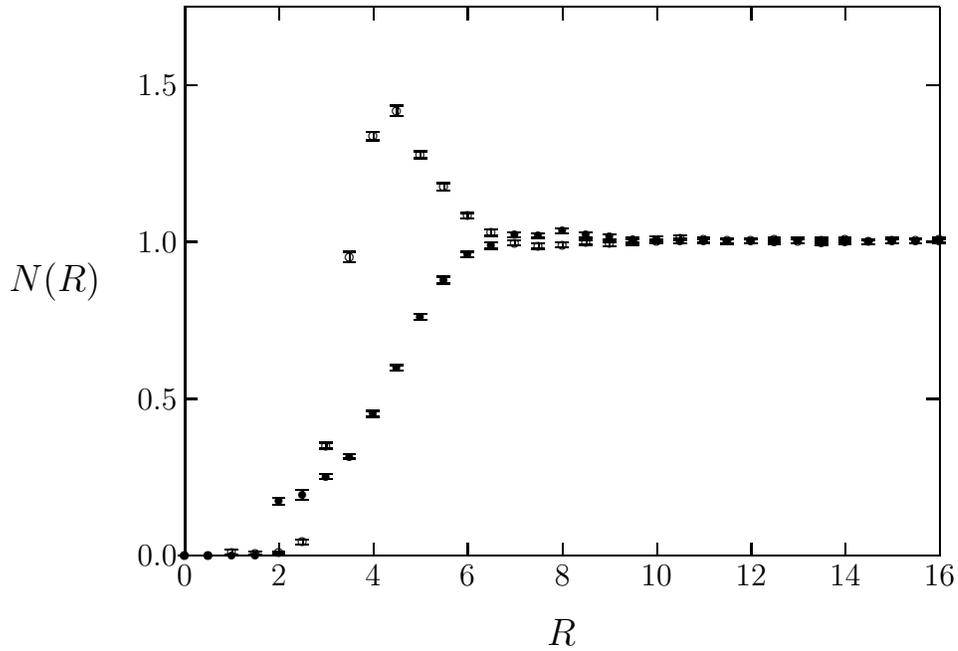}
\end	{center}
\vskip 0.15in
\caption{Number of same sign ($\bullet$) and opposite sign 
($\circ$) charges, per unit volume, as a function
of distance $R$ from the reference charge. At $\beta=6.2$
after 23 cooling sweeps.}
\label{fig-dist-un-like}
\end 	{figure}

In Fig.\ref{fig-dist-un-like} 
we show the distributions we obtain after 23 cooling sweeps
at $\beta=6.2$. These have been normalised so that they
go to unity at large $R$. There are three features one
immediately notes. At small $R$ there is a strong suppression.
Just after that there is a strong enhancement of unlike
charges and a slight enhancement of like charges. Finally 
at large $R$ the distributions are constant as expected.
(The slight enhancement of like charges at very small
$R$ is likely to be an artifact of our procedures.)

The suppression at short distances extends much too far to
be related to the fact that our definition of a peak uses 
$3^4$ hypercubes. In addition, it also occurs on the unfiltered 
data and so is not a product of our filtering procedure.

We note that the like distribution is suppressed to larger 
distances than the unlike one. That is to say, the nearest
neighbour is more likely to be a charge of the opposite
sign. This means that 
topological charges are `screened' by neighbouring charges.
This is reasonable: an $I\bar I$ pair will usually have
a lower action than an $II$ pair. Not so expected is the fact
that $N_{unlike}(R)$ shows a slight dip just after the
enhancement. This coincides with the enhancement in
$N_{like}(R)$ as we see in Fig.\ref{fig-dist-un-like}.
It indicates that there is
a region of $R$ where we have `anti-screening'. 
We shall return to a more detailed investigation of this 
potentially interesting phenomenon shortly.

The suppression at small $R$ and the immediate subsequent
enhancement are best analysed by focussing on the
nearest neighbours to each charge. In Table~\ref{table-nn-sameopp}
we list the average distances to the nearest charges of the
same sign and of the opposite sign. These are presented
in physical units using eqn(\ref{A18}). We note that 
these distances increase with the number of cooling sweeps.
One might try to explain this by arguing that under cooling
the nearest unlike charges should annihilate 
and disappear; while like charges should repel
each other since that lowers the action. Of course this 
argument disregards the complicated nature of the actual
environment around each charge.

If we look at an equivalent number of cooling sweeps,
as given in eqn(\ref{A20a}), we see that the distances
look nearly independent of $a$. This reassures us that
$a$ is small enough that we can extrapolate to the
continuum limit using only the leading $O(a^2)$ correction
just as we did in eqn(\ref{A21}). Doing so we find
that the distance to the nearest like and unlike charges
is 
\begin{equation}
{\bar R}_{like} = 1.081(15) {1\over{\surd\sigma}} \sim 0.49 fm
\label{A26}
\end{equation}
and 
\begin{equation}
{\bar R}_{unlike} = 0.993(13) {1\over{\surd\sigma}} \sim 0.45 fm
\label{A27}
\end{equation}
where we have used eqn(\ref{A19}) to introduce fermi units.

From eqns(\ref{A22},\ref{A23}) and eqns(\ref{A26},\ref{A27})
we see that 
\begin{equation}
{{\bar\rho}\over{\bar R}} \sim 1.2
\label{A28}
\end{equation}
and this confirms our previous conclusion that what we
have is certainly not a dilute gas. 

Although we see from Table~\ref{table-nn-sameopp} that there is some 
variation of ${\bar R}$ with the number of cooling sweeps,
we note a similar variation for ${\bar\rho}$ in 
Table~\ref{table-sizes-all}. Thus eqn(\ref{A28}) is
robust against cooling and is presumably also a property of the
uncooled fields.

\subsection{how dilute a gas?}

The fact that ${{\bar\rho}/{\bar R}} \sim 1$ and that
nearest neighbours are much more likely to be of the
opposite sign, tells us that the topological charges
do not form a dilute gas. It is probable however that
the smaller charges are dilute; if they are weakly
correlated to the large instantons, then they might
still lead to some physics that one would associate with
a dilute gas.

To investigate this possibility we note that in
a dilute gas we have $\langle Q^2\rangle = N_{tot}$, where 
$N_{tot}$ is the total number of charges.
Thus a measure of how close we are to a dilute gas
is provided by seeing how close the quantity 
$\langle Q^2 \rangle /\langle N_{tot}\rangle$, or
the quantity $\langle Q^2/N_{tot}\rangle$, is to unity.
Since we are interested in seeing whether the
smaller instantons form such a dilute gas,
we define the quantity
\begin{equation}
P(\rho_c) \equiv \Biggl\langle
{{Q^2(\rho\leq\rho_c)}\over
{N_{tot}(\rho\leq\rho_c)}}
 \Biggr\rangle.
\label{A29}
\end{equation}
Here 
$Q(\rho\leq\rho_c)=n_I(\rho\leq\rho_c)-n_{\bar I}(\rho\leq\rho_c)$
is the total topological charge of those charges that have
a size less than $\rho_c$; and $N_{tot}(\rho\leq\rho_c)$ is the
corresponding total number of charges.
So how close $P$ is to unity, provides a measure 
of how much these charges behave like a dilute gas. 

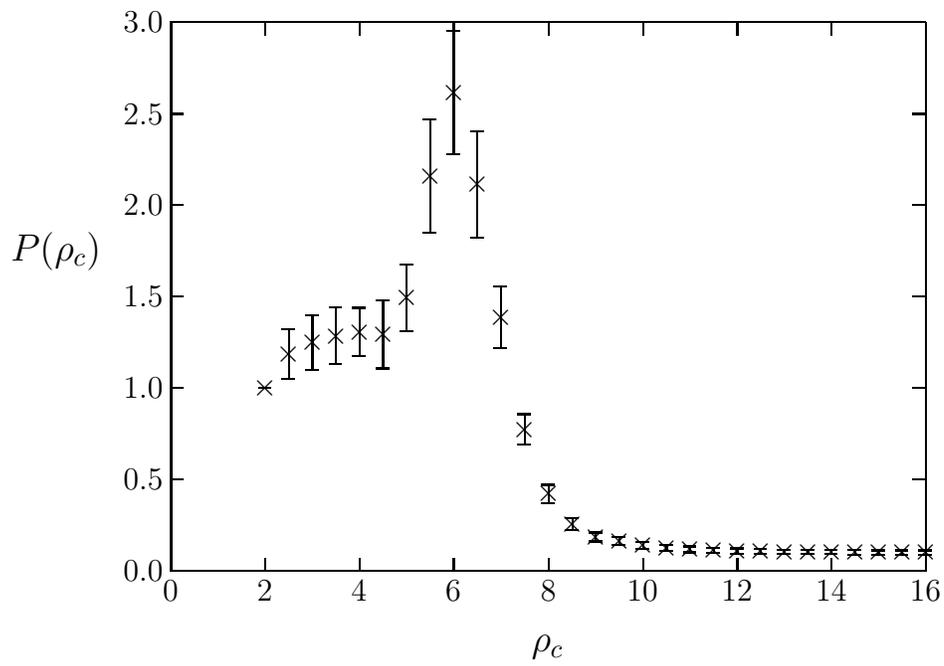
\begin	{figure}[p]
\begin	{center}
\leavevmode
% GNUPLOT: LaTeX picture
\setlength{\unitlength}{0.240900pt}
\ifx\plotpoint\undefined\newsavebox{\plotpoint}\fi
\sbox{\plotpoint}{\rule[-0.200pt]{0.400pt}{0.400pt}}%
\begin{picture}(1349,900)(0,0)
\font\gnuplot=cmr10 at 12pt
\gnuplot
\sbox{\plotpoint}{\rule[-0.200pt]{0.400pt}{0.400pt}}%
\put(120.0,31.0){\rule[-0.200pt]{4.818pt}{0.400pt}}
\put(108,31){\makebox(0,0)[r]{{$0.0$}}}
\put(1285.0,31.0){\rule[-0.200pt]{4.818pt}{0.400pt}}
\put(120.0,175.0){\rule[-0.200pt]{4.818pt}{0.400pt}}
\put(108,175){\makebox(0,0)[r]{{$0.5$}}}
\put(1285.0,175.0){\rule[-0.200pt]{4.818pt}{0.400pt}}
\put(120.0,318.0){\rule[-0.200pt]{4.818pt}{0.400pt}}
\put(108,318){\makebox(0,0)[r]{{$1.0$}}}
\put(1285.0,318.0){\rule[-0.200pt]{4.818pt}{0.400pt}}
\put(120.0,462.0){\rule[-0.200pt]{4.818pt}{0.400pt}}
\put(108,462){\makebox(0,0)[r]{{$1.5$}}}
\put(1285.0,462.0){\rule[-0.200pt]{4.818pt}{0.400pt}}
\put(120.0,606.0){\rule[-0.200pt]{4.818pt}{0.400pt}}
\put(108,606){\makebox(0,0)[r]{{$2.0$}}}
\put(1285.0,606.0){\rule[-0.200pt]{4.818pt}{0.400pt}}
\put(120.0,749.0){\rule[-0.200pt]{4.818pt}{0.400pt}}
\put(108,749){\makebox(0,0)[r]{{$2.5$}}}
\put(1285.0,749.0){\rule[-0.200pt]{4.818pt}{0.400pt}}
\put(120.0,893.0){\rule[-0.200pt]{4.818pt}{0.400pt}}
\put(108,893){\makebox(0,0)[r]{{$3.0$}}}
\put(1285.0,893.0){\rule[-0.200pt]{4.818pt}{0.400pt}}
\put(120.0,31.0){\rule[-0.200pt]{0.400pt}{4.818pt}}
\put(120,19){\makebox(0,0){\shortstack{\\ \\ \\ {$0$}}}}
\put(120.0,873.0){\rule[-0.200pt]{0.400pt}{4.818pt}}
\put(268.0,31.0){\rule[-0.200pt]{0.400pt}{4.818pt}}
\put(268,19){\makebox(0,0){\shortstack{\\ \\ \\ {$2$}}}}
\put(268.0,873.0){\rule[-0.200pt]{0.400pt}{4.818pt}}
\put(416.0,31.0){\rule[-0.200pt]{0.400pt}{4.818pt}}
\put(416,19){\makebox(0,0){\shortstack{\\ \\ \\ {$4$}}}}
\put(416.0,873.0){\rule[-0.200pt]{0.400pt}{4.818pt}}
\put(564.0,31.0){\rule[-0.200pt]{0.400pt}{4.818pt}}
\put(564,19){\makebox(0,0){\shortstack{\\ \\ \\ {$6$}}}}
\put(564.0,873.0){\rule[-0.200pt]{0.400pt}{4.818pt}}
\put(713.0,31.0){\rule[-0.200pt]{0.400pt}{4.818pt}}
\put(713,19){\makebox(0,0){\shortstack{\\ \\ \\ {$8$}}}}
\put(713.0,873.0){\rule[-0.200pt]{0.400pt}{4.818pt}}
\put(861.0,31.0){\rule[-0.200pt]{0.400pt}{4.818pt}}
\put(861,19){\makebox(0,0){\shortstack{\\ \\ \\ {$10$}}}}
\put(861.0,873.0){\rule[-0.200pt]{0.400pt}{4.818pt}}
\put(1009.0,31.0){\rule[-0.200pt]{0.400pt}{4.818pt}}
\put(1009,19){\makebox(0,0){\shortstack{\\ \\ \\ {$12$}}}}
\put(1009.0,873.0){\rule[-0.200pt]{0.400pt}{4.818pt}}
\put(1157.0,31.0){\rule[-0.200pt]{0.400pt}{4.818pt}}
\put(1157,19){\makebox(0,0){\shortstack{\\ \\ \\ {$14$}}}}
\put(1157.0,873.0){\rule[-0.200pt]{0.400pt}{4.818pt}}
\put(1305.0,31.0){\rule[-0.200pt]{0.400pt}{4.818pt}}
\put(1305,19){\makebox(0,0){\shortstack{\\ \\ \\ {$16$}}}}
\put(1305.0,873.0){\rule[-0.200pt]{0.400pt}{4.818pt}}
\put(120.0,31.0){\rule[-0.200pt]{285.466pt}{0.400pt}}
\put(1305.0,31.0){\rule[-0.200pt]{0.400pt}{207.656pt}}
\put(120.0,893.0){\rule[-0.200pt]{285.466pt}{0.400pt}}
\put(-60,534){\makebox(0,0){{\large{$P(\rho_c)$}}}}
\put(712,-89){\makebox(0,0){{\large{$\rho_c$}}}}
\put(120.0,31.0){\rule[-0.200pt]{0.400pt}{207.656pt}}
\put(268,318){\makebox(0,0){$\times$}}
\put(305,372){\makebox(0,0){$\times$}}
\put(342,390){\makebox(0,0){$\times$}}
\put(379,400){\makebox(0,0){$\times$}}
\put(416,406){\makebox(0,0){$\times$}}
\put(453,402){\makebox(0,0){$\times$}}
\put(490,460){\makebox(0,0){$\times$}}
\put(527,651){\makebox(0,0){$\times$}}
\put(564,782){\makebox(0,0){$\times$}}
\put(601,638){\makebox(0,0){$\times$}}
\put(638,429){\makebox(0,0){$\times$}}
\put(675,253){\makebox(0,0){$\times$}}
\put(713,152){\makebox(0,0){$\times$}}
\put(750,104){\makebox(0,0){$\times$}}
\put(787,84){\makebox(0,0){$\times$}}
\put(824,78){\makebox(0,0){$\times$}}
\put(861,71){\makebox(0,0){$\times$}}
\put(898,67){\makebox(0,0){$\times$}}
\put(935,65){\makebox(0,0){$\times$}}
\put(972,63){\makebox(0,0){$\times$}}
\put(1009,62){\makebox(0,0){$\times$}}
\put(1046,62){\makebox(0,0){$\times$}}
\put(1083,61){\makebox(0,0){$\times$}}
\put(1120,60){\makebox(0,0){$\times$}}
\put(1157,60){\makebox(0,0){$\times$}}
\put(1194,60){\makebox(0,0){$\times$}}
\put(1231,60){\makebox(0,0){$\times$}}
\put(1268,60){\makebox(0,0){$\times$}}
\put(1305,60){\makebox(0,0){$\times$}}
\put(268,318){\usebox{\plotpoint}}
\put(258.0,318.0){\rule[-0.200pt]{4.818pt}{0.400pt}}
\put(258.0,318.0){\rule[-0.200pt]{4.818pt}{0.400pt}}
\put(305.0,333.0){\rule[-0.200pt]{0.400pt}{18.790pt}}
\put(295.0,333.0){\rule[-0.200pt]{4.818pt}{0.400pt}}
\put(295.0,411.0){\rule[-0.200pt]{4.818pt}{0.400pt}}
\put(342.0,347.0){\rule[-0.200pt]{0.400pt}{20.717pt}}
\put(332.0,347.0){\rule[-0.200pt]{4.818pt}{0.400pt}}
\put(332.0,433.0){\rule[-0.200pt]{4.818pt}{0.400pt}}
\put(379.0,356.0){\rule[-0.200pt]{0.400pt}{21.440pt}}
\put(369.0,356.0){\rule[-0.200pt]{4.818pt}{0.400pt}}
\put(369.0,445.0){\rule[-0.200pt]{4.818pt}{0.400pt}}
\put(416.0,368.0){\rule[-0.200pt]{0.400pt}{18.308pt}}
\put(406.0,368.0){\rule[-0.200pt]{4.818pt}{0.400pt}}
\put(406.0,444.0){\rule[-0.200pt]{4.818pt}{0.400pt}}
\put(453.0,349.0){\rule[-0.200pt]{0.400pt}{25.776pt}}
\put(443.0,349.0){\rule[-0.200pt]{4.818pt}{0.400pt}}
\put(443.0,456.0){\rule[-0.200pt]{4.818pt}{0.400pt}}
\put(490.0,408.0){\rule[-0.200pt]{0.400pt}{25.054pt}}
\put(480.0,408.0){\rule[-0.200pt]{4.818pt}{0.400pt}}
\put(480.0,512.0){\rule[-0.200pt]{4.818pt}{0.400pt}}
\put(527.0,562.0){\rule[-0.200pt]{0.400pt}{42.880pt}}
\put(517.0,562.0){\rule[-0.200pt]{4.818pt}{0.400pt}}
\put(517.0,740.0){\rule[-0.200pt]{4.818pt}{0.400pt}}
\put(564.0,686.0){\rule[-0.200pt]{0.400pt}{46.494pt}}
\put(554.0,686.0){\rule[-0.200pt]{4.818pt}{0.400pt}}
\put(554.0,879.0){\rule[-0.200pt]{4.818pt}{0.400pt}}
\put(601.0,554.0){\rule[-0.200pt]{0.400pt}{40.230pt}}
\put(591.0,554.0){\rule[-0.200pt]{4.818pt}{0.400pt}}
\put(591.0,721.0){\rule[-0.200pt]{4.818pt}{0.400pt}}
\put(638.0,381.0){\rule[-0.200pt]{0.400pt}{23.367pt}}
\put(628.0,381.0){\rule[-0.200pt]{4.818pt}{0.400pt}}
\put(628.0,478.0){\rule[-0.200pt]{4.818pt}{0.400pt}}
\put(675.0,229.0){\rule[-0.200pt]{0.400pt}{11.563pt}}
\put(665.0,229.0){\rule[-0.200pt]{4.818pt}{0.400pt}}
\put(665.0,277.0){\rule[-0.200pt]{4.818pt}{0.400pt}}
\put(713.0,137.0){\rule[-0.200pt]{0.400pt}{6.986pt}}
\put(703.0,137.0){\rule[-0.200pt]{4.818pt}{0.400pt}}
\put(703.0,166.0){\rule[-0.200pt]{4.818pt}{0.400pt}}
\put(750.0,95.0){\rule[-0.200pt]{0.400pt}{4.577pt}}
\put(740.0,95.0){\rule[-0.200pt]{4.818pt}{0.400pt}}
\put(740.0,114.0){\rule[-0.200pt]{4.818pt}{0.400pt}}
\put(787.0,77.0){\rule[-0.200pt]{0.400pt}{3.373pt}}
\put(777.0,77.0){\rule[-0.200pt]{4.818pt}{0.400pt}}
\put(777.0,91.0){\rule[-0.200pt]{4.818pt}{0.400pt}}
\put(824.0,71.0){\rule[-0.200pt]{0.400pt}{3.132pt}}
\put(814.0,71.0){\rule[-0.200pt]{4.818pt}{0.400pt}}
\put(814.0,84.0){\rule[-0.200pt]{4.818pt}{0.400pt}}
\put(861.0,65.0){\rule[-0.200pt]{0.400pt}{2.650pt}}
\put(851.0,65.0){\rule[-0.200pt]{4.818pt}{0.400pt}}
\put(851.0,76.0){\rule[-0.200pt]{4.818pt}{0.400pt}}
\put(898.0,62.0){\rule[-0.200pt]{0.400pt}{2.168pt}}
\put(888.0,62.0){\rule[-0.200pt]{4.818pt}{0.400pt}}
\put(888.0,71.0){\rule[-0.200pt]{4.818pt}{0.400pt}}
\put(935.0,60.0){\rule[-0.200pt]{0.400pt}{2.168pt}}
\put(925.0,60.0){\rule[-0.200pt]{4.818pt}{0.400pt}}
\put(925.0,69.0){\rule[-0.200pt]{4.818pt}{0.400pt}}
\put(972.0,59.0){\rule[-0.200pt]{0.400pt}{1.927pt}}
\put(962.0,59.0){\rule[-0.200pt]{4.818pt}{0.400pt}}
\put(962.0,67.0){\rule[-0.200pt]{4.818pt}{0.400pt}}
\put(1009.0,58.0){\rule[-0.200pt]{0.400pt}{1.927pt}}
\put(999.0,58.0){\rule[-0.200pt]{4.818pt}{0.400pt}}
\put(999.0,66.0){\rule[-0.200pt]{4.818pt}{0.400pt}}
\put(1046.0,58.0){\rule[-0.200pt]{0.400pt}{1.686pt}}
\put(1036.0,58.0){\rule[-0.200pt]{4.818pt}{0.400pt}}
\put(1036.0,65.0){\rule[-0.200pt]{4.818pt}{0.400pt}}
\put(1083.0,57.0){\rule[-0.200pt]{0.400pt}{1.686pt}}
\put(1073.0,57.0){\rule[-0.200pt]{4.818pt}{0.400pt}}
\put(1073.0,64.0){\rule[-0.200pt]{4.818pt}{0.400pt}}
\put(1120.0,57.0){\rule[-0.200pt]{0.400pt}{1.686pt}}
\put(1110.0,57.0){\rule[-0.200pt]{4.818pt}{0.400pt}}
\put(1110.0,64.0){\rule[-0.200pt]{4.818pt}{0.400pt}}
\put(1157.0,57.0){\rule[-0.200pt]{0.400pt}{1.686pt}}
\put(1147.0,57.0){\rule[-0.200pt]{4.818pt}{0.400pt}}
\put(1147.0,64.0){\rule[-0.200pt]{4.818pt}{0.400pt}}
\put(1194.0,56.0){\rule[-0.200pt]{0.400pt}{1.927pt}}
\put(1184.0,56.0){\rule[-0.200pt]{4.818pt}{0.400pt}}
\put(1184.0,64.0){\rule[-0.200pt]{4.818pt}{0.400pt}}
\put(1231.0,56.0){\rule[-0.200pt]{0.400pt}{1.686pt}}
\put(1221.0,56.0){\rule[-0.200pt]{4.818pt}{0.400pt}}
\put(1221.0,63.0){\rule[-0.200pt]{4.818pt}{0.400pt}}
\put(1268.0,56.0){\rule[-0.200pt]{0.400pt}{1.686pt}}
\put(1258.0,56.0){\rule[-0.200pt]{4.818pt}{0.400pt}}
\put(1258.0,63.0){\rule[-0.200pt]{4.818pt}{0.400pt}}
\put(1305.0,56.0){\rule[-0.200pt]{0.400pt}{1.686pt}}
\put(1295.0,56.0){\rule[-0.200pt]{4.818pt}{0.400pt}}
\put(1295.0,63.0){\rule[-0.200pt]{4.818pt}{0.400pt}}
\end{picture}
\end	{center}
\vskip 0.15in
\caption{$P(\rho_c)$, as defined in eqn(\ref{A29}), 
versus $\rho_c$ at $\beta=6.2$ after 23 cools.}
\label{fig-dga-23}
\end 	{figure}

In Fig.\ref{fig-dga-23} we show how $P(\rho_c)$ varies 
with $\rho_c$ for the $\beta=6.2$ ensemble after 23 cooling 
sweeps. We observe that if we include charges with widths
up to $\rho \simeq 5$ the value of $P$ remains close to unity
indicating a dilute gas structure. As we increase $\rho_c$
beyond this value, $P$ begins to increase rapidly, becoming 
much larger than unity. Around 
$\rho_c \simeq {\bar\rho} \simeq 6$ the value of $P$ begins to
fall and continues falling to values $\ll 1$. The value
for $\rho_c \to \infty$ is the value one gets for $P$
when one includes all the charges. 

The way $P$ behaves at small and at large $\rho_c$ is
not too surprising. For sufficiently small instantons the
combination of low density and small sizes would make
them behave like a dilute gas. For large overlapping
instantons, on the other hand, we would expect a dominance of 
pairs of opposite sign which would suppress the fluctuations of
$Q = n_I - n_{\bar I}$ for a given value of $n_I + n_{\bar I}$,
thus leading to $P<1$. What is much more puzzling is the
$P \gg 1$ peak for $\rho_c \simeq {\bar\rho}$. One can only
have fluctuations of $Q$ that are larger than those of a dilute 
gas if the charges tend to have the same sign. That is to say,
what we are seeing is some kind of charge coherence phenomenon:
there is some interaction that ensures that charges of
a size just less than ${\bar\rho}$ tend to have the same sign.
This is in contrast to the evidence we saw in the 
previous subsection that on the average the nearest neighbour
has the opposite sign. We shall examine and resolve this
puzzle in the next section.

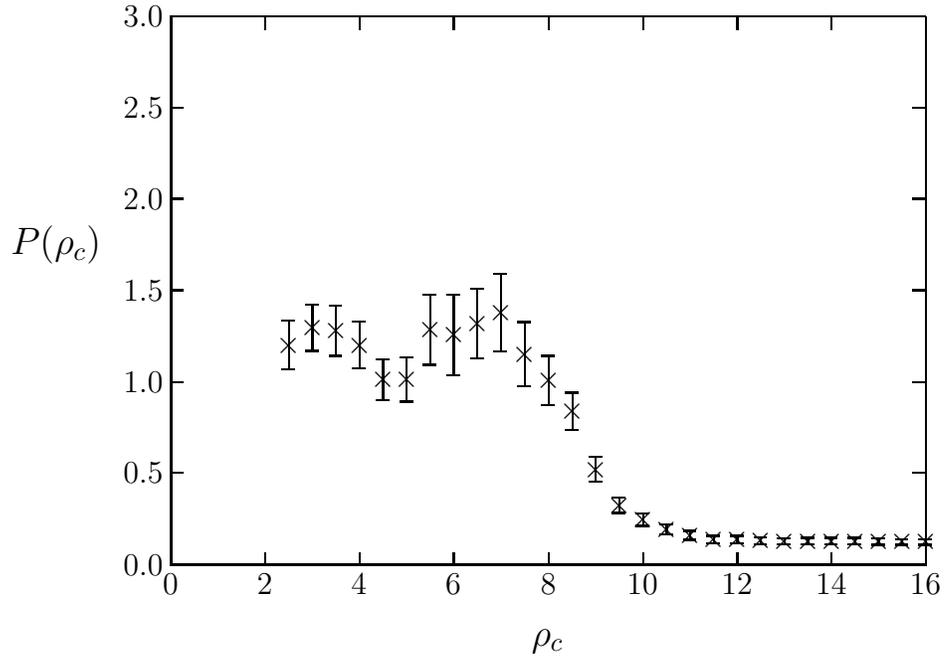
\begin	{figure}[p]
\begin	{center}
\leavevmode
% GNUPLOT: LaTeX picture
\setlength{\unitlength}{0.240900pt}
\ifx\plotpoint\undefined\newsavebox{\plotpoint}\fi
\sbox{\plotpoint}{\rule[-0.200pt]{0.400pt}{0.400pt}}%
\begin{picture}(1349,900)(0,0)
\font\gnuplot=cmr10 at 12pt
\gnuplot
\sbox{\plotpoint}{\rule[-0.200pt]{0.400pt}{0.400pt}}%
\put(120.0,31.0){\rule[-0.200pt]{4.818pt}{0.400pt}}
\put(108,31){\makebox(0,0)[r]{{$0.0$}}}
\put(1285.0,31.0){\rule[-0.200pt]{4.818pt}{0.400pt}}
\put(120.0,175.0){\rule[-0.200pt]{4.818pt}{0.400pt}}
\put(108,175){\makebox(0,0)[r]{{$0.5$}}}
\put(1285.0,175.0){\rule[-0.200pt]{4.818pt}{0.400pt}}
\put(120.0,318.0){\rule[-0.200pt]{4.818pt}{0.400pt}}
\put(108,318){\makebox(0,0)[r]{{$1.0$}}}
\put(1285.0,318.0){\rule[-0.200pt]{4.818pt}{0.400pt}}
\put(120.0,462.0){\rule[-0.200pt]{4.818pt}{0.400pt}}
\put(108,462){\makebox(0,0)[r]{{$1.5$}}}
\put(1285.0,462.0){\rule[-0.200pt]{4.818pt}{0.400pt}}
\put(120.0,606.0){\rule[-0.200pt]{4.818pt}{0.400pt}}
\put(108,606){\makebox(0,0)[r]{{$2.0$}}}
\put(1285.0,606.0){\rule[-0.200pt]{4.818pt}{0.400pt}}
\put(120.0,749.0){\rule[-0.200pt]{4.818pt}{0.400pt}}
\put(108,749){\makebox(0,0)[r]{{$2.5$}}}
\put(1285.0,749.0){\rule[-0.200pt]{4.818pt}{0.400pt}}
\put(120.0,893.0){\rule[-0.200pt]{4.818pt}{0.400pt}}
\put(108,893){\makebox(0,0)[r]{{$3.0$}}}
\put(1285.0,893.0){\rule[-0.200pt]{4.818pt}{0.400pt}}
\put(120.0,31.0){\rule[-0.200pt]{0.400pt}{4.818pt}}
\put(120,19){\makebox(0,0){\shortstack{\\ \\ \\ {$0$}}}}
\put(120.0,873.0){\rule[-0.200pt]{0.400pt}{4.818pt}}
\put(268.0,31.0){\rule[-0.200pt]{0.400pt}{4.818pt}}
\put(268,19){\makebox(0,0){\shortstack{\\ \\ \\ {$2$}}}}
\put(268.0,873.0){\rule[-0.200pt]{0.400pt}{4.818pt}}
\put(416.0,31.0){\rule[-0.200pt]{0.400pt}{4.818pt}}
\put(416,19){\makebox(0,0){\shortstack{\\ \\ \\ {$4$}}}}
\put(416.0,873.0){\rule[-0.200pt]{0.400pt}{4.818pt}}
\put(564.0,31.0){\rule[-0.200pt]{0.400pt}{4.818pt}}
\put(564,19){\makebox(0,0){\shortstack{\\ \\ \\ {$6$}}}}
\put(564.0,873.0){\rule[-0.200pt]{0.400pt}{4.818pt}}
\put(713.0,31.0){\rule[-0.200pt]{0.400pt}{4.818pt}}
\put(713,19){\makebox(0,0){\shortstack{\\ \\ \\ {$8$}}}}
\put(713.0,873.0){\rule[-0.200pt]{0.400pt}{4.818pt}}
\put(861.0,31.0){\rule[-0.200pt]{0.400pt}{4.818pt}}
\put(861,19){\makebox(0,0){\shortstack{\\ \\ \\ {$10$}}}}
\put(861.0,873.0){\rule[-0.200pt]{0.400pt}{4.818pt}}
\put(1009.0,31.0){\rule[-0.200pt]{0.400pt}{4.818pt}}
\put(1009,19){\makebox(0,0){\shortstack{\\ \\ \\ {$12$}}}}
\put(1009.0,873.0){\rule[-0.200pt]{0.400pt}{4.818pt}}
\put(1157.0,31.0){\rule[-0.200pt]{0.400pt}{4.818pt}}
\put(1157,19){\makebox(0,0){\shortstack{\\ \\ \\ {$14$}}}}
\put(1157.0,873.0){\rule[-0.200pt]{0.400pt}{4.818pt}}
\put(1305.0,31.0){\rule[-0.200pt]{0.400pt}{4.818pt}}
\put(1305,19){\makebox(0,0){\shortstack{\\ \\ \\ {$16$}}}}
\put(1305.0,873.0){\rule[-0.200pt]{0.400pt}{4.818pt}}
\put(120.0,31.0){\rule[-0.200pt]{285.466pt}{0.400pt}}
\put(1305.0,31.0){\rule[-0.200pt]{0.400pt}{207.656pt}}
\put(120.0,893.0){\rule[-0.200pt]{285.466pt}{0.400pt}}
\put(-60,534){\makebox(0,0){{\large{$P(\rho_c)$}}}}
\put(712,-89){\makebox(0,0){{\large{$\rho_c$}}}}
\put(120.0,31.0){\rule[-0.200pt]{0.400pt}{207.656pt}}
\put(305,376){\makebox(0,0){$\times$}}
\put(342,403){\makebox(0,0){$\times$}}
\put(379,399){\makebox(0,0){$\times$}}
\put(416,376){\makebox(0,0){$\times$}}
\put(453,322){\makebox(0,0){$\times$}}
\put(490,322){\makebox(0,0){$\times$}}
\put(527,400){\makebox(0,0){$\times$}}
\put(564,392){\makebox(0,0){$\times$}}
\put(601,410){\makebox(0,0){$\times$}}
\put(638,427){\makebox(0,0){$\times$}}
\put(675,361){\makebox(0,0){$\times$}}
\put(713,321){\makebox(0,0){$\times$}}
\put(750,272){\makebox(0,0){$\times$}}
\put(787,180){\makebox(0,0){$\times$}}
\put(824,124){\makebox(0,0){$\times$}}
\put(861,102){\makebox(0,0){$\times$}}
\put(898,86){\makebox(0,0){$\times$}}
\put(935,77){\makebox(0,0){$\times$}}
\put(972,70){\makebox(0,0){$\times$}}
\put(1009,70){\makebox(0,0){$\times$}}
\put(1046,69){\makebox(0,0){$\times$}}
\put(1083,67){\makebox(0,0){$\times$}}
\put(1120,68){\makebox(0,0){$\times$}}
\put(1157,68){\makebox(0,0){$\times$}}
\put(1194,68){\makebox(0,0){$\times$}}
\put(1231,67){\makebox(0,0){$\times$}}
\put(1268,67){\makebox(0,0){$\times$}}
\put(1305,67){\makebox(0,0){$\times$}}
\put(305.0,338.0){\rule[-0.200pt]{0.400pt}{18.308pt}}
\put(295.0,338.0){\rule[-0.200pt]{4.818pt}{0.400pt}}
\put(295.0,414.0){\rule[-0.200pt]{4.818pt}{0.400pt}}
\put(342.0,367.0){\rule[-0.200pt]{0.400pt}{17.345pt}}
\put(332.0,367.0){\rule[-0.200pt]{4.818pt}{0.400pt}}
\put(332.0,439.0){\rule[-0.200pt]{4.818pt}{0.400pt}}
\put(379.0,359.0){\rule[-0.200pt]{0.400pt}{19.031pt}}
\put(369.0,359.0){\rule[-0.200pt]{4.818pt}{0.400pt}}
\put(369.0,438.0){\rule[-0.200pt]{4.818pt}{0.400pt}}
\put(416.0,340.0){\rule[-0.200pt]{0.400pt}{17.586pt}}
\put(406.0,340.0){\rule[-0.200pt]{4.818pt}{0.400pt}}
\put(406.0,413.0){\rule[-0.200pt]{4.818pt}{0.400pt}}
\put(453.0,290.0){\rule[-0.200pt]{0.400pt}{15.177pt}}
\put(443.0,290.0){\rule[-0.200pt]{4.818pt}{0.400pt}}
\put(443.0,353.0){\rule[-0.200pt]{4.818pt}{0.400pt}}
\put(490.0,287.0){\rule[-0.200pt]{0.400pt}{16.863pt}}
\put(480.0,287.0){\rule[-0.200pt]{4.818pt}{0.400pt}}
\put(480.0,357.0){\rule[-0.200pt]{4.818pt}{0.400pt}}
\put(527.0,345.0){\rule[-0.200pt]{0.400pt}{26.499pt}}
\put(517.0,345.0){\rule[-0.200pt]{4.818pt}{0.400pt}}
\put(517.0,455.0){\rule[-0.200pt]{4.818pt}{0.400pt}}
\put(564.0,329.0){\rule[-0.200pt]{0.400pt}{30.353pt}}
\put(554.0,329.0){\rule[-0.200pt]{4.818pt}{0.400pt}}
\put(554.0,455.0){\rule[-0.200pt]{4.818pt}{0.400pt}}
\put(601.0,355.0){\rule[-0.200pt]{0.400pt}{26.499pt}}
\put(591.0,355.0){\rule[-0.200pt]{4.818pt}{0.400pt}}
\put(591.0,465.0){\rule[-0.200pt]{4.818pt}{0.400pt}}
\put(638.0,366.0){\rule[-0.200pt]{0.400pt}{29.390pt}}
\put(628.0,366.0){\rule[-0.200pt]{4.818pt}{0.400pt}}
\put(628.0,488.0){\rule[-0.200pt]{4.818pt}{0.400pt}}
\put(675.0,311.0){\rule[-0.200pt]{0.400pt}{24.331pt}}
\put(665.0,311.0){\rule[-0.200pt]{4.818pt}{0.400pt}}
\put(665.0,412.0){\rule[-0.200pt]{4.818pt}{0.400pt}}
\put(713.0,282.0){\rule[-0.200pt]{0.400pt}{18.549pt}}
\put(703.0,282.0){\rule[-0.200pt]{4.818pt}{0.400pt}}
\put(703.0,359.0){\rule[-0.200pt]{4.818pt}{0.400pt}}
\put(750.0,243.0){\rule[-0.200pt]{0.400pt}{13.972pt}}
\put(740.0,243.0){\rule[-0.200pt]{4.818pt}{0.400pt}}
\put(740.0,301.0){\rule[-0.200pt]{4.818pt}{0.400pt}}
\put(787.0,161.0){\rule[-0.200pt]{0.400pt}{9.395pt}}
\put(777.0,161.0){\rule[-0.200pt]{4.818pt}{0.400pt}}
\put(777.0,200.0){\rule[-0.200pt]{4.818pt}{0.400pt}}
\put(824.0,112.0){\rule[-0.200pt]{0.400pt}{5.782pt}}
\put(814.0,112.0){\rule[-0.200pt]{4.818pt}{0.400pt}}
\put(814.0,136.0){\rule[-0.200pt]{4.818pt}{0.400pt}}
\put(861.0,92.0){\rule[-0.200pt]{0.400pt}{4.577pt}}
\put(851.0,92.0){\rule[-0.200pt]{4.818pt}{0.400pt}}
\put(851.0,111.0){\rule[-0.200pt]{4.818pt}{0.400pt}}
\put(898.0,78.0){\rule[-0.200pt]{0.400pt}{3.854pt}}
\put(888.0,78.0){\rule[-0.200pt]{4.818pt}{0.400pt}}
\put(888.0,94.0){\rule[-0.200pt]{4.818pt}{0.400pt}}
\put(935.0,70.0){\rule[-0.200pt]{0.400pt}{3.373pt}}
\put(925.0,70.0){\rule[-0.200pt]{4.818pt}{0.400pt}}
\put(925.0,84.0){\rule[-0.200pt]{4.818pt}{0.400pt}}
\put(972.0,65.0){\rule[-0.200pt]{0.400pt}{2.650pt}}
\put(962.0,65.0){\rule[-0.200pt]{4.818pt}{0.400pt}}
\put(962.0,76.0){\rule[-0.200pt]{4.818pt}{0.400pt}}
\put(1009.0,65.0){\rule[-0.200pt]{0.400pt}{2.650pt}}
\put(999.0,65.0){\rule[-0.200pt]{4.818pt}{0.400pt}}
\put(999.0,76.0){\rule[-0.200pt]{4.818pt}{0.400pt}}
\put(1046.0,63.0){\rule[-0.200pt]{0.400pt}{2.650pt}}
\put(1036.0,63.0){\rule[-0.200pt]{4.818pt}{0.400pt}}
\put(1036.0,74.0){\rule[-0.200pt]{4.818pt}{0.400pt}}
\put(1083.0,63.0){\rule[-0.200pt]{0.400pt}{2.168pt}}
\put(1073.0,63.0){\rule[-0.200pt]{4.818pt}{0.400pt}}
\put(1073.0,72.0){\rule[-0.200pt]{4.818pt}{0.400pt}}
\put(1120.0,63.0){\rule[-0.200pt]{0.400pt}{2.409pt}}
\put(1110.0,63.0){\rule[-0.200pt]{4.818pt}{0.400pt}}
\put(1110.0,73.0){\rule[-0.200pt]{4.818pt}{0.400pt}}
\put(1157.0,63.0){\rule[-0.200pt]{0.400pt}{2.409pt}}
\put(1147.0,63.0){\rule[-0.200pt]{4.818pt}{0.400pt}}
\put(1147.0,73.0){\rule[-0.200pt]{4.818pt}{0.400pt}}
\put(1194.0,63.0){\rule[-0.200pt]{0.400pt}{2.409pt}}
\put(1184.0,63.0){\rule[-0.200pt]{4.818pt}{0.400pt}}
\put(1184.0,73.0){\rule[-0.200pt]{4.818pt}{0.400pt}}
\put(1231.0,62.0){\rule[-0.200pt]{0.400pt}{2.409pt}}
\put(1221.0,62.0){\rule[-0.200pt]{4.818pt}{0.400pt}}
\put(1221.0,72.0){\rule[-0.200pt]{4.818pt}{0.400pt}}
\put(1268.0,62.0){\rule[-0.200pt]{0.400pt}{2.168pt}}
\put(1258.0,62.0){\rule[-0.200pt]{4.818pt}{0.400pt}}
\put(1258.0,71.0){\rule[-0.200pt]{4.818pt}{0.400pt}}
\put(1305.0,62.0){\rule[-0.200pt]{0.400pt}{2.168pt}}
\put(1295.0,62.0){\rule[-0.200pt]{4.818pt}{0.400pt}}
\put(1295.0,71.0){\rule[-0.200pt]{4.818pt}{0.400pt}}
\end{picture}
\end	{center}
\vskip 0.15in
\caption{$P(\rho_c)$, as defined in eqn(\ref{A29}),
versus $\rho_c$ at $\beta=6.2$ after 46 cools.}
\label{fig-dga-46}
\end 	{figure}

Is this an artifact of cooling? In Fig.\ref{fig-dga-46}
we show how $P(\rho_c)$ varies with $\rho_c$ after 46 cooling 
sweeps. For large and small $\rho_c$ things are much the same 
as after 23 cools. However the peak near $\bar\rho$ has all
but disappeared. This indicates that cooling erases this 
interesting effect: it thus appears that this is 
a feature of the uncooled vacuum.
By comparing comparable plots at different $\beta$ we
find that, as long as the comparison is performed at
equivalent numbers of cooling sweeps (in the sense of
eqn(\ref{A20})), this phemonenon seems to roughly
scale.

\subsection{screening and polarisation}

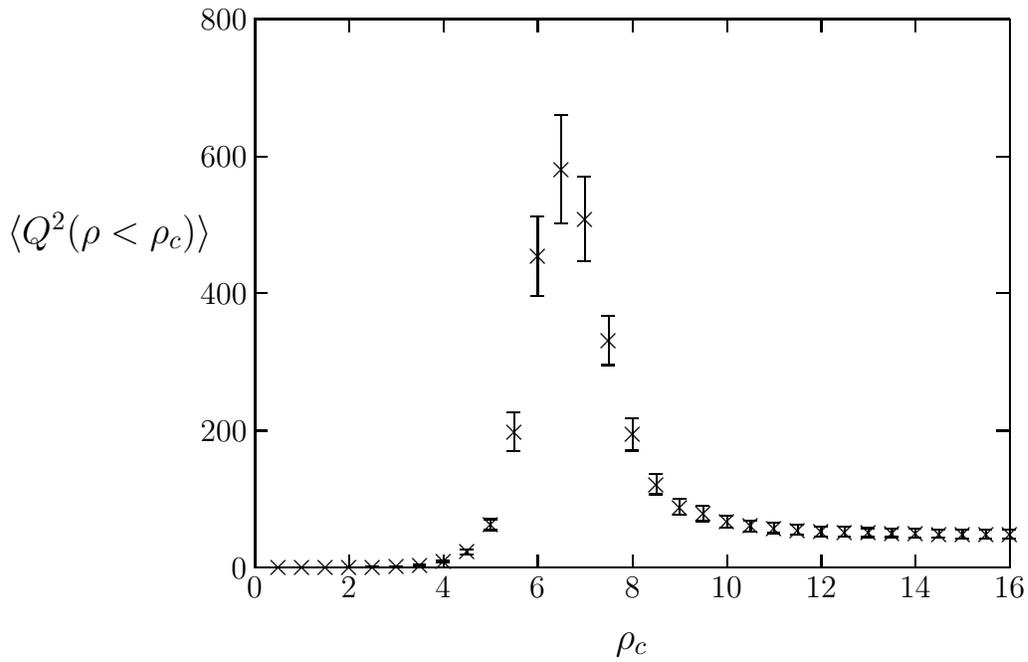
\begin	{figure}[p]
\begin	{center}
\leavevmode
% GNUPLOT: LaTeX picture
\setlength{\unitlength}{0.240900pt}
\ifx\plotpoint\undefined\newsavebox{\plotpoint}\fi
\sbox{\plotpoint}{\rule[-0.200pt]{0.400pt}{0.400pt}}%
\begin{picture}(1349,900)(0,0)
\font\gnuplot=cmr10 at 12pt
\gnuplot
\sbox{\plotpoint}{\rule[-0.200pt]{0.400pt}{0.400pt}}%
\put(120.0,31.0){\rule[-0.200pt]{4.818pt}{0.400pt}}
\put(108,31){\makebox(0,0)[r]{{$0$}}}
\put(1285.0,31.0){\rule[-0.200pt]{4.818pt}{0.400pt}}
\put(120.0,246.0){\rule[-0.200pt]{4.818pt}{0.400pt}}
\put(108,246){\makebox(0,0)[r]{{$200$}}}
\put(1285.0,246.0){\rule[-0.200pt]{4.818pt}{0.400pt}}
\put(120.0,462.0){\rule[-0.200pt]{4.818pt}{0.400pt}}
\put(108,462){\makebox(0,0)[r]{{$400$}}}
\put(1285.0,462.0){\rule[-0.200pt]{4.818pt}{0.400pt}}
\put(120.0,677.0){\rule[-0.200pt]{4.818pt}{0.400pt}}
\put(108,677){\makebox(0,0)[r]{{$600$}}}
\put(1285.0,677.0){\rule[-0.200pt]{4.818pt}{0.400pt}}
\put(120.0,893.0){\rule[-0.200pt]{4.818pt}{0.400pt}}
\put(108,893){\makebox(0,0)[r]{{$800$}}}
\put(1285.0,893.0){\rule[-0.200pt]{4.818pt}{0.400pt}}
\put(120.0,31.0){\rule[-0.200pt]{0.400pt}{4.818pt}}
\put(120,19){\makebox(0,0){\shortstack{\\ \\ \\ {$0$}}}}
\put(120.0,873.0){\rule[-0.200pt]{0.400pt}{4.818pt}}
\put(268.0,31.0){\rule[-0.200pt]{0.400pt}{4.818pt}}
\put(268,19){\makebox(0,0){\shortstack{\\ \\ \\ {$2$}}}}
\put(268.0,873.0){\rule[-0.200pt]{0.400pt}{4.818pt}}
\put(416.0,31.0){\rule[-0.200pt]{0.400pt}{4.818pt}}
\put(416,19){\makebox(0,0){\shortstack{\\ \\ \\ {$4$}}}}
\put(416.0,873.0){\rule[-0.200pt]{0.400pt}{4.818pt}}
\put(564.0,31.0){\rule[-0.200pt]{0.400pt}{4.818pt}}
\put(564,19){\makebox(0,0){\shortstack{\\ \\ \\ {$6$}}}}
\put(564.0,873.0){\rule[-0.200pt]{0.400pt}{4.818pt}}
\put(713.0,31.0){\rule[-0.200pt]{0.400pt}{4.818pt}}
\put(713,19){\makebox(0,0){\shortstack{\\ \\ \\ {$8$}}}}
\put(713.0,873.0){\rule[-0.200pt]{0.400pt}{4.818pt}}
\put(861.0,31.0){\rule[-0.200pt]{0.400pt}{4.818pt}}
\put(861,19){\makebox(0,0){\shortstack{\\ \\ \\ {$10$}}}}
\put(861.0,873.0){\rule[-0.200pt]{0.400pt}{4.818pt}}
\put(1009.0,31.0){\rule[-0.200pt]{0.400pt}{4.818pt}}
\put(1009,19){\makebox(0,0){\shortstack{\\ \\ \\ {$12$}}}}
\put(1009.0,873.0){\rule[-0.200pt]{0.400pt}{4.818pt}}
\put(1157.0,31.0){\rule[-0.200pt]{0.400pt}{4.818pt}}
\put(1157,19){\makebox(0,0){\shortstack{\\ \\ \\ {$14$}}}}
\put(1157.0,873.0){\rule[-0.200pt]{0.400pt}{4.818pt}}
\put(1305.0,31.0){\rule[-0.200pt]{0.400pt}{4.818pt}}
\put(1305,19){\makebox(0,0){\shortstack{\\ \\ \\ {$16$}}}}
\put(1305.0,873.0){\rule[-0.200pt]{0.400pt}{4.818pt}}
\put(120.0,31.0){\rule[-0.200pt]{285.466pt}{0.400pt}}
\put(1305.0,31.0){\rule[-0.200pt]{0.400pt}{207.656pt}}
\put(120.0,893.0){\rule[-0.200pt]{285.466pt}{0.400pt}}
\put(-108,558){\makebox(0,0){{\large{$\langle Q^2(\rho<\rho_c)\rangle$}}}}
\put(712,-89){\makebox(0,0){{\large{$\rho_c$}}}}
\put(120.0,31.0){\rule[-0.200pt]{0.400pt}{207.656pt}}
\put(157,31){\makebox(0,0){$\times$}}
\put(194,31){\makebox(0,0){$\times$}}
\put(231,31){\makebox(0,0){$\times$}}
\put(268,31){\makebox(0,0){$\times$}}
\put(305,31){\makebox(0,0){$\times$}}
\put(342,33){\makebox(0,0){$\times$}}
\put(379,35){\makebox(0,0){$\times$}}
\put(416,41){\makebox(0,0){$\times$}}
\put(453,56){\makebox(0,0){$\times$}}
\put(490,99){\makebox(0,0){$\times$}}
\put(527,244){\makebox(0,0){$\times$}}
\put(564,520){\makebox(0,0){$\times$}}
\put(601,657){\makebox(0,0){$\times$}}
\put(638,579){\makebox(0,0){$\times$}}
\put(675,388){\makebox(0,0){$\times$}}
\put(713,241){\makebox(0,0){$\times$}}
\put(750,162){\makebox(0,0){$\times$}}
\put(787,126){\makebox(0,0){$\times$}}
\put(824,116){\makebox(0,0){$\times$}}
\put(861,104){\makebox(0,0){$\times$}}
\put(898,97){\makebox(0,0){$\times$}}
\put(935,93){\makebox(0,0){$\times$}}
\put(972,90){\makebox(0,0){$\times$}}
\put(1009,88){\makebox(0,0){$\times$}}
\put(1046,87){\makebox(0,0){$\times$}}
\put(1083,86){\makebox(0,0){$\times$}}
\put(1120,85){\makebox(0,0){$\times$}}
\put(1157,85){\makebox(0,0){$\times$}}
\put(1194,84){\makebox(0,0){$\times$}}
\put(1231,84){\makebox(0,0){$\times$}}
\put(1268,84){\makebox(0,0){$\times$}}
\put(1305,84){\makebox(0,0){$\times$}}
\put(157,31){\usebox{\plotpoint}}
\put(147.0,31.0){\rule[-0.200pt]{4.818pt}{0.400pt}}
\put(147.0,31.0){\rule[-0.200pt]{4.818pt}{0.400pt}}
\put(194,31){\usebox{\plotpoint}}
\put(184.0,31.0){\rule[-0.200pt]{4.818pt}{0.400pt}}
\put(184.0,31.0){\rule[-0.200pt]{4.818pt}{0.400pt}}
\put(231,31){\usebox{\plotpoint}}
\put(221.0,31.0){\rule[-0.200pt]{4.818pt}{0.400pt}}
\put(221.0,31.0){\rule[-0.200pt]{4.818pt}{0.400pt}}
\put(268,31){\usebox{\plotpoint}}
\put(258.0,31.0){\rule[-0.200pt]{4.818pt}{0.400pt}}
\put(258.0,31.0){\rule[-0.200pt]{4.818pt}{0.400pt}}
\put(305.0,31.0){\usebox{\plotpoint}}
\put(295.0,31.0){\rule[-0.200pt]{4.818pt}{0.400pt}}
\put(295.0,32.0){\rule[-0.200pt]{4.818pt}{0.400pt}}
\put(342.0,32.0){\usebox{\plotpoint}}
\put(332.0,32.0){\rule[-0.200pt]{4.818pt}{0.400pt}}
\put(332.0,33.0){\rule[-0.200pt]{4.818pt}{0.400pt}}
\put(379.0,34.0){\rule[-0.200pt]{0.400pt}{0.482pt}}
\put(369.0,34.0){\rule[-0.200pt]{4.818pt}{0.400pt}}
\put(369.0,36.0){\rule[-0.200pt]{4.818pt}{0.400pt}}
\put(416.0,40.0){\rule[-0.200pt]{0.400pt}{0.482pt}}
\put(406.0,40.0){\rule[-0.200pt]{4.818pt}{0.400pt}}
\put(406.0,42.0){\rule[-0.200pt]{4.818pt}{0.400pt}}
\put(453.0,52.0){\rule[-0.200pt]{0.400pt}{1.686pt}}
\put(443.0,52.0){\rule[-0.200pt]{4.818pt}{0.400pt}}
\put(443.0,59.0){\rule[-0.200pt]{4.818pt}{0.400pt}}
\put(490.0,90.0){\rule[-0.200pt]{0.400pt}{4.095pt}}
\put(480.0,90.0){\rule[-0.200pt]{4.818pt}{0.400pt}}
\put(480.0,107.0){\rule[-0.200pt]{4.818pt}{0.400pt}}
\put(527.0,214.0){\rule[-0.200pt]{0.400pt}{14.695pt}}
\put(517.0,214.0){\rule[-0.200pt]{4.818pt}{0.400pt}}
\put(517.0,275.0){\rule[-0.200pt]{4.818pt}{0.400pt}}
\put(564.0,458.0){\rule[-0.200pt]{0.400pt}{30.112pt}}
\put(554.0,458.0){\rule[-0.200pt]{4.818pt}{0.400pt}}
\put(554.0,583.0){\rule[-0.200pt]{4.818pt}{0.400pt}}
\put(601.0,572.0){\rule[-0.200pt]{0.400pt}{40.953pt}}
\put(591.0,572.0){\rule[-0.200pt]{4.818pt}{0.400pt}}
\put(591.0,742.0){\rule[-0.200pt]{4.818pt}{0.400pt}}
\put(638.0,512.0){\rule[-0.200pt]{0.400pt}{32.040pt}}
\put(628.0,512.0){\rule[-0.200pt]{4.818pt}{0.400pt}}
\put(628.0,645.0){\rule[-0.200pt]{4.818pt}{0.400pt}}
\put(675.0,349.0){\rule[-0.200pt]{0.400pt}{18.790pt}}
\put(665.0,349.0){\rule[-0.200pt]{4.818pt}{0.400pt}}
\put(665.0,427.0){\rule[-0.200pt]{4.818pt}{0.400pt}}
\put(713.0,215.0){\rule[-0.200pt]{0.400pt}{12.286pt}}
\put(703.0,215.0){\rule[-0.200pt]{4.818pt}{0.400pt}}
\put(703.0,266.0){\rule[-0.200pt]{4.818pt}{0.400pt}}
\put(750.0,146.0){\rule[-0.200pt]{0.400pt}{7.709pt}}
\put(740.0,146.0){\rule[-0.200pt]{4.818pt}{0.400pt}}
\put(740.0,178.0){\rule[-0.200pt]{4.818pt}{0.400pt}}
\put(787.0,114.0){\rule[-0.200pt]{0.400pt}{6.022pt}}
\put(777.0,114.0){\rule[-0.200pt]{4.818pt}{0.400pt}}
\put(777.0,139.0){\rule[-0.200pt]{4.818pt}{0.400pt}}
\put(824.0,104.0){\rule[-0.200pt]{0.400pt}{5.782pt}}
\put(814.0,104.0){\rule[-0.200pt]{4.818pt}{0.400pt}}
\put(814.0,128.0){\rule[-0.200pt]{4.818pt}{0.400pt}}
\put(861.0,94.0){\rule[-0.200pt]{0.400pt}{4.577pt}}
\put(851.0,94.0){\rule[-0.200pt]{4.818pt}{0.400pt}}
\put(851.0,113.0){\rule[-0.200pt]{4.818pt}{0.400pt}}
\put(898.0,88.0){\rule[-0.200pt]{0.400pt}{4.095pt}}
\put(888.0,88.0){\rule[-0.200pt]{4.818pt}{0.400pt}}
\put(888.0,105.0){\rule[-0.200pt]{4.818pt}{0.400pt}}
\put(935.0,85.0){\rule[-0.200pt]{0.400pt}{3.854pt}}
\put(925.0,85.0){\rule[-0.200pt]{4.818pt}{0.400pt}}
\put(925.0,101.0){\rule[-0.200pt]{4.818pt}{0.400pt}}
\put(972.0,83.0){\rule[-0.200pt]{0.400pt}{3.613pt}}
\put(962.0,83.0){\rule[-0.200pt]{4.818pt}{0.400pt}}
\put(962.0,98.0){\rule[-0.200pt]{4.818pt}{0.400pt}}
\put(1009.0,80.0){\rule[-0.200pt]{0.400pt}{3.613pt}}
\put(999.0,80.0){\rule[-0.200pt]{4.818pt}{0.400pt}}
\put(999.0,95.0){\rule[-0.200pt]{4.818pt}{0.400pt}}
\put(1046.0,80.0){\rule[-0.200pt]{0.400pt}{3.613pt}}
\put(1036.0,80.0){\rule[-0.200pt]{4.818pt}{0.400pt}}
\put(1036.0,95.0){\rule[-0.200pt]{4.818pt}{0.400pt}}
\put(1083.0,79.0){\rule[-0.200pt]{0.400pt}{3.373pt}}
\put(1073.0,79.0){\rule[-0.200pt]{4.818pt}{0.400pt}}
\put(1073.0,93.0){\rule[-0.200pt]{4.818pt}{0.400pt}}
\put(1120.0,79.0){\rule[-0.200pt]{0.400pt}{3.132pt}}
\put(1110.0,79.0){\rule[-0.200pt]{4.818pt}{0.400pt}}
\put(1110.0,92.0){\rule[-0.200pt]{4.818pt}{0.400pt}}
\put(1157.0,78.0){\rule[-0.200pt]{0.400pt}{3.373pt}}
\put(1147.0,78.0){\rule[-0.200pt]{4.818pt}{0.400pt}}
\put(1147.0,92.0){\rule[-0.200pt]{4.818pt}{0.400pt}}
\put(1194.0,78.0){\rule[-0.200pt]{0.400pt}{3.132pt}}
\put(1184.0,78.0){\rule[-0.200pt]{4.818pt}{0.400pt}}
\put(1184.0,91.0){\rule[-0.200pt]{4.818pt}{0.400pt}}
\put(1231.0,77.0){\rule[-0.200pt]{0.400pt}{3.373pt}}
\put(1221.0,77.0){\rule[-0.200pt]{4.818pt}{0.400pt}}
\put(1221.0,91.0){\rule[-0.200pt]{4.818pt}{0.400pt}}
\put(1268.0,77.0){\rule[-0.200pt]{0.400pt}{3.373pt}}
\put(1258.0,77.0){\rule[-0.200pt]{4.818pt}{0.400pt}}
\put(1258.0,91.0){\rule[-0.200pt]{4.818pt}{0.400pt}}
\put(1305.0,77.0){\rule[-0.200pt]{0.400pt}{3.373pt}}
\put(1295.0,77.0){\rule[-0.200pt]{4.818pt}{0.400pt}}
\put(1295.0,91.0){\rule[-0.200pt]{4.818pt}{0.400pt}}
\end{picture}
\end	{center}
\vskip 0.15in
\caption{$\langle Q^2(\rho<\rho_c)\rangle$ 
against $\rho_c$ at $\beta=6.2$ after 23 cools.}
\label{fig-Q2-rho}
\end 	{figure}

As we have seen, $P(\infty) \ll 1$; i.e. if we include all the
charges one finds 
$\langle Q^2 \rangle \ll \langle n_I + n_{\bar I} \rangle $. However
we have also seen that if we take the $\sim 50\%$ of the charges
with $\rho \leq {\bar\rho}$, then one finds that
$\langle Q^2\rangle > \langle n_I + n_{\bar I}\rangle $. 
This suggests that if we 
look at $\langle Q^2 \rangle$ as a function of instanton size 
we will see some dramatic effects. In Fig.\ref{fig-Q2-rho} we plot
the value of $\langle Q^2(\rho\leq\rho_c) \rangle$ versus $\rho_c$,
and indeed we do find a dramatic effect: the charge
fluctuations for charges with  $\rho\leq\bar\rho$
are huge compared to the total $Q^2$; about 20 times
as large, in the case shown of $\beta=6.2$ after 23 cools.
 
What is the origin of this phenomenon? In Fig.\ref{fig-Q2-rho}
we see that charges with widths up to $\sim \bar\rho$
tend on the average to have the same sign; that is to say their
total charge is typically large. However as we include larger
charges we see that the typical total charge rapidly becomes
much smaller. That is to say: the smaller instantons tend to
have the opposite charge to the larger instantons --
the former are screened by the latter (and vice-versa).

To highlight this effect we define the following quantity:
\begin{equation}
C(\rho) \equiv \Biggl\langle
{Q \over {|Q|}}. {{Q(\rho)} \over {N(\rho)}} \Biggr\rangle.
\label{A30}
\end{equation}
where $Q(\rho)$ is the total topological charge of objects with 
widths in the bin centered on $\rho$, and $N(\rho)$ is their
corresponding total number. What $C(\rho)$ measures
is the correlation of the average charge of instantons of 
size $\rho$ with the sign of the total charge $Q$.
In Fig.\ref{fig-Q-Qrho} we show how $C(\rho)$ varies 
with $\rho$. We now see explicitly that the smaller
and larger charges tend to have opposite signs and,
moreover that it is the smaller charges that tend to
have the same sign as $Q$. 
What the latter tells us is that the net charge of
the smaller charges is greater (in modulus) than the
net charge of the larger charges. The large charges
are over-screened by the smaller charges. The boundary
between `large' and `small' is $\rho \simeq \bar\rho$, and 
scales roughly like a physical quantity when we change $\beta$.

To explore this phenomenon further, we calculate for
each reference instanton, with width $\rho_{ref}$,
the number of charges within a distance $R$ that have the 
same sign as the reference charge and whose widths fall
into a prescribed range e.g. $\rho>\rho_0$.
We call this $N_{same}(R;\rho>\rho_0)$. Similarly
for opposite sign charges we have $N_{opp}(R;\rho>\rho_0)$.
In Fig.\ref{fig-sameopp-nocut23} we show how $N_{same}-N_{opp}$
varies with $\rho_{ref}$ in the case when we include all 
the charges, i.e. $R=\infty$ and $\rho>0$. This is for the 
$\beta=6.2$ ensemble after 23 cooling sweeps. We note
that total screening corresponds to
$N_{same}-N_{opp} = -1$ and a value $< -1$ indicates
over-screening. 

The first thing that we observe in 
Fig.\ref{fig-sameopp-nocut23} is that the smallest
instantons are almost completely unscreened; this is  
what one would expect in a dilute gas and is consistent with
the fact that for these values of $\rho$ the
quantity $P(\rho_c)$ defined in eqn(\ref{A29})
also shows dilute gas behaviour. As we increase $\rho_{ref}$
we start to see screening. At $\rho_{ref} \simeq \bar\rho$
the screening is total and for larger sizes the instantons
are overscreened -- quite dramatically so for the 
very largest ones.

As always we have to ask ourselves whether what we see 
might not be a product of the cooling rather than
a property of the uncooled fields.
In  Fig.\ref{fig-sameopp-nocut46} we show the
corresponding plot after 46 cools. Although there
is a significant remnant of the under/overscreening
that we saw in  Fig.\ref{fig-sameopp-nocut23}, 
there is a clear trend towards
the much less interesting situation of total screening 
at all $\rho$. We conclude that cooling erodes rather 
than enhances the effect we have found, indicating that it 
is indeed a property of the original uncooled fields.

In Fig.\ref{fig-sameopp-smallcut} we show what happens 
when we include in $N_{same}-N_{opp}$ only the smaller charges, 
i.e. those with $\rho \leq \bar\rho$, and if we only 
count those charges that lie within distances $R=7$, 8, or 9 
of the reference charge. Let us first focus on small sizes;
say $\rho_{ref} \sim 3$ to $4$.
We observe from the $R=7$ data that there is some screening
of small charges by other nearby small charges. This is not
surprising: overlapping charges can reduce their total
action if they can partially annihilate. However we also observe
that as we increase $R$, so including charges that do
not significantly overlap with the reference instanton,
the screening disappears. This is odd: it tells us
that these more distant charges must tend to have the same charge
as the reference charge, despite being so far away that one
would naively expect very little correlation. If we
now turn to the large instantons, we observe that
the screening gets rapidly stronger as we increase $R$;
and indeed that large charges are overscreened
by the small charges under consideration here.

Fig.\ref{fig-sameopp-largecut} is the complement of 
Fig.\ref{fig-sameopp-smallcut}: now $N_{same}-N_{opp}$
includes only the larger charges,
i.e. those with widths $\rho \geq \bar\rho$.
We observe that the screening of small charges by
large charges increases as $R\uparrow$; the `normal'
screening behaviour. Indeed if we go to large $R$
it is clear from these two figures that the screening
of small charges is entirely driven by the large
charges. For large $\rho_{ref}$ the situation is
entirely different: large charges are strongly antiscreened
by other large charges i.e. these quite strongly 
overlapping large charges tend to have the same sign.

Again we find that all these effects weaken with increasing 
cooling suggest that they are properties of the original
uncooled fields. 

We now have enough information to hazard a model of the
topological structure that embodies all these features.
We think of the vacuum as being composed of small and
large charges. Small charges are superimposed on
the broad backs of the large charges because the
latter are everywhere: they densely pack Euclidean 
space-time. The small charges will tend to have the 
opposite sign to the large charge in which they are embedded 
and so, except when 
they are close enough for their mutual overlap to 
outweigh this effect, will tend to have the same sign as
each other. At the same time broad charges
overlap so these small charges will be simultaneously
embedded in more than one broad charge. They will thus 
tend align the charges of such overlapping broad charges.
So we have a picture of the broad charges tending,
on the average, to have the same sign throughout the vacuum
and the small charges also having the same sign as each other 
on the average but opposite to that of the broad charges. 
This is driven by the mutual interaction of 
the small, mutually non-overlapping charges with the large,
mutually overlapping charges. The net charge is that of
the smaller charges because, since they overlap less,
their charge polarisation is stronger.

This picture of the charge structure of the vacuum
goes somewhat beyond what our numerical evidence
demands but seems plausible. Presumably some
of the observed breakdown of our naive screening
intuition is not surprising: it arises
from the fact that while the latter is based on 
a $\propto 1/r$ potential between pointlike charges, the
effective potential here will have a more complicated
form when the charges overlap, and will fall much more
rapidly with $r$ when they do not. 

Although it is not possible to simply 
guess at the consequences of these non-trivial
long-range charge correlations for light quark physics,
it would be surprising if there were none.

\section{Topological susceptibility}

While the identification of the topological structure of the
vacuum is, as we have seen, a complicated and sometimes
ambiguous task, the total topological charge, $Q$, is
quite straightforward to extract. 
Moreover as we saw in the Introduction, this quantity
is related to the masses of the pseudoscalar mesons
via eqn(\ref{A1}). Although it was not the primary
purpose of our calculations, we have accumulated values of
$\langle Q^2 \rangle$ and hence of the susceptibility $\chi_t$
over the range $6.0 \leq \beta \leq 6.4$.
Our statistics is not very high but we do get closer
to the continuum limit than any other calculation that we
are aware of. It is therefore worthwhile extracting
a prediction for $\chi_t$ in physical units. 

There are various ways one can manipulate the raw,
non-integer lattice topological charge so as to obtain
an estimate of the `true' integer topological charge
of the field configuration. These definitions will
differ by lattice corrections which should vanish as
$a \to 0$. In Table~\ref{table-susc} we show our calculated
values of the susceptibility using two such definitions.
The first, $Q_L$, is 
simply the integral of the lattice topological charge 
density rounded to the nearest integer.
The second, $Q$, is our best estimate of 
the integer topological charge, obtained by applying
lattice corrections as described in the Appendix and
then rounding to the nearest (and usually nearby) integer.
We obtain the susceptibility in lattice units, $a^4\chi_t$,
if we divide $\langle Q^2 \rangle$ by the volume in lattice units;
and similarly  $a^4\chi_{t,L}$ from $\langle Q_L^2 \rangle$.
Since these quantities differ by lattice corrections,
they should possess a common continuum limit. This is
something we shall investigate below.

The first thing we observe is that if there is any
variation with the number of cooling sweeps it is
much less than the statistical error and so can be
ignored. We also note that the two lattice sizes
at $\beta=6.0$ give susceptibilities that are within 
$2\sigma$ of each other. We take this as evidence of
no significant finite-volume effects. Finally we remark
that the $\beta=6.4$ ensemble consists of only
20, albeit well-separated, field configurations and so one
should treat the corresponding error estimate with some caution.

\begin	{figure}[p]
\begin	{center}
\leavevmode
\input	{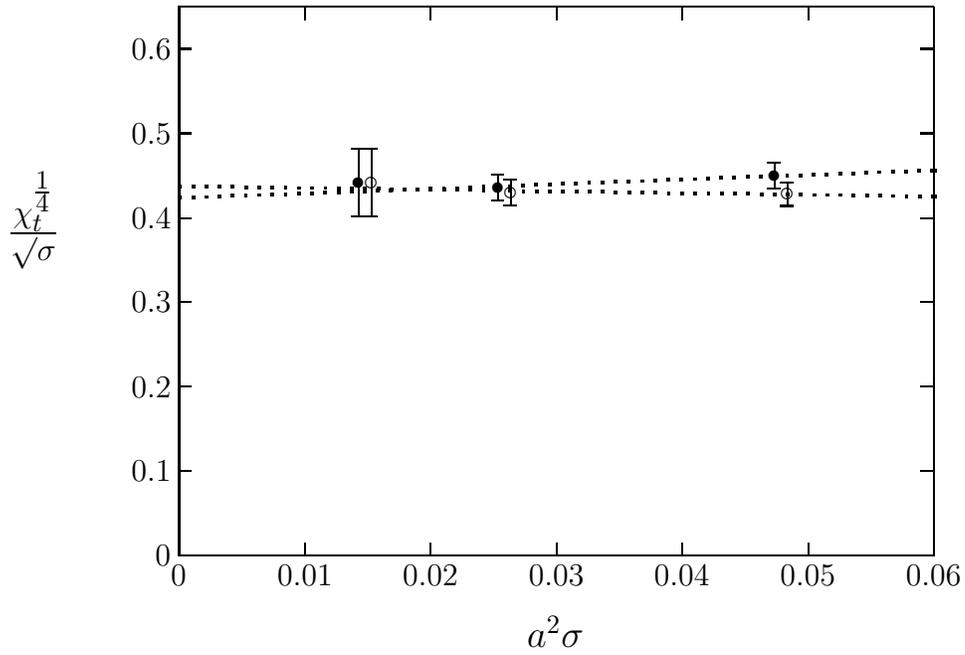}
\end	{center}
\vskip 0.15in
\caption{Plots of $\chi_t^{1/4}/\surd\sigma$ ($\bullet$) 
and  $\chi_{t,L}^{1/4}/\surd\sigma$($\circ$) against $a^2\sigma$
with continuum extrapolations.}
\label{fig-QSU3}
\end 	{figure}

Using the values for the string tension in eqn(\ref{A18})
we can form the
dimensionless mass ratio $\chi_t^{1/4}/\surd\sigma$.
In Fig.~\ref{fig-QSU3} we plot $\chi_t^{1/4}/\surd\sigma$ 
against the string tension 
in lattice units, $a^2 \sigma$. We expect the leading
lattice corrections to this dimensionless mass ratio to be
$O(a^2)$
\cite{Sym},    
so we can attempt a continuum extrapolation of the form
\begin{equation}
{{\chi_t^{1\over 4}(a)}\over{\surd\sigma(a)}}
= {{\chi_t^{1\over 4}(0)}\over{\surd\sigma(0)}} + ca^2\sigma      
\label{A31}
\end{equation}
This will be a simple straight line in our Figure. As we see the 
calculated values are consistent with this functional form.
In addition to the susceptibility calculated from
$\langle Q^2 \rangle$ we show the susceptibility calculated 
from $\langle Q_L^2 \rangle$.
In Table~\ref{table-chi} we show the continuum values of the
these ratios, together with earlier 
\cite{MT-newton}
results obtained over the range $5.7 \leq \beta \leq 6.2$.
We see that all the results are entirely consistent with 
each other.

To obtain $\chi_t$ in physical units, we use
the value for $\surd\sigma$ in eqn(\ref{A19}).
Substituting this value we obtain
\begin{equation}
 \chi^{1\over 4} = 187 \pm 14 \pm 16 MeV 
\label{A32}
\end{equation}       
where the second error reflects the uncertainty in assigning a value
to the string tension in MeV units. This, we note, is in satisfactory 
agreement with our expectations from eqn(\ref{A1}).

\section{Conclusions}

The influence of topological structure on the physics of QCD 
arises most directly from the near-zero modes
that it induces in the ${\not\!{\rm D}}[A]$ quark operator. 
Without performing the appropriate eigenvalue calculations on 
the fields it is not possible to be certain which aspects
of that structure are physically important and which are
not. For example, suppose that the broader
instantons belonged entirely to strongly overlapping
$I\bar I$ pairs. In that case they would not contribute
small modes to  ${\not\!{\rm D}}[A]$ and so would be essentially
irrelevant. As we have seen, the smaller instantons
form an approximate dilute gas. Such a dilute gas is
what is assumed in the Instanton Liquid Model
\cite{Shuryak}
and so in that case our apparently very different
topological structure would in fact be consistent with that model.

Given that we do not, in this paper, calculate the contribution 
of the topological structure to quark propagators and hadron 
physics, it has seemed
to us that the sensible approach is to expose all the
structure that is there without any prejudice as to what might
be important or not.

Exposing the topological structure
is a non-trivial task. First one needs to separate the
uninteresting high-frequency modes from the interesting
modes on physical length scales. We have chosen to use
the `cooling' technique; and we performed some studies to
find the variant which appeared to distort the interesting 
long-distance structure the least. There are other
approaches
\cite{DEF-SU2,DEG-SU2}
and the relationship between all these methods needs to be 
better understood; particularly as they appear 
to lead to significantly different results. 

Secondly one needs to categorise the topological structure
in some useful way. We have decomposed it into topological
charges of various sizes and locations. To do so requires a
rather complicated pattern-recognition algorithm. Although
we have tested the robustness of this algorithm to
small variations within the general scheme, this can 
only be a first step in exploring its reliability.

It is of course not at all certain that the topological
structure of the uncooled fields can be usefully
described in terms of (overlapping) charges that
are localised within some width $\rho$ of a particular
location. While there is some evidence from older calculations
that instanton collective coordinates are the appropriate
degrees of freedom to use, this really needs a much more 
careful study. Cooling will, of course, deform the field
into such a superposition of charges and so it makes
sense to perform the analysis in these terms on the
cooled fields. One might take the sophisticated
point of view that even if the uncooled topological 
structure is not really  a superposition
of approximately classical charges, cooling provides us with
the distribution of the latter charges that most closely
reproduces the true structure - and hence its physics as well.
This would then be the appropriate way to test the assumptions 
of instanton models.

In any case, it is clear that one must always check 
how robust under cooling are any conclusions that 
one wishes to draw. This is something we have attempted 
to do. The fact is that there is indeed a significant variation 
under cooling for many quantities and this introduces
some uncertainty into how one should interpret the calculated 
values. Unfortunately our `pattern recognition' algorithms
must break down as $n_c \to 0$ (because of the increase in the
apparent density) and so we are not able to attempt an
extrapolation to $n_c =0$. In this paper we have focussed on
developing techniques to reveal the structure of the
cooled fields: the problem of how precisely this relates
to the uncooled fields still awaits a convincing resolution.

At the same time, because the separation between
ultraviolet and physical frequencies is approximate,
but improves as $a \to 0$, it is important to perform 
scaling studies whenever possible. This we have also done:
the numbers we quote here for the mean size, density etc. are
the values that one obtains after an extrapolation to the continuum 
limit. Finally, we have explicitly checked that any finite-volume
effects are essentially within our statistical errors.

The simplest quantity to calculate, and one to which many
of the above caveats do not apply, is the topological
susceptibility, $\chi_t$. Our calculation goes to
smaller values of $a$ than any previous SU(3)
calculation and confirms previous claims that
eqn(\ref{A1}) is well satisfied; indeed, better than one
could expect.

The topological charge distribution we obtain is
characterised by a mean width $\bar\rho\simeq 0.56(5) fm$,
which is significantly larger than that which is typically 
assumed in Instanton Liquid Model calculations
\cite{Shuryak}.
The average separation between nearest neighbour
charges is comparable to the average width: so the vacuum 
is densely packed. 

The distribution is quite broad: the full-width is
$2\sigma_{\rho} \simeq 0.22(2) fm$. Our fits to the small-$\rho$
tail of this distribution do show some signs of a trend for 
it to approach
the predicted $D(\rho)\propto\rho^6$ behaviour when
$\beta$ is increased
and when the number of cooling sweeps is reduced,
but the evidence for this is rather rough. At large-$\rho$
we find a rapid fall-off that can be represented by
something like $D(\rho) \propto 1/\rho^{11}$. This is much
more severe than the $1/\rho^5$ behaviour that one might
argue on the basis of a coupling that freezes at large
distances. 

The values of $\bar\rho$ and $\sigma_\rho$ appear to
become smaller as the number of cooling sweeps is
decreased. However it is not clear whether this is
a real effect or whether it is a reflection of the
increasing unreliability of our pattern recognition
algorithms in the denser vacuum at smaller $n_c$. 
Neither is it clear how pronounced this effect would
be after extrapolation to the continuum limit.

We have found an interesting pattern of correlations amongst
the instantons. As expected, nearest neighbour charges are more 
likely to have opposite signs. There seems to be, in addition,
something like a `hard core' repulsion: instantons very close to
each other are suppressed much more than one would expect on the 
simple basis of phase space. Very small instantons do
behave like a dilute gas, but the bulk of charges with
$\rho < \bar\rho$ do not. Instead they seem to have charges that
are biased, on the average, towards being the same. 
Thus the fluctuations in topological 
charge, when restricted to sizes less than the mean, are
hugely amplified: 
$\langle Q^2(\rho < \bar\rho)\rangle \gg \langle Q^2 \rangle$.
At the same time the very large charges also tend to have the
same sign, and this is opposite to that of the smaller instantons.
It is the charge of the smaller instantons that is the
greater (in modulus) and they determine the sign of $Q$. 
So the smaller instantons are (under)screened by the larger
instantons; and the larger instantons are (over)screened by the
smaller instantons. That is to say, the 
vacuum has a long-range charge coherence
that depends on the scale. This effect becomes more pronounced
as the number of cools is decreased and so it seems reasonable
to infer that it is a real property of the uncooled vacuum.

We recall that topological 
charge fluctuations are physically important; e.g they 
play an important role in generating the $\eta^{\prime}$ 
mass. For this reason the effect we have identified 
might well have an impact on the physics. In particular if a 
quantity were sensitive to fluctuations in $Q$ but, for some
dynamical reason, was insensitive to $\rho > \bar\rho$ (or the
reverse), then
it would be affected far more strongly than if one simply
used the total $Q^2$ as a measure of the relevant fluctuations.
However to go further along these lines one really needs
to consider the fermionic physics in the background of these
fields; a topic that lies beyond the scope of this paper.

\vspace{0.50cm}
\noindent {\large {\bf Acknowledgements}}

\vspace{0.25cm}

{\noindent}DS would like to thank Hubert Simma for many useful discussions.
The work of DS was supported by the Carnegie Trust,
the UKQCD travel grant from PPARC, and by computing grants,
for T3D and J90 time, from PPARC and the University of
Edinburgh. MT has been supported by PPARC 
grants GR/K55752 and GR/K95338. Both DS and MT thank the
Newton Institute for hospitality during part of this work,
and DS also thanks Oxford Theoretical Physics for its
hospitality on various occasions.

\vspace{1.0cm}
\newpage

\noindent{\Large{\bf Appendix}}

\vspace{0.25cm}
{\noindent}In this Appendix we go briefly into some of the
details that it would have been tedious to leave in
the main body of the text.

\vspace{0.20cm}

\noindent{\bf (a) lattice corrections to ${\bf Q_L}$}

\vspace{0.15cm}

{\noindent}The lattice topological charge $Q_L = \int Q_L(x)dx$ 
of an instanton will not be exactly unity, but, as we
see in eqn(\ref{A8}), will suffer $O(a^2)$ corrections.
The relevant scale is $\rho$ and so if this is
expressed in lattice units, we have
\begin{equation}
Q_L = 1 + {c\over{\rho^2}} + O\biggl({1\over{\rho^4}}\biggr)
\label{B1}
\end{equation}       

Such corrections are relevant in several parts of our 
calculations. For example, we tune our filters so as
to minimise the discrepancy 
$\delta_Q = |Q-(n_I-n_{\bar I})|$. If we use $Q=Q_L$
this may introduce a significant error. Another example 
concerns the topological susceptibility: $\langle Q_L^2 \rangle$ 
will differ from $\langle Q^2 \rangle$ and it is the latter that 
we want. Since we determine the widths of all the charges in 
each field configuration, we can easily correct for all
the lattice artifacts once we have determined the
coefficients in the expression in eqn(\ref{B1}).
  
To determine $Q_L(\rho)$ for a single instanton, we take a 
large discretised instanton on a very large lattice and
cool it. As we cool it, it will gradually shrink. (Recall 
that we use a plaquette `action' to drive the cooling.)
As it shrinks we calculate $Q_L$ and $\rho$; we
then find that we can fit these with the simple form:
\begin{equation}
Q_L = 1 - {{0.65}\over{\rho^2}} - {{5.344}\over{\rho^4}}
\label{B2}
\end{equation}       
This heuristic fit is in practice only needed down to 
$\rho \sim 2$, and for that it is adequate.

There is one important detail we have skated over.
How does one determines the value of $\rho$ that
is used in eqn(\ref{B2})? This is intrinsically 
ambiguous since the shape of a small lattice
instanton differs qualitatively from the continuum
shape. It should however be apparent that this 
ambiguity does not matter; the role of $\rho$ is 
simply to act as a label parametrising instantons
of different sizes. What is important is that once we 
pick some definition of $\rho$, we then use it consistently
throughout our calculations. 

A simple way to define $\rho$ is to use a continuum
relation between some aspect of the instanton 
topological charge density and its width. For example, 
we could use the relation between $Q_p$ and $\rho$
as given in eqn(\ref{A12}). On the lattice we would
calculate $Q_{L,p}$ and then use this relation to
define $\rho$ for us. This might be improved by
replacing $Q_{p}$ with $Q_{L,p}/Q_L$ in eqn(\ref{A12}).
(To avoid the double-valuedness that potentially arises 
when $\rho\sim 1$.) In practice we have chosen to use
a different relation. We recall that for a continuum
instanton the total topological charge within a distance
$\rho$ of the centre is 1/2. We therefore define 
$\rho$ for our lattice instanton to be the distance
within which it contains a topological charge of $Q_L/2$.
This is used throughout our calculations. Of course 
the discrete nature of the lattice means that one
has to make some particular choices as to how to
implement this criterion. (There are some unsatisfactory 
features in the choices we actually made -- but in 
practice, because really small instantons are very rare,
these do not matter.)

As we increase $\beta$, the number of instantons small 
enough for these corrections to be important rapidly
decreases and so, as long as one performs a scaling study,
all methods lead to equivalent results.

\vspace{0.20cm}

\noindent{\bf (b) lattice correction to ${\bf Q_p}$} 

\vspace{0.15cm}

{\noindent}The continuum expression for  $Q_p$ in eqn(\ref{A12})
will also have $O(a^2)$ lattice corrections. The scale for
these is set by $\rho$ and so, if $\rho$ is expressed in 
lattice units, such a correction corresponds to an extra 
term of $O(1/\rho^6)$. To determine
the corrections we cooled a classical instanton, calculating
$\rho$ as described above. The resulting relationship
between the peak lattice topological charge and $\rho$
may be parameterised as
\begin{equation}
Q_{L,p} = {6\over{\pi^2 \rho^4}}
\biggl(1 - {{1.962}\over{\rho^2}} + {{1.198}\over{\rho^3}}\biggr).
\label{B3}
\end{equation}       
We have included a heuristic $1/\rho^3$ correction to mimic
all the higher (even) powers that will become important
at very small $\rho$. Eqn(\ref{B3}) has been used in this paper
for calculating $\rho$ from the peaks.

\vspace{0.20cm}

\noindent{\bf (c) lattice correction to ${\bf Q(|x|\leq R)}$}

\vspace{0.15cm}

{\noindent}In our filters we have used the topological charge
within a distance $\rho$ to define a width, $\rho_R$.
This is then compared to the width calculated
from the peak. The continuum expression for $Q(|x| \leq R)$
is given in eqn(\ref{A16}). For small $\rho$ this will 
need lattice corrections and these can be found just as
for $Q_p$. For example, for the case $R=2$ which we use in
practice in our filters, we find that one can parametrise the
lattice corrections by
\begin{equation}
Q_L(|x| \leq 2) = Q(|x| \leq 2)
\biggl(1 - {{1.66}\over{\rho^2}} - {{1.26}\over{\rho^4}}\biggr).
\label{B4}
\end{equation}       
This can then be used to extract a width, $\rho_2\equiv\rho$, 
for each peak in the vacuum and this can be used in the filters.
(In fact we used a slightly different, but considerably
less elegant, functional form which is essentially the same
for the values of $\rho$ that arise in practice.)

\vspace{0.20cm}

%\noindent{\bf (d) generalising eqn(\ref{A13}) to get $\rho_R$}
% -- see eqn(\ref{A16}) .

\vskip 1.0in

\newpage

\begin{table}
\begin{center}
\begin{tabular}{|c|c|c|c|}\hline
cools & $\langle \delta_Q \rangle$ & $\surd \langle Q^2 \rangle$ & 
$\langle N_{tot} \rangle$ \\ \hline
 23   &   12.45    & 4.1  & 493.8  \\ \hline
 32   &   10.39    & 4.0  & 273.6  \\ \hline
 46   &    7.08    & 4.0  & 151.1  \\ \hline
\end{tabular}
\caption{\label{table-deltaQ} 
Mismatch, $\delta_Q$, typical total charge, $\surd \langle Q^2
\rangle$, and total number of charges: against
number of cooling sweeps at $\beta=6.2$.}
\end{center}
\end{table}

\begin{table}
\begin{center}
\begin{tabular}{|c|c|c|c|c|c|}\hline
$\beta$ & cools & $\epsilon_2$ & $\rho_c$ & $\epsilon_c$ & 
$\delta_{Q}$ \\ \hline
6.0   &  23   &    0.19  &    6.84  & 0.20 &   2.90 \\ \hline
      &  28   &    0.17  &    6.50  & 0.25 &   2.25 \\ \hline
      &  32   &    0.11  &    8.46  & 0.20 &   2.75 \\ \hline
      &  46   &    0.10  &   16.00  & 0.10 &   1.70 \\ \hline
large &  46   &    0.10  &    6.89  & 0.25 &   6.25 \\ \hline\hline
6.2   &  23   &    0.27  &    7.80  & 0.30 &   4.40 \\ \hline
      &  32   &    0.20  &    6.83  & 0.25 &   3.55 \\ \hline
      &  46   &    0.11  &    9.31  & 0.20 &   2.25 \\ \hline\hline
6.4   &  30   &    0.24  &   10.89  & 0.20 &   4.00 \\ \hline
      &  50   &    0.21  &    9.43  & 0.30 &   2.80 \\ \hline
      &  70   &    0.10  &   11.25  & 0.30 &   1.83 \\ \hline
      &  80   &    0.05  &   10.83  & 0.15 &   1.92 \\ \hline
\end{tabular}
\caption{\label{table-deltaQ-filter1} 
The filter parameters used in this paper, with 
corresponding charge mismatch, $\delta_{Q}$.}
\end{center}
\end{table}

\begin{table}
\begin{center}
\begin{tabular}{|c|c|c|c|c|}\hline
cools & $\epsilon_3$ & $\rho_c$ & $\epsilon_c$ & $\delta_{Q}$ \\ \hline
 23   &    0.62    &    7.60    & 0.3 &  3.65     \\ \hline
 32   &    0.46    &    6.83    & 0.25&   3.85      \\ \hline
 46   &    0.24    &    7.18    & 0.20&   1.90      \\ \hline
\end{tabular}
\caption{\label{table-deltaQ-filter2} 
Filter parameters for $\beta=6.2$, using $\rho_3$ in the filter}
\end{center}
\end{table}

\begin{table}
\begin{center}
\begin{tabular}{|c|c|c|c|}\hline
cools &  \#peaks     & \#peaks  & \#peaks  \\ 
      & (unfiltered) & ($\rho_2$) & ($\rho_3$) \\ \hline
 23   &        540           &  495     &   489   \\ \hline
 32   &        302           &  271     &   267   \\ \hline
 46   &        164           &  158     &   148   \\ \hline
\end{tabular}
\caption{\label{table-num-filter}
Number of peaks, at $\beta=6.2$, before and after applying filters
based on $\rho_2$ and $\rho_3$ respectively.}
\end{center}
\end{table}

\begin{table}
\begin{center}
\begin{tabular}{|c|c|c|c|c|}\hline
cools & \#peaks  & \#peaks  & R & \#peaks with \\
      & in $Q(x)$& in $S(x)$&   & $| x_{action} - x_{charge} | \le R$ \\ \hline
 23   &   168    &   95     & 0 &      54      \\ \hline
 23   &   168    &   95     &$\surd{2}$ & 86     \\ \hline
 46   &    63    &   46     & 0 &    33        \\ \hline
 46   &    63    &   46     &$\surd{2}$ & 43   \\ \hline
\end{tabular}
\caption{\label{table-peak-SQ}
Number of peaks in $S(x)$ that can be
indentified with a peak in $Q(x)$.}
\end{center}
\end{table}

\begin{table}
\begin{center}
\begin{tabular}{|c|c|c|c|c|c|}\hline
$\beta$ & cools & $\bar\rho$ & $\sigma_{\rho}$ & $\langle N_{tot} \rangle$ &
$\langle N_{tot}\rangle /V$ \\ \hline
6.0    & 23 & 5.25(1) & 1.20(4)  &  147.5 & 0.328(8)  \\ \hline
       & 28 & 5.40(2) & 1.29(5)  &  113.0 & 0.251(6)  \\ \hline
       & 32 & 5.43(2) & 1.36(5)  &   86.2 & 0.192(5)  \\ \hline
       & 46 & 5.65(2) & 1.51(6)  &   57.5 & 0.128(3)  \\ \hline
large  & 46 & 5.63(1) & 1.51(3)  &  613.6 & 0.128(3)  \\ \hline\hline
6.2    & 23 & 6.41(1) & 1.17(3)  &  493.8 & 1.112(28) \\ \hline
       & 32 & 6.94(1) & 1.35(4)  &  273.6 & 0.616(15) \\ \hline
       & 46 & 7.44(2) & 1.57(5)  &  151.1 & 0.340(9)  \\ \hline\hline
6.4    & 30 & 7.86(1) & 1.32(5)  & 1005.3 & 2.195(82) \\ \hline
       & 50 & 9.00(2) & 1.61(8)  &  369.7 & 0.807(30) \\ \hline
       & 70 & 9.70(4) & 1.92(11) &  205.2 & 0.448(17) \\ \hline
       & 80 & 9.84(5) & 2.04(12) &  156.6 & 0.342(13) \\ \hline
\end{tabular}
\caption{\label{table-sizes-all}
Average and width of size distribution; also the total number of
charges and the number per unit physical volume.} 
\end{center}
\end{table}

\begin{table}
\begin{center}
\begin{tabular}{|c|c|c|c|c|}\hline
cools & $\bar\rho$ & $\sigma_{\rho}$ & $\langle N_{tot}
\rangle$ & $f$ \\ \hline
 20 &  6.92 & 1.29  & 2177 & 15.6 \\ \hline
 30 &  7.96 & 1.52  & 1063 & 13.1 \\ \hline
 40 &  8.74 & 1.95  &  604 & 12.0 \\ \hline
 50 &  9.27 & 1.98  &  395 &  9.1 \\ \hline
 60 &  9.70 & 2.23  &  286 & 26.2 \\ \hline
 70 & 10.06 & 2.27  &  224 & 22.8 \\ \hline
\end{tabular}
\caption{\label{table-cool-b64}
Variation with cooling, $n_c$, of a single configuration at
$\beta=6.4$ prior to any filtering.}
\end{center}
\end{table}

\begin{table}
\begin{center}
\begin{tabular}{|c|c|c|c|}\hline
$\beta$ & cools &  range &  $\gamma_s$ \\ \hline
  6.0   &  23   & 0.44-0.77  & 3.3(2)  \\ \hline
  6.0   &  28   & 0.44-0.99  & 3.1(1)  \\ \hline
  6.0   &  32   & 0.44-0.77  & 2.9(2)  \\ \hline
  6.0   &  46   & 0.44-0.99  & 2.6(1)  \\ \hline
  6.2   &  23   & 0.48-0.72  & 6.9(2)  \\ \hline
  6.2   &  32   & 0.48-0.88  & 5.4(2)  \\ \hline
  6.2   &  46   & 0.48-0.72  & 4.8(3)  \\ \hline
  6.4   &  30   & 0.55-0.73  & 9.2(4)  \\ \hline
  6.4   &  50   & 0.55-0.97  & 5.7(2)  \\ \hline
  6.4   &  70   & 0.55-0.97  & 4.7(3)  \\ \hline
  6.4   &  80   & 0.61-1.03  & 4.0(3)  \\ \hline
\end{tabular}
\caption{\label{table-tail-small}
Fits to the small-$\rho$ tails of the size distributions.
The range fitted is in units of $1/\surd\sigma$.}
\end{center}
\end{table}

\begin{table}
\begin{center}
\begin{tabular}{|c|c|c|c|}\hline
$\beta$ & cools &  range &  $\gamma_l$ \\ \hline
  6.0   &  23   &  1.43-1.87 & 11.2(4) \\ \hline
  6.0   &  28   &  1.54-2.20 & 10.8(4) \\ \hline
  6.0   &  32   &  1.54-2.20 & 10.1(4) \\ \hline
  6.0   &  46   &  1.65-2.31 & 10.0(5) \\ \hline
  6.2   &  23   &  1.20-1.44 & 11.9(3) \\ \hline
  6.2   &  32   &  1.36-1.68 & 11.9(3) \\ \hline
  6.2   &  46   &  1.44-2.00 & 10.2(3) \\ \hline
  6.4   &  30   &  1.09-1.46 & 11.0(3) \\ \hline
  6.4   &  50   &  1.46-1.64 & 12.2(2) \\ \hline
  6.4   &  70   &  1.52-1.64 & 10.0(2) \\ \hline
  6.4   &  80   &  1.46-1.82 & 10.4(8) \\ \hline
\end{tabular}
\caption{\label{table-tail-large}
Fits to the large-$\rho$ tails of the size distributions.
The range fitted is in units of $1/\surd\sigma$.}
\end{center}
\end{table}

\begin{table}
\begin{center}
\begin{tabular}{|c|c|c|c|}\hline
$\beta$ & cools &  ${\bar R}_{like}\surd\sigma$ &  
${\bar R}_{unlike}\surd\sigma$ \\ \hline
  6.0   &  23   &  1.072(6)  & 0.931(6)  \\ \hline
  6.0   &  28   &  1.152(7)  & 1.008(6)  \\ \hline
  6.0   &  32   &  1.233(8)  & 1.077(7)  \\ \hline
  6.0   &  46   &  1.359(9)  & 1.223(9)  \\ \hline
  6.2   &  23   &  0.791(5)  & 0.700(5)  \\ \hline
  6.2   &  32   &  0.935(6)  & 0.818(6)  \\ \hline
  6.2   &  46   &  1.075(7)  & 0.958(7)  \\ \hline
  6.4   &  30   &  0.667(6)  & 0.604(6)  \\ \hline
  6.4   &  50   &  0.884(7)  & 0.775(8)  \\ \hline
  6.4   &  70   &  1.013(11) & 0.911(9)  \\ \hline
  6.4   &  80   &  1.080(12) & 0.976(10) \\ \hline
\end{tabular}
\caption{\label{table-nn-sameopp}
Average distance to nearest charge of the
same and opposite sign; in units of $1/\surd\sigma$.}
\end{center}
\end{table}

\begin{table}
\begin{center}
\begin{tabular}{|c|c|c|c|}\hline
$\beta$ & cools &  $\langle Q_L^2\rangle$ &  $\langle Q^2\rangle$ \\ \hline
  6.0   &  23   & 14.4(2.0)   & 18.8(2.8)   \\ \hline
  6.0   &  28   & 15.0(2.0)   & 18.6(2.6)   \\ \hline
  6.0   &  32   & 15.1(2.1)   & 18.5(2.6)   \\ \hline
  6.0   &  46   & 15.7(2.1)   & 17.9(2.6)   \\ \hline
 large  &  46   & 97.3(18.6)  & 120.2(23.1) \\ \hline\hline
  6.2   &  23   & 14.1(2.0)   & 17.1(2.6)   \\ \hline
  6.2   &  32   & 14.5(2.1)   & 16.1(2.4)   \\ \hline
  6.2   &  46   & 15.2(2.2)   & 16.0(2.4)   \\ \hline\hline
  6.4   &  30   & 16.4(6.6)   & 17.4(7.1)   \\ \hline
  6.4   &  50   & 17.4(7.1)   & 17.4(7.1)   \\ \hline
  6.4   &  70   & 17.4(7.1)   & 17.4(7.1)   \\ \hline
  6.4   &  80   & 17.4(7.1)   & 17.4(7.1)   \\ \hline
\end{tabular}
\caption{\label{table-susc}
Average values of $Q^2$ for corrected ($Q$) and uncorrected ($Q_L$)
chrges respectively.}
\end{center}
\end{table}

\begin{table}
\begin{center}
\begin{tabular}{|c|c|c|}\hline
%${\chi_t^{1/4}}\over{\surd\sigma}$  &  
%${\chi_{t,L}^{1/4}}\over{\surd\sigma}$ & 
%${\chi_{t,old}^{1\over 4}}\over{\surd\sigma}$ \\ \hline
${\chi_t^{1/4}}/{\surd\sigma}$  &  
${\chi_{t,L}^{1/4}}/{\surd\sigma}$ & 
${\chi_{t,old}^{1/4}}/{\surd\sigma}$ \\ \hline
0.424(32)    & 0.437(34)   & 0.437(25)   \\ \hline
\end{tabular}
\caption{\label{table-chi}
Continuum limit of susceptibility in units of the string tension:
for corrected ($\chi_t$) and uncorrected ($\chi_{t,L}$) charges; 
and a previous calculation \cite{MT-newton} for comparison.}
\end{center}
\end{table}

\clearpage
\newpage

%\begin{figure}
%\begin{center}
%\leavevmode
%\epsfxsize = 1.00\figsize
%\epsfysize = 0.75\figsize
%\epsffile{reffig-pair-cool.ps}
%\end{center}
%\caption{\label{fig-pair-cool}
%Action of $I\bar I$ pair against number of calibrated cooling sweeps.}
%\end{figure} 

\clearpage

\begin	{figure}[p]
\begin	{center}
\leavevmode
\input	{plot_fig_size_SQ23}
\end	{center}
\vskip 0.15in
\caption{Size distributions from $S(x)$ ($\circ$) and 
from $Q(x)$ ($\bullet$) after 23 cooling sweeps at $\beta=6.0$.}
\label{fig-size-SQ23}
\end 	{figure}

\begin	{figure}[p]
\begin	{center}
\leavevmode
\input	{plot_fig_size_SQ46}
\end	{center}
\vskip 0.15in
\caption{Size distributions from $S(x)$ ($\circ$) and 
from $Q(x)$ ($\bullet$) after 46 cooling sweeps at $\beta=6.0$.}
\label{fig-size-SQ46}
\end 	{figure}

\clearpage

\clearpage

\begin	{figure}[p]
\begin	{center}
\leavevmode
% GNUPLOT: LaTeX picture
\setlength{\unitlength}{0.240900pt}
\ifx\plotpoint\undefined\newsavebox{\plotpoint}\fi
\sbox{\plotpoint}{\rule[-0.200pt]{0.400pt}{0.400pt}}%
\begin{picture}(1349,900)(0,0)
\font\gnuplot=cmr10 at 12pt
\gnuplot
\sbox{\plotpoint}{\rule[-0.200pt]{0.400pt}{0.400pt}}%
\put(120.0,31.0){\rule[-0.200pt]{4.818pt}{0.400pt}}
\put(108,31){\makebox(0,0)[r]{{$-0.4$}}}
\put(1285.0,31.0){\rule[-0.200pt]{4.818pt}{0.400pt}}
\put(120.0,247.0){\rule[-0.200pt]{4.818pt}{0.400pt}}
\put(108,247){\makebox(0,0)[r]{{$-0.2$}}}
\put(1285.0,247.0){\rule[-0.200pt]{4.818pt}{0.400pt}}
\put(120.0,462.0){\rule[-0.200pt]{4.818pt}{0.400pt}}
\put(108,462){\makebox(0,0)[r]{{$0$}}}
\put(1285.0,462.0){\rule[-0.200pt]{4.818pt}{0.400pt}}
\put(120.0,678.0){\rule[-0.200pt]{4.818pt}{0.400pt}}
\put(108,678){\makebox(0,0)[r]{{$0.2$}}}
\put(1285.0,678.0){\rule[-0.200pt]{4.818pt}{0.400pt}}
\put(120.0,893.0){\rule[-0.200pt]{4.818pt}{0.400pt}}
\put(108,893){\makebox(0,0)[r]{{$0.4$}}}
\put(1285.0,893.0){\rule[-0.200pt]{4.818pt}{0.400pt}}
\put(120.0,31.0){\rule[-0.200pt]{0.400pt}{4.818pt}}
\put(120,19){\makebox(0,0){\shortstack{\\ \\ \\ {$0$}}}}
\put(120.0,873.0){\rule[-0.200pt]{0.400pt}{4.818pt}}
\put(268.0,31.0){\rule[-0.200pt]{0.400pt}{4.818pt}}
\put(268,19){\makebox(0,0){\shortstack{\\ \\ \\ {$2$}}}}
\put(268.0,873.0){\rule[-0.200pt]{0.400pt}{4.818pt}}
\put(416.0,31.0){\rule[-0.200pt]{0.400pt}{4.818pt}}
\put(416,19){\makebox(0,0){\shortstack{\\ \\ \\ {$4$}}}}
\put(416.0,873.0){\rule[-0.200pt]{0.400pt}{4.818pt}}
\put(564.0,31.0){\rule[-0.200pt]{0.400pt}{4.818pt}}
\put(564,19){\makebox(0,0){\shortstack{\\ \\ \\ {$6$}}}}
\put(564.0,873.0){\rule[-0.200pt]{0.400pt}{4.818pt}}
\put(713.0,31.0){\rule[-0.200pt]{0.400pt}{4.818pt}}
\put(713,19){\makebox(0,0){\shortstack{\\ \\ \\ {$8$}}}}
\put(713.0,873.0){\rule[-0.200pt]{0.400pt}{4.818pt}}
\put(861.0,31.0){\rule[-0.200pt]{0.400pt}{4.818pt}}
\put(861,19){\makebox(0,0){\shortstack{\\ \\ \\ {$10$}}}}
\put(861.0,873.0){\rule[-0.200pt]{0.400pt}{4.818pt}}
\put(1009.0,31.0){\rule[-0.200pt]{0.400pt}{4.818pt}}
\put(1009,19){\makebox(0,0){\shortstack{\\ \\ \\ {$12$}}}}
\put(1009.0,873.0){\rule[-0.200pt]{0.400pt}{4.818pt}}
\put(1157.0,31.0){\rule[-0.200pt]{0.400pt}{4.818pt}}
\put(1157,19){\makebox(0,0){\shortstack{\\ \\ \\ {$14$}}}}
\put(1157.0,873.0){\rule[-0.200pt]{0.400pt}{4.818pt}}
\put(1305.0,31.0){\rule[-0.200pt]{0.400pt}{4.818pt}}
\put(1305,19){\makebox(0,0){\shortstack{\\ \\ \\ {$16$}}}}
\put(1305.0,873.0){\rule[-0.200pt]{0.400pt}{4.818pt}}
\put(120.0,31.0){\rule[-0.200pt]{285.466pt}{0.400pt}}
\put(1305.0,31.0){\rule[-0.200pt]{0.400pt}{207.656pt}}
\put(120.0,893.0){\rule[-0.200pt]{285.466pt}{0.400pt}}
\put(-84,558){\makebox(0,0){{\large{$C(\rho)$}}}}
\put(712,-89){\makebox(0,0){{\large{$\rho$}}}}
\put(120.0,31.0){\rule[-0.200pt]{0.400pt}{207.656pt}}
\put(120,462){\makebox(0,0){$\times$}}
\put(157,462){\makebox(0,0){$\times$}}
\put(194,462){\makebox(0,0){$\times$}}
\put(231,474){\makebox(0,0){$\times$}}
\put(268,510){\makebox(0,0){$\times$}}
\put(305,674){\makebox(0,0){$\times$}}
\put(342,556){\makebox(0,0){$\times$}}
\put(379,725){\makebox(0,0){$\times$}}
\put(416,578){\makebox(0,0){$\times$}}
\put(453,547){\makebox(0,0){$\times$}}
\put(490,561){\makebox(0,0){$\times$}}
\put(527,507){\makebox(0,0){$\times$}}
\put(564,459){\makebox(0,0){$\times$}}
\put(601,447){\makebox(0,0){$\times$}}
\put(638,398){\makebox(0,0){$\times$}}
\put(675,434){\makebox(0,0){$\times$}}
\put(713,403){\makebox(0,0){$\times$}}
\put(750,359){\makebox(0,0){$\times$}}
\put(787,369){\makebox(0,0){$\times$}}
\put(824,283){\makebox(0,0){$\times$}}
\put(861,385){\makebox(0,0){$\times$}}
\put(898,206){\makebox(0,0){$\times$}}
\put(935,406){\makebox(0,0){$\times$}}
\put(972,224){\makebox(0,0){$\times$}}
\put(1009,351){\makebox(0,0){$\times$}}
\put(1046,446){\makebox(0,0){$\times$}}
\put(1083,474){\makebox(0,0){$\times$}}
\put(1120,462){\makebox(0,0){$\times$}}
\put(1157,474){\makebox(0,0){$\times$}}
\put(1194,474){\makebox(0,0){$\times$}}
\put(1231,438){\makebox(0,0){$\times$}}
\put(1268,438){\makebox(0,0){$\times$}}
\put(120,462){\usebox{\plotpoint}}
\put(110.0,462.0){\rule[-0.200pt]{4.818pt}{0.400pt}}
\put(110.0,462.0){\rule[-0.200pt]{4.818pt}{0.400pt}}
\put(157,462){\usebox{\plotpoint}}
\put(147.0,462.0){\rule[-0.200pt]{4.818pt}{0.400pt}}
\put(147.0,462.0){\rule[-0.200pt]{4.818pt}{0.400pt}}
\put(194,462){\usebox{\plotpoint}}
\put(184.0,462.0){\rule[-0.200pt]{4.818pt}{0.400pt}}
\put(184.0,462.0){\rule[-0.200pt]{4.818pt}{0.400pt}}
\put(231.0,453.0){\rule[-0.200pt]{0.400pt}{10.118pt}}
\put(221.0,453.0){\rule[-0.200pt]{4.818pt}{0.400pt}}
\put(221.0,495.0){\rule[-0.200pt]{4.818pt}{0.400pt}}
\put(268.0,456.0){\rule[-0.200pt]{0.400pt}{26.258pt}}
\put(258.0,456.0){\rule[-0.200pt]{4.818pt}{0.400pt}}
\put(258.0,565.0){\rule[-0.200pt]{4.818pt}{0.400pt}}
\put(305.0,592.0){\rule[-0.200pt]{0.400pt}{39.267pt}}
\put(295.0,592.0){\rule[-0.200pt]{4.818pt}{0.400pt}}
\put(295.0,755.0){\rule[-0.200pt]{4.818pt}{0.400pt}}
\put(342.0,468.0){\rule[-0.200pt]{0.400pt}{42.398pt}}
\put(332.0,468.0){\rule[-0.200pt]{4.818pt}{0.400pt}}
\put(332.0,644.0){\rule[-0.200pt]{4.818pt}{0.400pt}}
\put(379.0,665.0){\rule[-0.200pt]{0.400pt}{28.908pt}}
\put(369.0,665.0){\rule[-0.200pt]{4.818pt}{0.400pt}}
\put(369.0,785.0){\rule[-0.200pt]{4.818pt}{0.400pt}}
\put(416.0,544.0){\rule[-0.200pt]{0.400pt}{16.381pt}}
\put(406.0,544.0){\rule[-0.200pt]{4.818pt}{0.400pt}}
\put(406.0,612.0){\rule[-0.200pt]{4.818pt}{0.400pt}}
\put(453.0,520.0){\rule[-0.200pt]{0.400pt}{12.768pt}}
\put(443.0,520.0){\rule[-0.200pt]{4.818pt}{0.400pt}}
\put(443.0,573.0){\rule[-0.200pt]{4.818pt}{0.400pt}}
\put(490.0,542.0){\rule[-0.200pt]{0.400pt}{9.154pt}}
\put(480.0,542.0){\rule[-0.200pt]{4.818pt}{0.400pt}}
\put(480.0,580.0){\rule[-0.200pt]{4.818pt}{0.400pt}}
\put(527.0,490.0){\rule[-0.200pt]{0.400pt}{7.950pt}}
\put(517.0,490.0){\rule[-0.200pt]{4.818pt}{0.400pt}}
\put(517.0,523.0){\rule[-0.200pt]{4.818pt}{0.400pt}}
\put(564.0,449.0){\rule[-0.200pt]{0.400pt}{4.818pt}}
\put(554.0,449.0){\rule[-0.200pt]{4.818pt}{0.400pt}}
\put(554.0,469.0){\rule[-0.200pt]{4.818pt}{0.400pt}}
\put(601.0,436.0){\rule[-0.200pt]{0.400pt}{5.059pt}}
\put(591.0,436.0){\rule[-0.200pt]{4.818pt}{0.400pt}}
\put(591.0,457.0){\rule[-0.200pt]{4.818pt}{0.400pt}}
\put(638.0,385.0){\rule[-0.200pt]{0.400pt}{6.504pt}}
\put(628.0,385.0){\rule[-0.200pt]{4.818pt}{0.400pt}}
\put(628.0,412.0){\rule[-0.200pt]{4.818pt}{0.400pt}}
\put(675.0,408.0){\rule[-0.200pt]{0.400pt}{12.527pt}}
\put(665.0,408.0){\rule[-0.200pt]{4.818pt}{0.400pt}}
\put(665.0,460.0){\rule[-0.200pt]{4.818pt}{0.400pt}}
\put(713.0,364.0){\rule[-0.200pt]{0.400pt}{18.549pt}}
\put(703.0,364.0){\rule[-0.200pt]{4.818pt}{0.400pt}}
\put(703.0,441.0){\rule[-0.200pt]{4.818pt}{0.400pt}}
\put(750.0,299.0){\rule[-0.200pt]{0.400pt}{28.908pt}}
\put(740.0,299.0){\rule[-0.200pt]{4.818pt}{0.400pt}}
\put(740.0,419.0){\rule[-0.200pt]{4.818pt}{0.400pt}}
\put(787.0,291.0){\rule[-0.200pt]{0.400pt}{37.339pt}}
\put(777.0,291.0){\rule[-0.200pt]{4.818pt}{0.400pt}}
\put(777.0,446.0){\rule[-0.200pt]{4.818pt}{0.400pt}}
\put(824.0,198.0){\rule[-0.200pt]{0.400pt}{40.712pt}}
\put(814.0,198.0){\rule[-0.200pt]{4.818pt}{0.400pt}}
\put(814.0,367.0){\rule[-0.200pt]{4.818pt}{0.400pt}}
\put(861.0,303.0){\rule[-0.200pt]{0.400pt}{39.748pt}}
\put(851.0,303.0){\rule[-0.200pt]{4.818pt}{0.400pt}}
\put(851.0,468.0){\rule[-0.200pt]{4.818pt}{0.400pt}}
\put(898.0,126.0){\rule[-0.200pt]{0.400pt}{38.303pt}}
\put(888.0,126.0){\rule[-0.200pt]{4.818pt}{0.400pt}}
\put(888.0,285.0){\rule[-0.200pt]{4.818pt}{0.400pt}}
\put(935.0,326.0){\rule[-0.200pt]{0.400pt}{38.303pt}}
\put(925.0,326.0){\rule[-0.200pt]{4.818pt}{0.400pt}}
\put(925.0,485.0){\rule[-0.200pt]{4.818pt}{0.400pt}}
\put(972.0,148.0){\rule[-0.200pt]{0.400pt}{36.617pt}}
\put(962.0,148.0){\rule[-0.200pt]{4.818pt}{0.400pt}}
\put(962.0,300.0){\rule[-0.200pt]{4.818pt}{0.400pt}}
\put(1009.0,284.0){\rule[-0.200pt]{0.400pt}{32.281pt}}
\put(999.0,284.0){\rule[-0.200pt]{4.818pt}{0.400pt}}
\put(999.0,418.0){\rule[-0.200pt]{4.818pt}{0.400pt}}
\put(1046.0,393.0){\rule[-0.200pt]{0.400pt}{25.535pt}}
\put(1036.0,393.0){\rule[-0.200pt]{4.818pt}{0.400pt}}
\put(1036.0,499.0){\rule[-0.200pt]{4.818pt}{0.400pt}}
\put(1083.0,434.0){\rule[-0.200pt]{0.400pt}{19.272pt}}
\put(1073.0,434.0){\rule[-0.200pt]{4.818pt}{0.400pt}}
\put(1073.0,514.0){\rule[-0.200pt]{4.818pt}{0.400pt}}
\put(1120.0,428.0){\rule[-0.200pt]{0.400pt}{16.381pt}}
\put(1110.0,428.0){\rule[-0.200pt]{4.818pt}{0.400pt}}
\put(1110.0,496.0){\rule[-0.200pt]{4.818pt}{0.400pt}}
\put(1157.0,442.0){\rule[-0.200pt]{0.400pt}{15.418pt}}
\put(1147.0,442.0){\rule[-0.200pt]{4.818pt}{0.400pt}}
\put(1147.0,506.0){\rule[-0.200pt]{4.818pt}{0.400pt}}
\put(1194.0,442.0){\rule[-0.200pt]{0.400pt}{15.418pt}}
\put(1184.0,442.0){\rule[-0.200pt]{4.818pt}{0.400pt}}
\put(1184.0,506.0){\rule[-0.200pt]{4.818pt}{0.400pt}}
\put(1231.0,421.0){\rule[-0.200pt]{0.400pt}{8.191pt}}
\put(1221.0,421.0){\rule[-0.200pt]{4.818pt}{0.400pt}}
\put(1221.0,455.0){\rule[-0.200pt]{4.818pt}{0.400pt}}
\put(1268.0,408.0){\rule[-0.200pt]{0.400pt}{14.213pt}}
\put(1258.0,408.0){\rule[-0.200pt]{4.818pt}{0.400pt}}
\put(1258.0,467.0){\rule[-0.200pt]{4.818pt}{0.400pt}}
\end{picture}
\end	{center}
\vskip 0.15in
\caption{$C(\rho)$, as defined in eqn(\ref{A30}), 
versus $\rho$ at $\beta=6.2$ after 23 cools.}
\label{fig-Q-Qrho}
\end 	{figure}
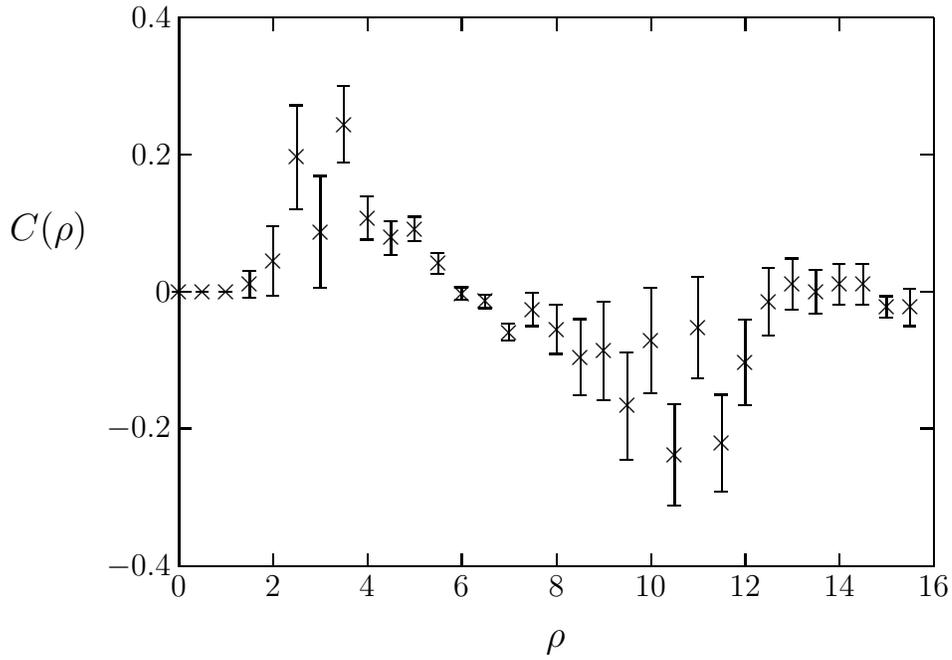

\begin	{figure}[p]
\begin	{center}
\leavevmode
% GNUPLOT: LaTeX picture
\setlength{\unitlength}{0.240900pt}
\ifx\plotpoint\undefined\newsavebox{\plotpoint}\fi
\begin{picture}(1200,900)(0,0)
\font\gnuplot=cmr10 at 12pt
\gnuplot
\sbox{\plotpoint}{\rule[-0.200pt]{0.400pt}{0.400pt}}%
\put(120.0,31.0){\rule[-0.200pt]{4.818pt}{0.400pt}}
\put(108,31){\makebox(0,0)[r]{{$-3$}}}
\put(1136.0,31.0){\rule[-0.200pt]{4.818pt}{0.400pt}}
\put(120.0,247.0){\rule[-0.200pt]{4.818pt}{0.400pt}}
\put(108,247){\makebox(0,0)[r]{{$-2$}}}
\put(1136.0,247.0){\rule[-0.200pt]{4.818pt}{0.400pt}}
\put(120.0,462.0){\rule[-0.200pt]{4.818pt}{0.400pt}}
\put(108,462){\makebox(0,0)[r]{{$-1$}}}
\put(1136.0,462.0){\rule[-0.200pt]{4.818pt}{0.400pt}}
\put(120.0,678.0){\rule[-0.200pt]{4.818pt}{0.400pt}}
\put(108,678){\makebox(0,0)[r]{{$0$}}}
\put(1136.0,678.0){\rule[-0.200pt]{4.818pt}{0.400pt}}
\put(120.0,893.0){\rule[-0.200pt]{4.818pt}{0.400pt}}
\put(108,893){\makebox(0,0)[r]{{$1$}}}
\put(1136.0,893.0){\rule[-0.200pt]{4.818pt}{0.400pt}}
\put(120.0,31.0){\rule[-0.200pt]{0.400pt}{4.818pt}}
\put(120,19){\makebox(0,0){\shortstack{\\ \\ \\ {$0$}}}}
\put(120.0,873.0){\rule[-0.200pt]{0.400pt}{4.818pt}}
\put(293.0,31.0){\rule[-0.200pt]{0.400pt}{4.818pt}}
\put(293,19){\makebox(0,0){\shortstack{\\ \\ \\ {$2$}}}}
\put(293.0,873.0){\rule[-0.200pt]{0.400pt}{4.818pt}}
\put(465.0,31.0){\rule[-0.200pt]{0.400pt}{4.818pt}}
\put(465,19){\makebox(0,0){\shortstack{\\ \\ \\ {$4$}}}}
\put(465.0,873.0){\rule[-0.200pt]{0.400pt}{4.818pt}}
\put(638.0,31.0){\rule[-0.200pt]{0.400pt}{4.818pt}}
\put(638,19){\makebox(0,0){\shortstack{\\ \\ \\ {$6$}}}}
\put(638.0,873.0){\rule[-0.200pt]{0.400pt}{4.818pt}}
\put(811.0,31.0){\rule[-0.200pt]{0.400pt}{4.818pt}}
\put(811,19){\makebox(0,0){\shortstack{\\ \\ \\ {$8$}}}}
\put(811.0,873.0){\rule[-0.200pt]{0.400pt}{4.818pt}}
\put(983.0,31.0){\rule[-0.200pt]{0.400pt}{4.818pt}}
\put(983,19){\makebox(0,0){\shortstack{\\ \\ \\ {$10$}}}}
\put(983.0,873.0){\rule[-0.200pt]{0.400pt}{4.818pt}}
\put(1156.0,31.0){\rule[-0.200pt]{0.400pt}{4.818pt}}
\put(1156,19){\makebox(0,0){\shortstack{\\ \\ \\ {$12$}}}}
\put(1156.0,873.0){\rule[-0.200pt]{0.400pt}{4.818pt}}
\put(120.0,31.0){\rule[-0.200pt]{249.572pt}{0.400pt}}
\put(1156.0,31.0){\rule[-0.200pt]{0.400pt}{207.656pt}}
\put(120.0,893.0){\rule[-0.200pt]{249.572pt}{0.400pt}}
\put(-168,558){\makebox(0,0){{\large{$N_{same}-N_{opp}$}}}}
\put(638,-89){\makebox(0,0){{\large{$\rho_{ref}$}}}}
\put(120.0,31.0){\rule[-0.200pt]{0.400pt}{207.656pt}}
\put(120,678){\makebox(0,0){$\times$}}
\put(163,678){\makebox(0,0){$\times$}}
\put(206,678){\makebox(0,0){$\times$}}
\put(250,662){\makebox(0,0){$\times$}}
\put(293,563){\makebox(0,0){$\times$}}
\put(336,707){\makebox(0,0){$\times$}}
\put(379,574){\makebox(0,0){$\times$}}
\put(422,664){\makebox(0,0){$\times$}}
\put(465,654){\makebox(0,0){$\times$}}
\put(509,582){\makebox(0,0){$\times$}}
\put(552,575){\makebox(0,0){$\times$}}
\put(595,552){\makebox(0,0){$\times$}}
\put(638,480){\makebox(0,0){$\times$}}
\put(681,470){\makebox(0,0){$\times$}}
\put(724,428){\makebox(0,0){$\times$}}
\put(768,396){\makebox(0,0){$\times$}}
\put(811,315){\makebox(0,0){$\times$}}
\put(854,253){\makebox(0,0){$\times$}}
\put(897,379){\makebox(0,0){$\times$}}
\put(940,170){\makebox(0,0){$\times$}}
\put(983,271){\makebox(0,0){$\times$}}
\put(1027,360){\makebox(0,0){$\times$}}
\put(1070,475){\makebox(0,0){$\times$}}
\put(1113,397){\makebox(0,0){$\times$}}
\put(1156,678){\makebox(0,0){$\times$}}
\put(120,678){\usebox{\plotpoint}}
\put(110.0,678.0){\rule[-0.200pt]{4.818pt}{0.400pt}}
\put(110.0,678.0){\rule[-0.200pt]{4.818pt}{0.400pt}}
\put(163,678){\usebox{\plotpoint}}
\put(153.0,678.0){\rule[-0.200pt]{4.818pt}{0.400pt}}
\put(153.0,678.0){\rule[-0.200pt]{4.818pt}{0.400pt}}
\put(206,678){\usebox{\plotpoint}}
\put(196.0,678.0){\rule[-0.200pt]{4.818pt}{0.400pt}}
\put(196.0,678.0){\rule[-0.200pt]{4.818pt}{0.400pt}}
\put(250.0,632.0){\rule[-0.200pt]{0.400pt}{14.695pt}}
\put(240.0,632.0){\rule[-0.200pt]{4.818pt}{0.400pt}}
\put(240.0,693.0){\rule[-0.200pt]{4.818pt}{0.400pt}}
\put(293.0,493.0){\rule[-0.200pt]{0.400pt}{33.967pt}}
\put(283.0,493.0){\rule[-0.200pt]{4.818pt}{0.400pt}}
\put(283.0,634.0){\rule[-0.200pt]{4.818pt}{0.400pt}}
\put(336.0,598.0){\rule[-0.200pt]{0.400pt}{52.275pt}}
\put(326.0,598.0){\rule[-0.200pt]{4.818pt}{0.400pt}}
\put(326.0,815.0){\rule[-0.200pt]{4.818pt}{0.400pt}}
\put(379.0,454.0){\rule[-0.200pt]{0.400pt}{57.816pt}}
\put(369.0,454.0){\rule[-0.200pt]{4.818pt}{0.400pt}}
\put(369.0,694.0){\rule[-0.200pt]{4.818pt}{0.400pt}}
\put(422.0,568.0){\rule[-0.200pt]{0.400pt}{46.494pt}}
\put(412.0,568.0){\rule[-0.200pt]{4.818pt}{0.400pt}}
\put(412.0,761.0){\rule[-0.200pt]{4.818pt}{0.400pt}}
\put(465.0,606.0){\rule[-0.200pt]{0.400pt}{22.885pt}}
\put(455.0,606.0){\rule[-0.200pt]{4.818pt}{0.400pt}}
\put(455.0,701.0){\rule[-0.200pt]{4.818pt}{0.400pt}}
\put(509.0,544.0){\rule[-0.200pt]{0.400pt}{18.067pt}}
\put(499.0,544.0){\rule[-0.200pt]{4.818pt}{0.400pt}}
\put(499.0,619.0){\rule[-0.200pt]{4.818pt}{0.400pt}}
\put(552.0,542.0){\rule[-0.200pt]{0.400pt}{15.899pt}}
\put(542.0,542.0){\rule[-0.200pt]{4.818pt}{0.400pt}}
\put(542.0,608.0){\rule[-0.200pt]{4.818pt}{0.400pt}}
\put(595.0,525.0){\rule[-0.200pt]{0.400pt}{12.768pt}}
\put(585.0,525.0){\rule[-0.200pt]{4.818pt}{0.400pt}}
\put(585.0,578.0){\rule[-0.200pt]{4.818pt}{0.400pt}}
\put(638.0,467.0){\rule[-0.200pt]{0.400pt}{6.263pt}}
\put(628.0,467.0){\rule[-0.200pt]{4.818pt}{0.400pt}}
\put(628.0,493.0){\rule[-0.200pt]{4.818pt}{0.400pt}}
\put(681.0,455.0){\rule[-0.200pt]{0.400pt}{6.986pt}}
\put(671.0,455.0){\rule[-0.200pt]{4.818pt}{0.400pt}}
\put(671.0,484.0){\rule[-0.200pt]{4.818pt}{0.400pt}}
\put(724.0,404.0){\rule[-0.200pt]{0.400pt}{11.804pt}}
\put(714.0,404.0){\rule[-0.200pt]{4.818pt}{0.400pt}}
\put(714.0,453.0){\rule[-0.200pt]{4.818pt}{0.400pt}}
\put(768.0,365.0){\rule[-0.200pt]{0.400pt}{14.936pt}}
\put(758.0,365.0){\rule[-0.200pt]{4.818pt}{0.400pt}}
\put(758.0,427.0){\rule[-0.200pt]{4.818pt}{0.400pt}}
\put(811.0,266.0){\rule[-0.200pt]{0.400pt}{23.367pt}}
\put(801.0,266.0){\rule[-0.200pt]{4.818pt}{0.400pt}}
\put(801.0,363.0){\rule[-0.200pt]{4.818pt}{0.400pt}}
\put(854.0,183.0){\rule[-0.200pt]{0.400pt}{33.485pt}}
\put(844.0,183.0){\rule[-0.200pt]{4.818pt}{0.400pt}}
\put(844.0,322.0){\rule[-0.200pt]{4.818pt}{0.400pt}}
\put(897.0,265.0){\rule[-0.200pt]{0.400pt}{54.925pt}}
\put(887.0,265.0){\rule[-0.200pt]{4.818pt}{0.400pt}}
\put(887.0,493.0){\rule[-0.200pt]{4.818pt}{0.400pt}}
\put(940.0,48.0){\rule[-0.200pt]{0.400pt}{58.780pt}}
\put(930.0,48.0){\rule[-0.200pt]{4.818pt}{0.400pt}}
\put(930.0,292.0){\rule[-0.200pt]{4.818pt}{0.400pt}}
\put(983.0,165.0){\rule[-0.200pt]{0.400pt}{51.312pt}}
\put(973.0,165.0){\rule[-0.200pt]{4.818pt}{0.400pt}}
\put(973.0,378.0){\rule[-0.200pt]{4.818pt}{0.400pt}}
\put(1027.0,248.0){\rule[-0.200pt]{0.400pt}{54.202pt}}
\put(1017.0,248.0){\rule[-0.200pt]{4.818pt}{0.400pt}}
\put(1017.0,473.0){\rule[-0.200pt]{4.818pt}{0.400pt}}
\put(1070.0,364.0){\rule[-0.200pt]{0.400pt}{53.480pt}}
\put(1060.0,364.0){\rule[-0.200pt]{4.818pt}{0.400pt}}
\put(1060.0,586.0){\rule[-0.200pt]{4.818pt}{0.400pt}}
\put(1113.0,282.0){\rule[-0.200pt]{0.400pt}{55.407pt}}
\put(1103.0,282.0){\rule[-0.200pt]{4.818pt}{0.400pt}}
\put(1103.0,512.0){\rule[-0.200pt]{4.818pt}{0.400pt}}
\put(1156,678){\usebox{\plotpoint}}
\put(1146.0,678.0){\rule[-0.200pt]{4.818pt}{0.400pt}}
\put(1146.0,678.0){\rule[-0.200pt]{4.818pt}{0.400pt}}
\end{picture}
\end	{center}
\vskip 0.15in
\caption{Net screening charge around a reference charge of
size $\rho_{ref}$: at $\beta=6.2$ after 23 cools.}
\label{fig-sameopp-nocut23}
\end 	{figure}
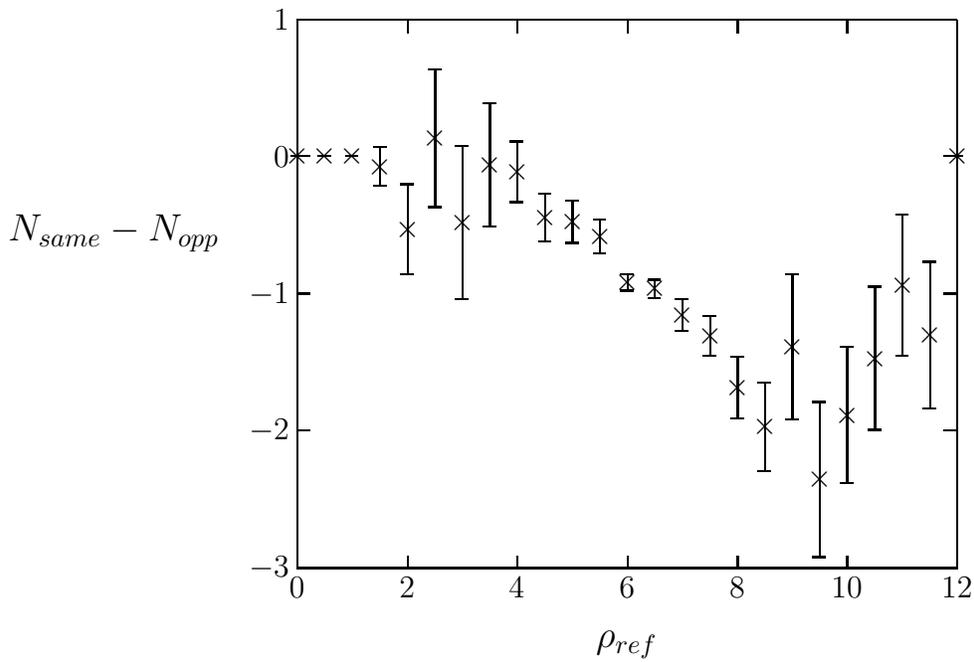

\begin	{figure}[p]
\begin	{center}
\leavevmode
% GNUPLOT: LaTeX picture
\setlength{\unitlength}{0.240900pt}
\ifx\plotpoint\undefined\newsavebox{\plotpoint}\fi
\begin{picture}(1200,900)(0,0)
\font\gnuplot=cmr10 at 12pt
\gnuplot
\sbox{\plotpoint}{\rule[-0.200pt]{0.400pt}{0.400pt}}%
\put(120.0,31.0){\rule[-0.200pt]{4.818pt}{0.400pt}}
\put(108,31){\makebox(0,0)[r]{{$-3$}}}
\put(1136.0,31.0){\rule[-0.200pt]{4.818pt}{0.400pt}}
\put(120.0,247.0){\rule[-0.200pt]{4.818pt}{0.400pt}}
\put(108,247){\makebox(0,0)[r]{{$-2$}}}
\put(1136.0,247.0){\rule[-0.200pt]{4.818pt}{0.400pt}}
\put(120.0,462.0){\rule[-0.200pt]{4.818pt}{0.400pt}}
\put(108,462){\makebox(0,0)[r]{{$-1$}}}
\put(1136.0,462.0){\rule[-0.200pt]{4.818pt}{0.400pt}}
\put(120.0,678.0){\rule[-0.200pt]{4.818pt}{0.400pt}}
\put(108,678){\makebox(0,0)[r]{{$0$}}}
\put(1136.0,678.0){\rule[-0.200pt]{4.818pt}{0.400pt}}
\put(120.0,893.0){\rule[-0.200pt]{4.818pt}{0.400pt}}
\put(108,893){\makebox(0,0)[r]{{$1$}}}
\put(1136.0,893.0){\rule[-0.200pt]{4.818pt}{0.400pt}}
\put(120.0,31.0){\rule[-0.200pt]{0.400pt}{4.818pt}}
\put(120,19){\makebox(0,0){\shortstack{\\ \\ \\ {$0$}}}}
\put(120.0,873.0){\rule[-0.200pt]{0.400pt}{4.818pt}}
\put(293.0,31.0){\rule[-0.200pt]{0.400pt}{4.818pt}}
\put(293,19){\makebox(0,0){\shortstack{\\ \\ \\ {$2$}}}}
\put(293.0,873.0){\rule[-0.200pt]{0.400pt}{4.818pt}}
\put(465.0,31.0){\rule[-0.200pt]{0.400pt}{4.818pt}}
\put(465,19){\makebox(0,0){\shortstack{\\ \\ \\ {$4$}}}}
\put(465.0,873.0){\rule[-0.200pt]{0.400pt}{4.818pt}}
\put(638.0,31.0){\rule[-0.200pt]{0.400pt}{4.818pt}}
\put(638,19){\makebox(0,0){\shortstack{\\ \\ \\ {$6$}}}}
\put(638.0,873.0){\rule[-0.200pt]{0.400pt}{4.818pt}}
\put(811.0,31.0){\rule[-0.200pt]{0.400pt}{4.818pt}}
\put(811,19){\makebox(0,0){\shortstack{\\ \\ \\ {$8$}}}}
\put(811.0,873.0){\rule[-0.200pt]{0.400pt}{4.818pt}}
\put(983.0,31.0){\rule[-0.200pt]{0.400pt}{4.818pt}}
\put(983,19){\makebox(0,0){\shortstack{\\ \\ \\ {$10$}}}}
\put(983.0,873.0){\rule[-0.200pt]{0.400pt}{4.818pt}}
\put(1156.0,31.0){\rule[-0.200pt]{0.400pt}{4.818pt}}
\put(1156,19){\makebox(0,0){\shortstack{\\ \\ \\ {$12$}}}}
\put(1156.0,873.0){\rule[-0.200pt]{0.400pt}{4.818pt}}
\put(120.0,31.0){\rule[-0.200pt]{249.572pt}{0.400pt}}
\put(1156.0,31.0){\rule[-0.200pt]{0.400pt}{207.656pt}}
\put(120.0,893.0){\rule[-0.200pt]{249.572pt}{0.400pt}}
\put(-168,558){\makebox(0,0){{\large{$N_{same}-N_{opp}$}}}}
\put(638,-89){\makebox(0,0){{\large{$\rho_{ref}$}}}}
\put(120.0,31.0){\rule[-0.200pt]{0.400pt}{207.656pt}}
\put(120,678){\makebox(0,0){$\times$}}
\put(163,678){\makebox(0,0){$\times$}}
\put(206,678){\makebox(0,0){$\times$}}
\put(250,678){\makebox(0,0){$\times$}}
\put(293,725){\makebox(0,0){$\times$}}
\put(336,701){\makebox(0,0){$\times$}}
\put(379,546){\makebox(0,0){$\times$}}
\put(422,500){\makebox(0,0){$\times$}}
\put(465,588){\makebox(0,0){$\times$}}
\put(509,567){\makebox(0,0){$\times$}}
\put(552,636){\makebox(0,0){$\times$}}
\put(595,474){\makebox(0,0){$\times$}}
\put(638,495){\makebox(0,0){$\times$}}
\put(681,519){\makebox(0,0){$\times$}}
\put(724,502){\makebox(0,0){$\times$}}
\put(768,503){\makebox(0,0){$\times$}}
\put(811,475){\makebox(0,0){$\times$}}
\put(854,439){\makebox(0,0){$\times$}}
\put(897,432){\makebox(0,0){$\times$}}
\put(940,443){\makebox(0,0){$\times$}}
\put(983,494){\makebox(0,0){$\times$}}
\put(1027,389){\makebox(0,0){$\times$}}
\put(1070,382){\makebox(0,0){$\times$}}
\put(1113,626){\makebox(0,0){$\times$}}
\put(1156,678){\makebox(0,0){$\times$}}
\put(120,678){\usebox{\plotpoint}}
\put(110.0,678.0){\rule[-0.200pt]{4.818pt}{0.400pt}}
\put(110.0,678.0){\rule[-0.200pt]{4.818pt}{0.400pt}}
\put(163,678){\usebox{\plotpoint}}
\put(153.0,678.0){\rule[-0.200pt]{4.818pt}{0.400pt}}
\put(153.0,678.0){\rule[-0.200pt]{4.818pt}{0.400pt}}
\put(206,678){\usebox{\plotpoint}}
\put(196.0,678.0){\rule[-0.200pt]{4.818pt}{0.400pt}}
\put(196.0,678.0){\rule[-0.200pt]{4.818pt}{0.400pt}}
\put(250,678){\usebox{\plotpoint}}
\put(240.0,678.0){\rule[-0.200pt]{4.818pt}{0.400pt}}
\put(240.0,678.0){\rule[-0.200pt]{4.818pt}{0.400pt}}
\put(293.0,689.0){\rule[-0.200pt]{0.400pt}{17.345pt}}
\put(283.0,689.0){\rule[-0.200pt]{4.818pt}{0.400pt}}
\put(283.0,761.0){\rule[-0.200pt]{4.818pt}{0.400pt}}
\put(336.0,643.0){\rule[-0.200pt]{0.400pt}{27.944pt}}
\put(326.0,643.0){\rule[-0.200pt]{4.818pt}{0.400pt}}
\put(326.0,759.0){\rule[-0.200pt]{4.818pt}{0.400pt}}
\put(379.0,481.0){\rule[-0.200pt]{0.400pt}{31.076pt}}
\put(369.0,481.0){\rule[-0.200pt]{4.818pt}{0.400pt}}
\put(369.0,610.0){\rule[-0.200pt]{4.818pt}{0.400pt}}
\put(422.0,427.0){\rule[-0.200pt]{0.400pt}{35.412pt}}
\put(412.0,427.0){\rule[-0.200pt]{4.818pt}{0.400pt}}
\put(412.0,574.0){\rule[-0.200pt]{4.818pt}{0.400pt}}
\put(465.0,521.0){\rule[-0.200pt]{0.400pt}{32.281pt}}
\put(455.0,521.0){\rule[-0.200pt]{4.818pt}{0.400pt}}
\put(455.0,655.0){\rule[-0.200pt]{4.818pt}{0.400pt}}
\put(509.0,504.0){\rule[-0.200pt]{0.400pt}{30.353pt}}
\put(499.0,504.0){\rule[-0.200pt]{4.818pt}{0.400pt}}
\put(499.0,630.0){\rule[-0.200pt]{4.818pt}{0.400pt}}
\put(552.0,584.0){\rule[-0.200pt]{0.400pt}{25.294pt}}
\put(542.0,584.0){\rule[-0.200pt]{4.818pt}{0.400pt}}
\put(542.0,689.0){\rule[-0.200pt]{4.818pt}{0.400pt}}
\put(595.0,438.0){\rule[-0.200pt]{0.400pt}{17.104pt}}
\put(585.0,438.0){\rule[-0.200pt]{4.818pt}{0.400pt}}
\put(585.0,509.0){\rule[-0.200pt]{4.818pt}{0.400pt}}
\put(638.0,468.0){\rule[-0.200pt]{0.400pt}{13.249pt}}
\put(628.0,468.0){\rule[-0.200pt]{4.818pt}{0.400pt}}
\put(628.0,523.0){\rule[-0.200pt]{4.818pt}{0.400pt}}
\put(681.0,490.0){\rule[-0.200pt]{0.400pt}{14.213pt}}
\put(671.0,490.0){\rule[-0.200pt]{4.818pt}{0.400pt}}
\put(671.0,549.0){\rule[-0.200pt]{4.818pt}{0.400pt}}
\put(724.0,485.0){\rule[-0.200pt]{0.400pt}{7.950pt}}
\put(714.0,485.0){\rule[-0.200pt]{4.818pt}{0.400pt}}
\put(714.0,518.0){\rule[-0.200pt]{4.818pt}{0.400pt}}
\put(768.0,485.0){\rule[-0.200pt]{0.400pt}{8.431pt}}
\put(758.0,485.0){\rule[-0.200pt]{4.818pt}{0.400pt}}
\put(758.0,520.0){\rule[-0.200pt]{4.818pt}{0.400pt}}
\put(811.0,453.0){\rule[-0.200pt]{0.400pt}{10.359pt}}
\put(801.0,453.0){\rule[-0.200pt]{4.818pt}{0.400pt}}
\put(801.0,496.0){\rule[-0.200pt]{4.818pt}{0.400pt}}
\put(854.0,412.0){\rule[-0.200pt]{0.400pt}{12.768pt}}
\put(844.0,412.0){\rule[-0.200pt]{4.818pt}{0.400pt}}
\put(844.0,465.0){\rule[-0.200pt]{4.818pt}{0.400pt}}
\put(897.0,398.0){\rule[-0.200pt]{0.400pt}{16.622pt}}
\put(887.0,398.0){\rule[-0.200pt]{4.818pt}{0.400pt}}
\put(887.0,467.0){\rule[-0.200pt]{4.818pt}{0.400pt}}
\put(940.0,401.0){\rule[-0.200pt]{0.400pt}{20.476pt}}
\put(930.0,401.0){\rule[-0.200pt]{4.818pt}{0.400pt}}
\put(930.0,486.0){\rule[-0.200pt]{4.818pt}{0.400pt}}
\put(983.0,419.0){\rule[-0.200pt]{0.400pt}{36.376pt}}
\put(973.0,419.0){\rule[-0.200pt]{4.818pt}{0.400pt}}
\put(973.0,570.0){\rule[-0.200pt]{4.818pt}{0.400pt}}
\put(1027.0,316.0){\rule[-0.200pt]{0.400pt}{35.412pt}}
\put(1017.0,316.0){\rule[-0.200pt]{4.818pt}{0.400pt}}
\put(1017.0,463.0){\rule[-0.200pt]{4.818pt}{0.400pt}}
\put(1070.0,313.0){\rule[-0.200pt]{0.400pt}{33.485pt}}
\put(1060.0,313.0){\rule[-0.200pt]{4.818pt}{0.400pt}}
\put(1060.0,452.0){\rule[-0.200pt]{4.818pt}{0.400pt}}
\put(1113.0,566.0){\rule[-0.200pt]{0.400pt}{28.908pt}}
\put(1103.0,566.0){\rule[-0.200pt]{4.818pt}{0.400pt}}
\put(1103.0,686.0){\rule[-0.200pt]{4.818pt}{0.400pt}}
\put(1156,678){\usebox{\plotpoint}}
\put(1146.0,678.0){\rule[-0.200pt]{4.818pt}{0.400pt}}
\put(1146.0,678.0){\rule[-0.200pt]{4.818pt}{0.400pt}}
\end{picture}
\end	{center}
\vskip 0.15in
\caption{As in Fig~\ref{fig-sameopp-nocut23} but after 46 cools.}
\label{fig-sameopp-nocut46}
\end 	{figure}
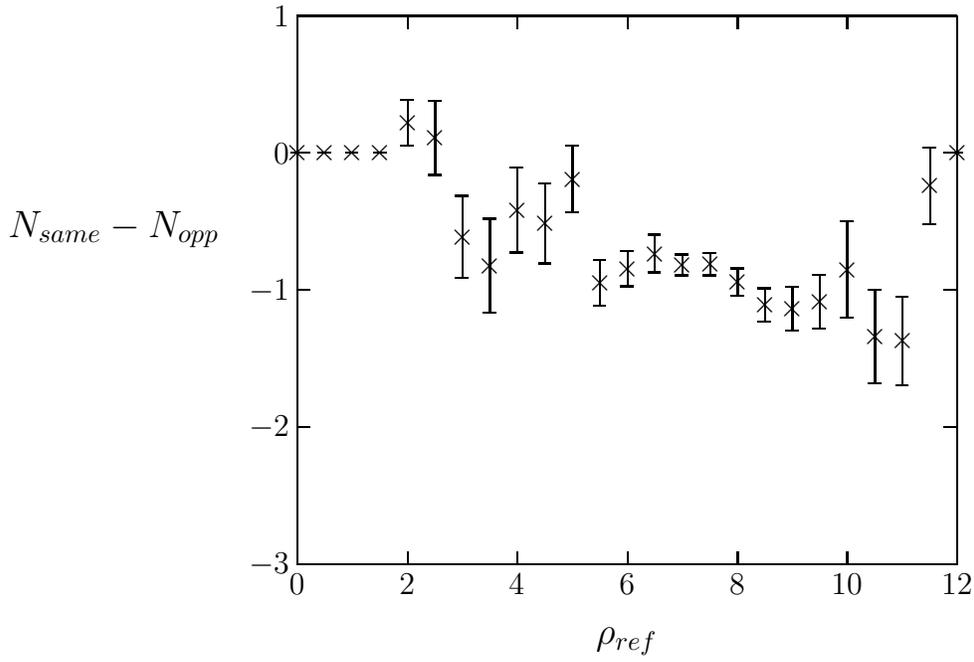

\begin	{figure}[p]
\begin	{center}
\leavevmode
\input	{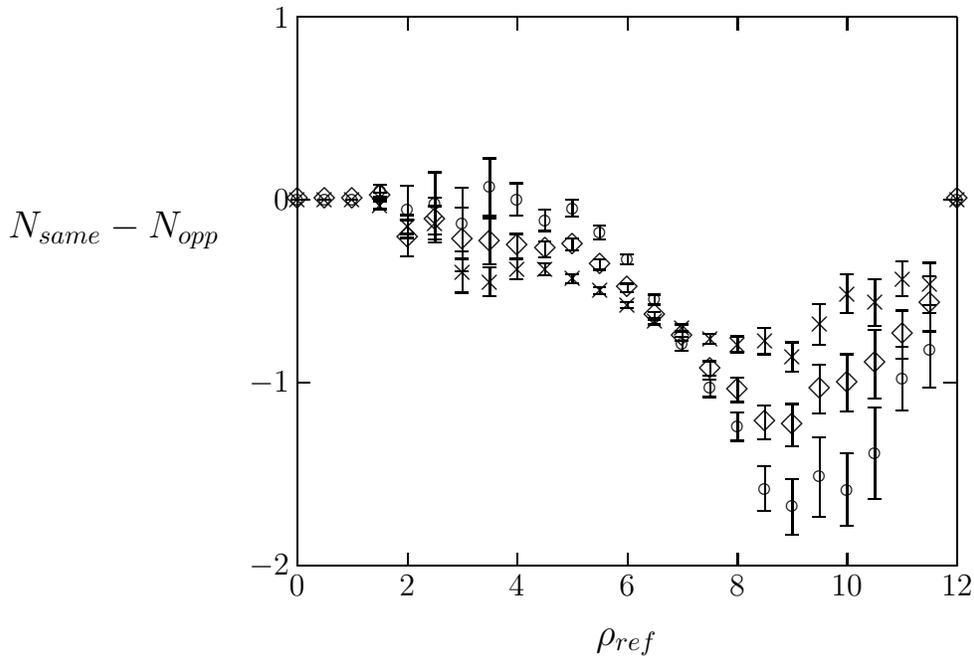}
\end	{center}
\vskip 0.15in
\caption{As in Fig~\ref{fig-sameopp-nocut23},
but only including screening charges with $\rho<6$ and 
within a distance $R=7(\times),8(\diamond)$ or $9(\circ)$ 
of the reference charge.}
\label{fig-sameopp-smallcut}
\end 	{figure}

\clearpage

\begin	{figure}[p]
\begin	{center}
\leavevmode
\input	{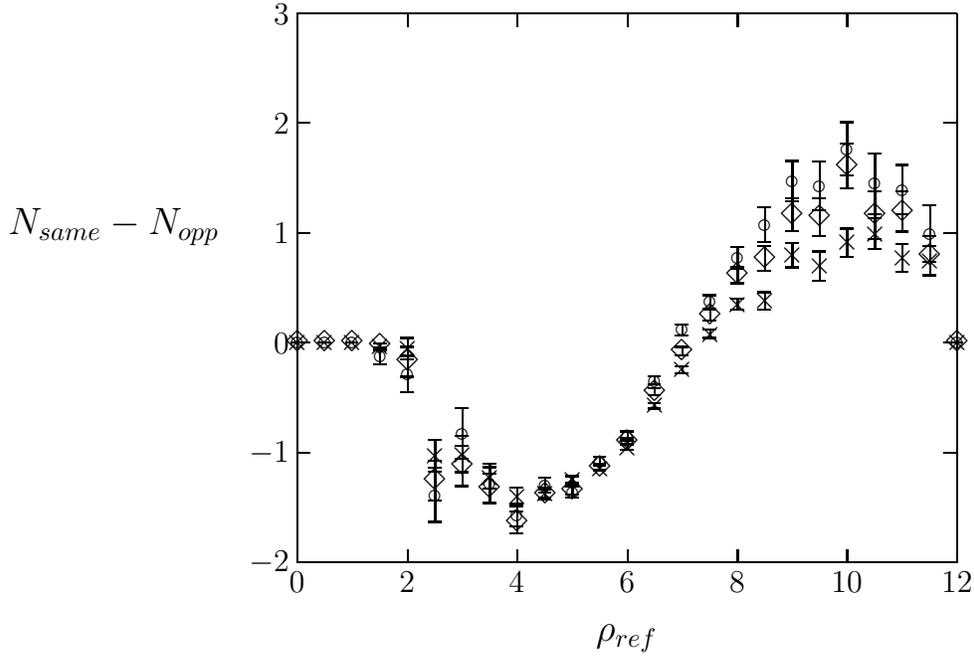}
\end	{center}
\vskip 0.15in
\caption{As in Fig~\ref{fig-sameopp-smallcut},
but only including charges with $\rho>6$.}
\label{fig-sameopp-largecut}
\end 	{figure}

\end{document}